\documentclass[12pt]{article}
\pdfoutput=1
\usepackage{Max}
\usepackage{subcaption}

\title{K3 metrics}
\author[1]{Shamit Kachru\thanks{skachru@stanford.edu}}
\author[2]{Arnav Tripathy\thanks{tripathy@math.harvard.edu}}
\author[3]{Max Zimet\thanks{mzimet@fas.harvard.edu}}
\affil[1]{Stanford Institute for Theoretical Physics,

Stanford University, Stanford, CA 94305 USA

~}
\affil[2]{Department of Mathematics,

Harvard University, Cambridge, MA 02138 USA

~}
\affil[3]{Black Hole Initiative and Department of Physics,

Harvard University, Cambridge, MA 02138 USA}

\date{}

\begin{document}

\maketitle

\begin{abstract}
We provide an explicit construction of Ricci-flat K3 metrics. It employs the technology of D-geometry, which in the case of interest is equivalent to a hyper-K\"ahler quotient. We relate it to the construction of \cite{mz:k3}, and in particular show that it contains the solution to the BPS state counting problem (that of computing the BPS index of a heterotic little string theory compactified on $T^2$) discussed therein, which is the data needed for this second construction of K3 metrics.

\end{abstract}

\newpage
\tableofcontents
\hypersetup{linkcolor=blue}

\newpage

\section{Introduction}

Compactifications on Calabi-Yau manifolds have been a central testing ground for string theory since their discovery in \cite{w:cy}.  The crucial ingredient in their existence and
stability is Yau's proof \cite{yau:CY} of the Calabi conjecture \cite{calabi:conjecture}, demonstrating that there exists a unique Ricci-flat 
metric on a K\"ahler manifold with vanishing first Chern class for each choice of the cohomology class of the K\"ahler form.

Through their role as 4d $\N=1$ ($\N=2$) supersymmetric compactifications in heterotic (type II) string theories, Calabi-Yau threefolds
have played important roles in studies of string duality and in attempts to use string theory to model elementary particle physics, cosmology, and black 
hole physics \cite{GSW2,greene:cy,BBS}.  

A particularly important role has been played by what is in some sense the simplest non-trivial (i.e., non-toroidal) Calabi-Yau
manifold: the K3 surface.  This is the only compact Calabi-Yau with generic ($SU(2)$) holonomy in two complex dimensions.  It appears,
singly or as an important factor in a $K3 \times T^2$ compactification, in numerous intricate and beautiful constructions in string duality \cite{Aspinwall:1996mn}.  

However, in none of these constructions -- whether based on Calabi-Yau threefolds or their simpler
K3 cousin -- has the main defining datum, namely the Ricci-flat metric proven to exist by Yau's theorem, been determined in an analytical manner.  Ingenious and
practically useful numerical approximation schemes do exist \cite{headrick:numericK3,donaldson:numericK3,douglas:cy3Numerics}.  

In this paper, we study two (dual) analytical expressions for Ricci-flat metrics on K3 surfaces, one of which is completely explicit.  We do this as follows.
In \cite{mz:k3}, we demonstrated that if one could determine the spectrum of BPS states of a certain compactified
little string theory, one would obtain an analytical formula for a Ricci-flat K3 metric expressed in terms of (integral) BPS state counts.
Roughly, the K3 surface arises as the Coulomb branch moduli space of the little string theory compactified to three space-time dimensions
on $T^2 \times S^1$, and the BPS states of the 4d theory obtained from $T^2$ compactification determine the desired metric
(as they provide BPS instantons correcting the 3d metric by traversing the $S^1$).
This is 
a particular application of a formalism developed in the context of 4d $\N=2$ supersymmetric field theory by Gaiotto, Moore, and Neitzke
\cite{GMN:walls}.

Here, we advance the story of \cite{mz:k3}: we provide a means of determining the relevant BPS state counts which determine the K3 metrics for a variety
of smooth K3 surfaces.  We do this by using duality with a different picture.  In this dual picture, the K3 surface arises as the {\it Higgs~branch}
moduli space of a (mirror dual) D-brane field theory.  By generalizing Taylor's construction of D-brane field theories with compact
moduli spaces first discussed in the context of M(atrix) theory in \cite{wati:dBraneT}, we obtain (building on similar work in
\cite{ho:noncomm,waldram:lol,greene:lol}) an infinite-dimensional hyper-K\"ahler quotient construction of this Higgs branch, which
yields the K3 metrics.  Matching the expansion of the K3 metrics obtained in this picture with the expansion of the Coulomb branch
metrics in terms of BPS instantons in our first picture should both determine the BPS state counts and yield a formula for the
metric in the picture of \cite{mz:k3}. We explicitly demonstrate this at first non-trivial order in a perturbative expansion, which allows us to extract part of the BPS spectrum of the little string theory. The BPS state counts so obtained have a clear interpretation in terms of
known results. We also explain how to continue this procedure to higher orders in order to extract the full spectrum; we relegate the implementation of this to \cite{mz:K3HK2}, as this paper is already lengthy enough.

Explicit analytical checks (in a controlled expansion) indicate that the K3 metrics so obtained are, indeed, Ricci-flat.

The organization of this paper is as follows.  In \S\ref{sec:3dmirror}, we introduce the two duality frames we consider by describing branes in various string and M-theory compactifications.  In \S\ref{sec:hk}, we slowly work up to the hyper-K\"ahler quotient construction of
K3, starting with simpler examples.  In \S\ref{sec:coulomb}, we review the method of \cite{mz:k3} for using little string theories in order to describe the geometry of K3 surfaces, warm up by considering an auxiliary field theory problem, and then solve for part of the BPS spectra governing a family of smooth K3 metrics by comparing
this picture to that of \S\ref{sec:hk}.  We describe related physics and mathematics problems which provide a number of ways of studying the BPS spectra of interest
in \S\ref{sec:counting}, explain in \S\ref{sec:moreStates} from a number of perspectives how we know that we have not yet found the full spectrum, and conclude with directions for future research in \S\ref{sec:conclusion}.

\section{Compactified little string theories and 3d mirror symmetry} \label{sec:3dmirror}

A starring role in this paper is played by compactifications of the $\N=(1,0)$ little string theories arising on heterotic NS5-branes.
These theories have different avatars.
One of the incarnations used in \cite{mz:k3} was that of the $SO(32)$ heterotic NS5-brane compactified on $T^3$.  In a limit where
$g_s \to 0$, one obtains a decoupled little string theory on the worldvolume \cite{s:hlst}.  By using string dualities and doing suitable
duality chasing, one can obtain dual
descriptions of this object.  For instance, by S-dualizing one can view it in terms of the D5-brane of type I string theory on $T^3$.  
After T-dualizing the circles of the $T^3$, one obtains a picture in terms of a D2-brane
probing the $T^3/Z_2$ orientifold.  Its moduli space of vacua 
has a Coulomb branch consisting of the $T^3/Z_2$ together with the (dual) photon.  The total space is a K3 surface; indeed, this manifold is visible in the lift of this orientifold to M-theory, where one finds an M2-brane probing K3 \cite{s:K3,sw:3d,intriligator:compact}. At generic points in the parameter space of the little string theory (i.e., the moduli space of the heterotic string theory), this K3 surface is smooth.

We can specialize to the case that the $T^3$ geometry takes the form of a 
 $T^2 \times S^1$ where the circle has some large radius $R$.
It was explained in \cite{mz:k3} that in this limit one can obtain a controlled expansion for the K3
metric (expanding around so-called ``semi-flat metrics").  The BPS states of the little string theory on $T^2$ give instanton effects
correcting the Coulomb branch metric in 3d, and at large radius $R$ these manifest in an expansion in $e^{-mR}$ (with $m$ the mass
of the relevant BPS states).
While summing up these instanton effects directly is beyond reach, the formalism of \cite{GMN:walls} lets us exchange this problem
for the easier problem of finding expectation values of supersymmetric Wilson-'t Hooft lines, which are piecewise holomorphic functions on the Coulomb branch
of vacua with known asymptotics. In this way, by exactly determining certain canonical coordinates on the 3d Coulomb branch, we are able to solve for the
metric.  The key non-trivial piece of data we require is the spectrum of BPS states of the little string theory, which determines the discontinuities of these coordinates.

To fix this, we will use a second, purely classical picture.  One can realize K3 as a moduli space of a D2-brane theory in another
way: by considering a D2-brane probe in type IIA string theory compactified on K3.
We specialize to the situation where the K3 geometry is close to an orbifold point; for definiteness we consider 
$T^4/Z_2$ here.  Then, we can obtain a gauge theory description of the low energy physics \cite{ho:noncomm,greene:lol,waldram:lol}, using
the theory of D-branes at orbifold singularities \cite{moore:ALEinst}.  This description is particularly useful because the 
K3 surface is arising as a Higgs branch, whose geometry enjoys a non-renormalisation theorem which implies that it is governed by a hyper-K\"ahler quotient. We extract from this description an explicit expansion of the metric around the orbifold point.
The hyper-K\"ahler quotient construction required is slightly unusual; it must incorporate an infinite-dimensional gauge group to
maintain knowledge of the compact moduli space, as discussed first in \cite{wati:dBraneT}.
We develop this line of thinking to the point where it provides useful expressions for the K3 metric in the next section.

This story reconnects with our first little string picture in a simple way.  
Happily, the expansions around the $T^4/Z_2$ orbifold locus and the large $R$ (``semi-flat") limit have overlapping regions of validity.  We
demonstrate in \S\ref{sec:coulomb}, after reviewing \cite{mz:k3}, that the hyper-K\"ahler
quotient construction around the orbifold point, which we will also call the `Higgs branch construction,' in fact gives us enough information to reconstruct part of the BPS spectrum
of the compactified little string theory (on a particular locus in parameter space, but everywhere in moduli space). (Indeed, it contains the entire BPS spectrum; we hope to extract this in \cite{mz:K3HK2}.) This provides us with a means of completing the specification of a second `Coulomb branch' expansion of
the K3 metric.  At the order to which we work in this paper (and presumably at all orders), the formulae of the two descriptions perfectly
match, and are in fact related by Poisson resummation.\footnote{The final story is somewhat reminiscent of closed string mirror symmetry, where instanton effects on one side of the duality are implicitly summed
up by finding special coordinates via a classical period computation in the mirror dual \cite{cogp}.}

We note that these two pictures can be related by 3d mirror symmetry \cite{s:3dmirror}, at least at the level of brane
probe field theories \cite{porrati:mirror}.  For instance, probing an O6-plane with a D2-brane in the $T^3/Z_2$ orientifold yields an 
$SU(2)$ gauge theory (with fundamental flavors arising if there are D6-branes coincident with the O6-plane).  The low energy solution
of this theory is visible in the lift of this orientifold to M-theory on K3; reducing on a transverse circle gives the IIA D2-brane
picture where the K3 surface arises as a Higgs branch moduli space, and this yields a mirror description of the 3d field theory physics.

An important subtlety of 3d mirror symmetry is also evident in the relationship between the D-brane mirrors. The perturbative IIA $T^4/Z_2$ 
orbifold has discrete $B$-flux threading the 16 collapsed two-cycles at the $Z_2$ fixed points \cite{aspinwall:theta}.  This is important in using the
orbifold technology of \cite{moore:ALEinst} to determine the D-brane worldvolume theory of a brane.
On the other hand, M-theory on $T^4/Z_2$ has an $A_1$ singularity at each of the fixed points,
and the physics is genuinely singular (in the sense that an enhanced $SU(2)$ gauge factor is associated with each $A_1$ singularity and appears as a global symmetry on a probe M2-brane).  
The relationship between these compactifications can be understood as follows: in M-theory on $T^4/Z_2\times S^1$, the $B$-flux is described by a three-form $C$-flux which dilutes away as the M-theory circle decompactifies, and so in this limit the gauge symmetry enhances.
This physics plays well with the relevant field theory mirror symmetry \cite{porrati:mirror}:
in one of the descriptions, there is an emergent enhanced $SU(2)$ global symmetry
in the IR, while the $SU(2)$ global symmetry is evident in the UV physics of the mirror description (here, captured by the physics of
a D2-brane probing an O6-plane with two coincident D6-branes).\footnote{More precisely, this $SU(2)$ $N_f=2$ theory has a Coulomb branch given by $(\RR^3\times S^1)/Z_2$, and if one stays near one of the fixed points and takes the IR limit then this becomes $\RR^4/Z_2$ and the mirror is given by the $U(1)$ $N_f=2$ theory. The former theory has a $\Spin(4)\cong SU(2)\times SU(2)$ flavor symmetry, where each $SU(2)$ is associated to a $Z_2$ fixed point in the IR, whereas the latter only has a $U(1)\subset SU(2)$ subgroup (associated to the symmetry of shifting the dual photon) in the UV.}

Importantly, the metric in the D2-brane probe theory on K3 is independent of the radius of the M-theory circle because of a
non-renormalization theorem \cite{s:moduliQCD} -- the radius enters in a background vector multiplet. We can therefore use D-brane technology to compute this Higgs branch metric
perturbatively near the IIA orbifold point and use that data to determine the BPS invariants we require in the little string computations.

\section{K3 as a hyper-K\"ahler quotient} \label{sec:hk}

In this section, we provide the `Higgs branch' construction of K3 surfaces via infinite-dimensional hyper-K\"ahler quotients. (As explained in \S9.3 of \cite{hitchin:HK}, compact manifolds cannot be realized as finite-dimensional hyper-K\"ahler quotients of vector spaces.) We will build up to it by warming up with a number of examples, each of which shares some features with the main objective.

\subsection{$\Sym^N \CC^2$} \label{sec:hkStart}

We begin by reviewing the hyper-K\"ahler quotient construction \cite{hitchin:hkSUSY}, as applied to maximally supersymmetric Yang-Mills theory. For definiteness, we consider the case of 3d $\N=8$ $U(N)$ gauge theory. Thought of as a 3d $\N=4$ theory, the matter content consists of a vector multiplet and an adjoint hypermultiplet. We denote the $\N=2$ adjoint chiral multiplet in the former by $\Phi$ and the chirals in the hypermultiplet by $U,V$. The moduli space is $\Sym^N (\RR^7\times S^1)$, as is clear from the fact that this gauge theory describes the worldvolume of $N$ parallel D2-branes. (The circle is the M-theory circle, which is parametrized in the gauge theory by the dual photon.) However, by focusing on the subset where the real scalar and the dual photon in the vector multiplet and the complex scalar in $\Phi$ vanish, one obtains the `Higgs branch' $\Sym^N \CC^2$ where only $U,V$ assume vacuum expectation values. We will explain how the gauge theory naturally constructs this space as a hyper-K\"ahler quotient of $\CC^{2N^2}$ (parametrized by $U,V$) by $U(N)$. Physically, this procedure simply consists of finding the classical moduli space of the gauge theory with infinite Yang-Mills coupling $g$ (i.e., no $\Tr F^2$ kinetic term, so the gauge field is not dynamical, and -- thanks to supersymmetry -- an infinitely large superpotential). A non-renormalization theorem \cite{s:moduliQCD} protects the result from quantum corrections.

Since we are interested in vacua with $\Phi=0$, the superpotential $W=g\Tr \Phi[U,V]$ simply yields
\be \partial_{\Phi^A} W = g \Tr T_A [U,V] = 0 \ . \ee
Here, $A$ is a $U(N)$ adjoint index, i.e. $\Phi = \Phi^A T_A$, where $T_A$ are Hermitian generators of $\mf{u}(N)$, normalized so that $\Tr T_A T_B=\delta_{AB}$. We thus find that
\be \mu_+ \equiv \mu_+^A T_A \equiv -2[U,V] = 0 \ , \ee
where the notation $\mu_+$ indicates that this is the holomorphic moment map of the hyper-K\"ahler quotient. We denote the anti-holomorphic moment map by
\be \mu_-=\mu_+^\dagger = \mu_-^A T_A \ , \quad \mu_-^A = (\mu_+^A)^* \ . \ee
The D-term equation is
\be \mu_\RR \equiv \mu_\RR^A T_A \equiv [U,U^\dagger] + [V, V^\dagger] = 0 \ ; \ee
we call this the real moment map. Gauge invariance allows us to add a complex FI parameter $\xi_+$ to the holomorphic moment map (i.e., to add $g\xi_+\Tr\Phi$ to the superpotential) and a real FI parameter $\xi_\RR$ to the real moment map, but these lead to the elimination of all supersymmetric vacua, since commutators are traceless. Lastly, it will be useful to establish notation for the action of $U(N)$ on $\CC^{2N^2}$. The adjoint action
\be \delta_A U = i \epsilon [T_A,U] = -\epsilon f_{AB}{}^C U^B T_C \ , \ee
where $f_{AB}{}^C$ are the structure constants, defined by $[T_A,T_B]=if_{AB}{}^C T_C$, yields the following Killing vector field associated to $T_A$:
\be k_A = -f_{AB}{}^C(U^B\partial_{U^C} + U^{\dagger B} \partial_{U^{\dagger C}} + V^B \partial_{V^C} + V^{\dagger B}\partial_{V^{\dagger C}}) \ . \ee
The moment maps are all equivariant, i.e. they also transform in the adjoint representation of $U(N)$:
\be k_A\, \mu = - f_{AB}{}^C \mu^B T_C \ ; \ee
here, $\mu$ is any of the moment maps.

The moduli space $M$ is simply the quotient by the $U(N)$ gauge group of the submanifold $\tilde M$ of $\CC^{2N^2}$ defined by $\mu_+=\mu_\RR=0$. Assuming that $\mu_+=0$, $U$ and $V$ may be simultaneously unitarily upper triangulized, thanks to the Schur decomposition. Writing $U=PAP^\dagger$ and $V=PBP^\dagger$, where $P$ is unitary and $A,B$ are upper triangular, the real moment map equation becomes
\be [A,A^\dagger] + [B,B^\dagger] = 0 \ . \ee
In particular,
\be e_i^T([A,A^\dagger] + [B,B^\dagger])e_i = \sum_{k>i}(|A_{ik}|^2+|B_{ik}|^2) - \sum_{k<i}(|A_{ki}|^2+|B_{ki}|^2) = 0 \ , \ee
where $e_i$ is the $i$-th vector in the standard basis, allows one to inductively reason that $A$ and $B$ are diagonal. So, $U$ and $V$ are actually simultaneously unitarily diagonalizable. We then fix most of the $U(N)$ gauge freedom by diagonalizing $U$ and $V$; what remains is the $S_N$ Weyl group of $U(N)$, which conjugates diagonal matrices to diagonal matrices, i.e. it reorders the eigenvalues in the diagonalized matrices. We thus see that $M$ is $\Sym^N \CC^2$, as promised. That the moduli space is parametrized by the eigenvalues of $U$ and $V$ is a standard aspect of D-brane probes.

Instead of imposing $\mu_\RR=0$ and quotienting by $U(N)$, we could have simply quotiented by the complexified group $GL(N,\CC)$. (This is natural in superspace, as gauge transformations are parametrized by chiral multiplets.) In the present example, this (along with $\mu_+=0$) would again allow us to simultaneously diagonalize $U$ and $V$, assuming that they are each diagonalizable. However, we will generally eschew this approach in this paper, as it entails some subtleties. Namely, before quotienting one must restrict to the open subset of `stable points,' i.e., those points whose $GL(N,\CC)$ orbits meet $\mu_\RR^{-1}(0)$. For example, in the present example this is necessary to deal with non-diagonalizable matrices. Introducing FI parameters in the real moment map requires care in this approach.

We now explain how the hyper-K\"ahler structure on $\CC^{2N^2}$ descends to $M$. Ignoring points of $\tilde M$ with non-trivial stabilizers, which are responsible for singularities in $M$, it is the case that $\tilde M$ is a principal $U(N)$-bundle over $M$. We denote the projection map of this bundle, or equivalently the quotient map, by $\pi : \tilde M\to M$. As above, we fix our gauge by specifying a section $s: M\to \tilde M$ of this bundle, i.e. an embedding of the quotient manifold into $\tilde M$. This allows us to parametrize $M$ using coordinates $t^i$ on $\tilde M$. There is a natural adjoint-valued 1-form $\theta^A$ on $\tilde M$:
\be \theta_i^A = \tilde g_{ij} k^j_B H^{AB} \ . \label{eq:connection} \ee
Here, $\tilde g_{ij}$ is the metric induced on $\tilde M$ by pulling back via the inclusion map $i:\tilde M\hookrightarrow \CC^{2N^2}$ and $H^{AB}$ is the inverse of (the restriction to $\tilde M$ of) $\tilde g_{ij}k^i_A k^j_B$. This provides what mathematicians call a principal connection on the bundle $\tilde M\to M$. Physicists will likely instead be more familiar with the gauge-dependent connection 1-form $s^*\theta^A$. Indeed, one way to see that \eqref{eq:connection} is natural is that the equation of motion of the gauge field sets it equal to the pullback to spacetime of $s^*\theta^A$.\footnote{That is, we schematically have $A_\mu = f(t) \partial_\mu t$. This does not spontaneously break Lorentz invariance, as $\avg{A_\mu}=0$. But, as we explain around \eqref{eq:gCorrected}, substituting this expression for $A_\mu$ back into the action has an important effect.} Mathematically, the connection is related to the projection of tangent vectors to $\tilde M$ to tangent vectors to $M$. For, $T\tilde M$ splits as the direct sum of the `vertical bundle' $V\tilde M$ spanned by the $k_A$ and its orthogonal complement, the `horizontal bundle' $H\tilde M$, and $TM$ is identified with the latter by the isomorphism $\pi_*|_{H\tilde M}:H\tilde M\to TM$; we henceforth abbreviate this by defining $p=\pi_*|_{H\tilde M}$. So, we see that the operations of projection to the vertical and horizontal subspaces are quite important; these are given, respectively, by
\be P_V(\partial_n) = H^{AB}\tilde g_{in} k^i_A k^j_B \partial_j = \theta^A_n k_A \ , \quad P_H = 1-P_V \ , \ee
where $1$ is the identity.

This is all the background that we need to state the induced metric from the hyper-K\"ahler quotient, but of course there is also a whole $\PP^1$-worth of complex structures and compatible K\"ahler forms. This is intimately tied to the following $SU(2)$ R-symmetry of the Yang-Mills theory:
\be \column{V^\dagger}{U}\mapsto e^{i \theta \, b\cdot \sigma}\column{V^\dagger}{U}\ , \quad \twoMatrix{\mu_\RR}{\mu_-}{\mu_+}{-\mu_\RR}\mapsto e^{i \theta \, b\cdot\sigma}\twoMatrix{\mu_\RR}{\mu_-}{\mu_+}{-\mu_\RR}e^{-i\theta \, b\cdot\sigma} \ . \label{eq:rSymm} \ee
Here, $\theta$ is an angle, $b$ is a unit vector in $\RR^3$, and $\sigma$ is the vector of Pauli matrices. Another way of thinking about this symmetry is to note that, assuming the on-shell condition $\mu_+=0$, we can reformulate the real moment map equation as
\be [U+i\zeta V^\dagger, U^\dagger + i\zeta^{-1} V] = 0 \label{eq:allMoments} \ee
for all $\zeta\in \CC^\times$. Additionally, $\mu_\pm=0$ arise as the limits of this equation as $\zeta\to 0,\infty$. So, \eqref{eq:allMoments} is equivalent to $\mu_+=\mu_\RR=0$ and makes clear that $(U+i\zeta V^\dagger, U^\dagger + i\zeta^{-1} V)$ are the holomorphic coordinates in the complex structure associated to $\zeta\in \PP^1$. We now show that the $SU(2)$ R-symmetry rotates the $\PP^1$ of complex structures of $\CC^{2N^2}$ by showing that it maps $\mu_+=0$ to \eqref{eq:allMoments}.

We denote by $K$ the complex structure on $\CC^{2N^2}$ in which $U,V$ are holomorphic. We then take $I$ and $J$ to be the two other canonical complex structures. That is, the triplet $J_\sigma=(I,J,K)$ satisfies the quaternion algebra
\be J_\sigma J_\tau = \epsilon_{\sigma\tau\upsilon} J_\upsilon -\delta_{\sigma\tau} \ . \label{eq:quats} \ee
A general complex structure then takes the form
\be J^{(c)}=c\cdot J \ , \label{eq:Jgen} \ee
where $c$ is a unit vector in $\RR^3$. To relate $\theta b$ and $\zeta$, we first find the image, $c$, of $(0,0,1)$, regarded as an $SU(2)$ triplet, under conjugation by $e^{-i\theta \, b\cdot\sigma}$:\footnote{That is, in the original R-symmetry frame we thought of ourselves as being in complex structure $K$, and we are now finding the coordinates of that complex structure in the new R-symmetry frame.}
\be c = (b_y \sin 2\theta + 2b_x b_z \sin^2\theta, \, -b_x\sin 2\theta + 2b_y b_z \sin^2\theta, \, \cos 2\theta + 2 b_z^2 \sin^2\theta ) \ . \ee
With these identifications, we find that the R-symmetry $e^{i\theta\, b\cdot\sigma}$ maps $[U,V]=0$ to \eqref{eq:allMoments}. This is clear if we write the latter as
\be \mu_\zeta\equiv -\frac{i}{2\zeta} \mu_+ + \mu_\RR - \frac{i\zeta}{2} \mu_- = 0 \ee
and note that the bottom-left entry of the last expression in \eqref{eq:rSymm} is proportional to $\mu_\zeta$ if we identify
\be \zeta = \frac{b_x+ib_y}{\cot\theta - ib_z} = \frac{ic_1-c_2}{1+c_3} \ . \ee
$c$ and $\zeta$ are related via stereographic projection:
\be c = \frac{1}{1+|\zeta|^2}(2\Imag \zeta, -2\Real \zeta, 1-|\zeta|^2) \ . \ee
We will often label complex structures using $\zeta\in \CC\cup\{\infty\}$, e.g. we will write $J^{(\zeta)}$ in addition to $J^{(c)}$. We also introduce the notation $\mu_\pm = \mu_I \pm i\mu_J$, $\mu_\RR=\mu_K$, where $\mu_{I,J,K}^\dagger = \mu_{I,J,K}$.

As the $U(N)$ action on $\CC^{2N^2}$ is triholomorphic (i.e., it preserves the complex structures of $\CC^{2N^2}$, or equivalently the gauge theory has at least $\N=4$ supersymmetry), this R-symmetry persists in the infrared non-linear sigma model with target space the moduli space, where it has the effect of rotating the complex structures into each other.

Lastly, for each complex structure $J^{(c)}$ on $\CC^{2N^2}$ there is a compatible K\"ahler form $\tilde\omega^{(c)}$:\footnote{This notation is slightly inconsistent, since we defined $\tilde g_{ij}$ to be the metric on $\tilde M$, rather than on $\CC^{2N^2}$. However, pulling back the metric from $\CC^{2N^2}$ to $\tilde M$ gives a metric on $\tilde M$, while pulling back the K\"ahler forms to $\tilde M$ does not give K\"ahler forms, as is clear from the fact that $\tilde M$ can be odd-dimensional. So, we will denote the pullback of the K\"ahler forms to $\tilde M$ by $i^*\tilde\omega^{(c)}$.}
\be \tilde\omega^{(c)}(v_1,v_2) = G(J^{(c)}v_1, v_2) \ , \quad G(v_1, v_2) = \tilde \omega^{(c)}(v_1, J^{(c)}v_2) \ , \quad J^{(c)} = - G^{-1} \tilde \omega^{(c)} \ . \label{eq:compat} \ee
Here, $v_1$ and $v_2$ are tangent vectors to $\CC^{2N^2}$, $G$ is the flat metric
\be ds^2 = 2 \Tr (dU dU^\dagger + dV dV^\dagger) \ee
on $\CC^{2N^2}$ (which is the same for all complex structures), and
\begin{align}
\tilde\omega^{(c)} &= c\cdot \tilde\omega \label{eq:wGen} \\
\tilde\omega_I &= \frac{i}{2}\Tr(d(U-V^\dagger)\wedge d(U^\dagger-V) + d(U^\dagger + V)\wedge d(U+V^\dagger)) \nonumber \\
&= i\Tr(-dU\wedge dV + dU^\dagger\wedge dV^\dagger) \label{eq:flatI} \\
\tilde\omega_J &= \frac{i}{2}\Tr(d(U-iV^\dagger)\wedge d(U^\dagger+iV) + d(U^\dagger - iV)\wedge d(U+iV^\dagger)) \nonumber \\
&= -\Tr(dU\wedge dV + dU^\dagger \wedge dV^\dagger) \label{eq:flatJ} \\
\tilde \omega_K &= i \Tr(dU\wedge dU^\dagger + dV\wedge dV^\dagger) \ . \label{eq:flatK}
\end{align}
Defining
\be \tilde\omega_\pm = \tilde \omega_I \pm i \tilde\omega_J \ , \label{eq:holoW} \ee
which in complex structure $\pm K$ is a holomorphic symplectic 2-form, we have
\be d\mu_\RR^A = - \iota_{k_A} \tilde\omega_K \ , \quad d\mu_\pm^A = - \iota_{k_A} \tilde\omega_\pm \ , \label{eq:mmGood} \ee
where $\iota$ denotes interior product, or contraction. More explicitly,
\be \tilde\omega_+ = -2i \Tr(dU\wedge dV) \ , \quad \tilde\omega_- = 2i\Tr(dU^\dagger\wedge dV^\dagger) \ . \ee
\eqref{eq:quats} and \eqref{eq:compat} allow one to determine the complex structures and metric in terms of the three K\"ahler forms $\tilde\omega_\sigma$:
\be G = -\tilde\omega_I\tilde\omega_J^{-1}\tilde\omega_K = -\tilde\omega_K\tilde\omega_I^{-1}\tilde\omega_J = -\tilde\omega_J\tilde\omega_K^{-1}\tilde\omega_I \ , \quad J_\sigma = \tilde\omega_\sigma^{-1} G = - G^{-1} \tilde\omega_\sigma \ . \label{eq:fromW} \ee
The K\"ahler forms may be canonically packaged into
\be \tilde\varpi(\zeta) = - \frac{i}{2\zeta} \tilde\omega_+ + \tilde\omega_K - \frac{i\zeta}{2} \tilde\omega_- \ , \label{eq:holoSymp} \ee
which is a holomorphic symplectic 2-form in complex structure $\zeta$.

With this geometric background, we can straightforwardly state the relationship between the hyper-K\"ahler structures on $\CC^{2N^2}$ and $M$. The metric, acting on tangent vectors to a point $\tau\in M$, is defined by
\be g(X,Y) = \tilde g(p^{-1}(X), p^{-1}(Y)) \ . \label{eq:quotMet1} \ee
Note that the ingredients in this equation are gauge-dependent, as $p^{-1}=P_H\circ s_*$ is a map from $T_\tau M$ to $H_{s(\tau)}\tilde M$, but the answer is gauge-independent, since $U(N)$ acts via isometries. We can make this explicit by writing
\be g(X,Y) = \tilde g(P_H(s_*(X)), P_H(s_*(Y))) = \tilde g(s_*(X), P_H(s_*(Y))) \equiv g'(s_*(X), s_*(Y)) \ , \ee
where $g'$ is the the following metric on the image of $s$:
\be ds^2 = (\tilde g_{ij} - \tilde g_{im} \tilde g_{jn} k^m_A k^n_B H^{AB}) dt^i dt^j \ , \label{eq:gCorrected} \ee
where $dt^i dt^j\equiv \frac{dt^i\otimes dt^j + dt^j\otimes dt^i}{2}$, and where $dt^i$ is restricted to be a one-form on the image of $s$. This metric is precisely what one obtains in the gauge theory by substituting the gauge field that extremizes the action back into the action. Similarly, the K\"ahler forms on $M$ are uniquely characterized by
\be \pi^*\omega_\sigma = i^*\tilde\omega_\sigma \ . \ee
That is,
\be \omega_\sigma(X,Y) = \tilde\omega_\sigma(i_*(\tilde X), i_*(\tilde Y)) \ , \ee
where $\tilde X$ and $\tilde Y$ are any vectors in $T\tilde M$ such that $\pi_*(\tilde X)=X$ and $\pi_*(\tilde Y)=Y$. This is well-defined because $\tilde\omega_\sigma(i_*(\tilde X), i_*(\tilde Y))=0$ whenever $\tilde X\in V\tilde M$,\footnote{This follows from \eqref{eq:mmGood}, as $d\mu_\zeta=0$ on $\tilde M$, for all $\zeta$.} so we can instead require $\tilde X$ and $\tilde Y$ to be in $H\tilde M$ and satisfy $p(\tilde X)=X$ and $p(\tilde Y)=Y$. If we pick a gauge then we can write the formula more explicitly:
\be \omega_\sigma(X,Y) = \tilde\omega_\sigma(i_*(s_*(X)), i_*(s_*(Y))) \label{eq:quotKF} \ . \ee
In summary, $g$ is the metric induced on $M$ from $(s(M), g')$, while $\omega_\sigma$ are the pullbacks of the K\"ahler forms on $\CC^{2N^2}$ via $i\circ s$. The complex structures $J^{(\zeta),M}$ are induced from $\CC^{2N^2}$ by noting that $i_*\circ P_H\circ \tilde P$, where $\tilde P$ projects from $T\CC^{2N^2}$ to $T\tilde M$, commutes with $J^{(\zeta)}$, i.e. that $H\tilde M$ is a complex subbundle of $T\CC^{2N^2}|_{\tilde M}$.\footnote{To see this, we show that the complement of $H\tilde M$ in $T\CC^{2N^2}|_{\tilde M}$, spanned by $k_A,\nabla \mu_I^A,\nabla \mu_J^A, \nabla\mu_K^A$, is preserved by each complex structure. Focusing, for concreteness, on the $K$ complex structure, we first note that for all $Y\in T\CC^{2N^2}$,
\be G(\nabla \mu_K^A, Y) = \iota_Y d\mu_K^A = -\tilde \omega_K(k_A, Y) = - G(Kk_A, Y) \ , \ee
which implies that $\nabla\mu_K^A = - K k_A$. We similarly have $\nabla\mu_I^A = - I k_A$ and  $\nabla\mu_J^A = - J k_A$, i.e. $\nabla \mu_I^A = KJ k_A = -K\nabla\mu_J^A$.} For, $p$ then translates this complex structure to $M$:
\be J^{(\zeta),M}(X) = \pi_*(\tilde P(J^{(\zeta)}(i_*(p^{-1}(X))))) \ , \label{eq:newJ} \ee
where $X\in TM$. ($\tilde P$ acts trivially here -- it simply relabels its argument as being in $T\tilde M$ -- since the image of $J^{(\zeta)}\circ i_*\circ p^{-1}$ is in $H\tilde M$.) In words, we lift a vector to $T\CC^{2N^2}$, act with the complex structure, and then project it back to $TM$.

In this subsection, we have presented many aspects of hyper-K\"ahler geometry -- such as the existence of a $\PP^1$ of K\"ahler structures satisfying \eqref{eq:quats}, \eqref{eq:Jgen}, \eqref{eq:compat}, \eqref{eq:wGen}, \eqref{eq:holoW}, \eqref{eq:fromW}, and \eqref{eq:holoSymp} -- in the context of a particular example. However, we emphasize that they hold for all hyper-K\"ahler manifolds. Similarly, features such as the existence of a triplet of equivariant moment maps satisfying \eqref{eq:mmGood} are common to all hyper-K\"ahler quotients, as is the procedure we explained for constructing the hyper-K\"ahler structure on the quotient. One caveat pertains to \eqref{eq:rSymm}: that the `names' of the complex structures do not matter does not imply that they are all equivalent, i.e. that there must be an $SU(2)$ R-symmetry. There is no problem with having coupling constants which transform under the R-`symmetry'. Physically, this means that the R-`symmetry' is explicitly broken, but the theory can nevertheless possess 3d $\N=4$ supersymmetry.

\subsection{$\CC^2/Z_2$} \label{sec:douglasMoore}

Our next example is associated to a D2-brane probing the orbifold $\CC^2/Z_2$. Following the prescription of \cite{polchinski:consistent,moore:ALEinst}, we begin on the covering space $\CC^2$, with a D2-brane and its image; this is simply the $N=2$ case of the previous section. We then impose the projections
\be U = - \sigma_z U \sigma_z \ , \quad V = - \sigma_z V \sigma_z \ , \quad g = \sigma_z g \sigma_z \ . \ee
Here, $g$ is an element of the gauge group, so the final equation restricts the gauge group to a subgroup of $U(2)$. There is nothing special about $\sigma_z$ here; any unitary matrix with eigenvalues $1,-1$ would suffice. The important point is that such a matrix represents the action of the non-trivial element of $Z_2$ on the `regular representation' of dimension $|Z_2|=2$. The relative minus signs in the first two equations arise from the fact that $U$ and $V$ are negated by the geometric orbifold action. These orbifold projection conditions are solved by
\be U=u_x\sigma_x+u_y\sigma_y=\twoMatrix{}{u_+}{u_-}{} \ , \quad u_\pm = u_x \mp i u_y \ , \ee
a similar equation for $V$, and
\be g=e^{i(\theta I + \alpha\sigma_z/2)}=e^{i\theta}\twoMatrix{e^{i\alpha/2}}{}{}{e^{-i\alpha/2}} \ . \label{eq:EHgauge} \ee
Note that $e^{i\theta}$ acts trivially on $U$ and $V$, and so the gauge group is effectively $U(1)$. $\alpha$ is valued in $\RR/2\pi\ZZ$, since when it equals $2\pi$, it can equivalently be taken to vanish by modifying $\theta$. Under the non-trivial $U(1)$ gauge transformations, $u_\pm$ has charge $\pm 1$, and similarly for $v_\pm$. The moment maps are obtained by substituting the projected forms of $U$ and $V$ into the moment maps of the previous section. However, thanks to the projection, there now exist new gauge-invariant FI parameters:
\begin{align}
\mu_+ &= -2\sigma_z(2i(u_x v_y-u_y v_x) - \xi_+)= -2\sigma^z(u_+v_- - u_- v_+ - \xi_+) \ , \\
\mu_\RR &= \sigma_z(2i(u_x u_y^* - u_y u_x^* + (U\mapsto V)) - \xi_\RR) \\
&= \sigma_z (|u_+|^2 + |v_+|^2 - |u_-|^2 - |v_-|^2 - \xi_\RR) \ . 
\end{align}
These FI parameters transform as a triplet under the $SU(2)$ R-symmetry, just like the moment maps. Per our usual convention, we define $\xi_-=\xi_+^*$. The FI parameters serve to resolve the orbifold singularity of the moduli space, producing the Eguchi-Hanson $A_1$ ALE space. Note that these moment maps make it clear that $(u_+,v_-^*)$ comprise a charge $+1$ hypermultiplet, while $(u_-,v_+^*)$ comprise a charge $-1$ hypermultiplet.

When the FI parameters vanish, $\mu_+=0$ implies that there exists $\lambda\in\CC$ such that
\be \column{u_+}{v_+}=\lambda \column{u_-}{v_-} \ , \ee
and $\mu_\RR=0$ then implies that $|\lambda|=1$. Lastly, $U(1)$ gauge transformations allow us to set $\lambda=1$, so $u_+=u_-=u_x$ and $u_y=0$, and similarly for $V$. However, there exists a residual gauge symmetry, since $g=i\sigma_z$ (i.e., $\alpha=\pi$) preserves the above gauge choice, while negating $u_-$ and $v_-$. We therefore see that the moduli space is $\CC^2/Z_2$, with coordinates $u_-,v_-$ which are well-defined up to the discrete $Z_2$ action.

The same approach works with few modifications if we turn on only $\xi_\RR$. We now simply need to allow $\lambda$ to be a function of $u_-$ and $v_-$. The real moment map equation implies that
\be |\lambda| = \sqrt{1 + \frac{\xi_\RR}{|u_-|^2+|v_-|^2}} \ ; \ee
thanks to the $U(1)$ gauge symmetry, we can take $\lambda$ to be real and positive. As before, this fixes all of the gauge symmetry except for the $Z_2$. For future reference, we note that
\begin{align}
u_x &= \frac{u_+ + u_-}{2} = \frac{u_-(\lambda+1)}{2} \ , \quad u_y = \frac{u_+ - u_-}{-2i} = \frac{u_-(\lambda - 1)}{-2i} \ , \\
&\qquad\qquad\qquad\frac{|u_x|^2}{|u_-|^2}=\frac{|v_x|^2}{|v_-|^2}=\frac{1}{4}(\lambda+1)^2 \ , \\
&\Rightarrow \frac{u_y}{u_x} = \frac{i(\lambda - 1)}{\lambda + 1} = \frac{i(\lambda^2-1)}{(\lambda+1)^2} = \frac{i\xi_\RR}{4 (|u_x|^2 + |v_x|^2)} \ .
\end{align}
In particular, since we have taken $\lambda$ to be real, we have $\Real \frac{u_y}{u_x} = 0$.

We now introduce a different approach, which will serve to warm the reader up for our upcoming perturbative approach to K3 metrics. The gauge transformation $\delta u_y=-\epsilon u_x$ allows us to adopt the gauge $\Real \frac{u_y}{u_x}=0$; indeed, this is the same gauge that we adopted in the last paragraph. Substituting $u_y=i\alpha u_x$, where $\alpha\in\RR$, into the moment map equations yields three linear equations in the three unknowns $\alpha,v_y,v_y^*$. We thus find the solution
\begin{align}
\alpha &= \frac{\xi_\RR |u_x|^2 + \xi_+ u_x^* v_x^* + \xi_- u_x v_x}{4|u_x|^2(|u_x|^2+|v_x|^2)} \\
v_y &= \frac{-\xi_\RR |u_x|^2 v_x + \xi_+ u_x^*(2|u_x|^2 + |v_x|^2) - \xi_- u_x v_x^2}{4i |u_x|^2(|u_x|^2+|v_x|^2)} \ .
\end{align}
Upon setting $\xi_+=\xi_-=0$, these results agree with those of the previous paragraph.

Lastly, we note that dealing with arbitrary gauge choices is rather straightforward in this approach. Before gauge fixing, we have three equations in the four unknowns $Q=(u_y,u_y^*,v_y,v_y^*)^T$. Writing this linear (in $Q$) system as
\be L(q) Q = \xi \ , \ee
where $q=(u_x,u_x^*,v_x,v_x^*)$, the rows of $L$ correspond, respectively, to $\mu_+,\mu_-,\mu_\RR$, and $\xi=(\xi_+,\xi_-,\xi_\RR)^T$ is the vector of FI parameters, we can write one solution (the `least norm solution') as
\be Q = L^\dagger(LL^\dagger)^{-1} \xi \ . \label{eq:leastNorm} \ee
However, $L$ has a one-dimensional nullspace, which corresponds to the $U(1)$ gauge freedom, and so a general solution is the sum of \eqref{eq:leastNorm} with an element of the nullspace. Gauge transformations act on $Q$ as
\be \delta Q = -\epsilon \begin{pmatrix} u_x \\ u_x^* \\ v_x \\ v_x^* \end{pmatrix} \ , \ee
and indeed one may verify that this vector spans the nullspace of $L$.\footnote{This vector is annihilated by $L$ for any complex $\epsilon$, but the conditions $(Q_1)^*=Q_2$ and $(Q_3)^*=Q_4$ are only preserved when $\epsilon\in\RR$.} So, the general solution is
\be Q = L^\dagger (L L^\dagger)^{-1}\xi - \epsilon(q) \begin{pmatrix} u_x \\ u_x^* \\ v_x \\ v_x^* \end{pmatrix} \ , \ee
where $\epsilon$ is an arbitrary real function of $q$. We note that gauge transformations modify $q$, in addition to $Q$, and so in different gauges, the coordinates $(u_x,v_x)$ refer to different points in moduli space.

\subsection{$\Sym^k T^4$}

We now introduce a compact example, following \cite{wati:dBraneT}. We obtain $\Sym^k T^4$ as the `Higgs branch' of $k$ D2-branes probing $T^4$. Thinking of this as the orbifold $\RR^4/\Lambda$, where $\Lambda$ is an embedded 4-dimensional lattice in $\RR^4$, allows us to employ the same approach as in the previous section. So, we consider our $k$ D-branes, and their images under the $\ZZ^4$ orbifold group, probing the covering space, $\RR^4$. These D-branes have a $U(k\infty^4)$ gauge group. Among their field content are Hermitian adjoint scalars $X^a_{im;jn}$, $a=1,\ldots,4$, $m,n\in\Lambda$, $i,j=1,\ldots,k$, whose eigenvalues serve as the positions of our D-brane probes. We can equivalently think of them as $k\times k$ matrices $X^a_{mn}$, which satisfy $(X^a_{mn})^\dagger = X^a_{nm}$. We now impose the orbifold projections
\begin{align}
X^a_{(m+\delta)(n+\delta)} &= X^a_{mn} + \delta^a I \delta_{mn} \ , \label{eq:trans} \\
A^\mu_{(m+\delta)(n+\delta)} &= A^\mu_{mn} \ . \label{eq:noTrans}
\end{align}
Here, $\delta^a$ are the coordinates in $\RR^4$ of a displacement $\delta\in\Lambda$, $I$ is the $k\times k$ identity matrix, and $A^\mu$, $\mu=0,1,2$, is the gauge field of the D2-brane. We note that the projection of $A$ is simpler than that of $X$ because the latter transforms under translations in $\RR^4$, while the former does not. Note that the projections can be written in a similar form as in the previous section if we define the shift matrices
\be e(n)_{\ell m} = \delta_{\ell,m-n} \ , \label{eq:eDef} \ee
which satisfy
\be e(n)^\dagger = e(-n) = e(n)^{-1} \ , \quad e(m)e(n) = e(m+n) \ , \ee
since $X^a_{(m+\delta)(n+\delta)} = (e(\delta) X^a e(\delta)^\dagger)_{mn}$. That is, the $e(m)$ represent $\ZZ^4$ in its regular representation; in particular, $e(0)$ is the identity. In words, $e(n)$ is the $k\times k$ identity on the $n$-th diagonal. The gauge group of our orbifold theory is the subgroup of $U(k\infty^4)$ which commutes with all $e(\delta)$; that is,
\be g=e(\delta)ge(\delta)^\dagger \quad \Leftrightarrow \quad g_{(m+\delta)(n+\delta)}=g_{mn} \quad \forall m,n \ . \label{eq:gProj} \ee
Writing $g\approx 1+i\epsilon h$ shows that Lie algebra elements $h$ obey the same condition as \eqref{eq:gProj}; of course, this had to be the case, thanks to \eqref{eq:noTrans}.

These projections mean that the only independent components of $X^a$ (before imposing the Hermiticity constraint) are
\be X^a_n \equiv X^a_{0n} \ ; \label{eq:from0} \ee
Hermiticity further demands that $(X_n^a)^\dagger = X_{-n}^a$. Indeed, we can write the most general solution of the constraints as
\be X^a = w^a + \sum_n X_n^a \, e(n) \ , \quad w^a_{mn} = \delta_{mn} m^a \ . \label{eq:constSol} \ee
$w^a$ is a diagonal matrix responsible for the shifts in \eqref{eq:trans}. Note the important properties
\be (w^a)^\dagger = w^a\ , \quad [e(n), w^a] = n^a e(n) \ ; \ee
we also have $[e(m),e(n)]=[w^a,w^b]=0$.

Similarly, the solutions of \eqref{eq:gProj} take the form
\be g = \sum_n g_n e(n) \ , \ee
with the additional condition $g^\dagger = g^{-1}$, which thanks to
\be g^\dagger g = \sum_{m,n}e(m) g_n^\dagger g_{m+n} \ee
is equivalent to $\sum_n g_n^\dagger g_{m+n} = \delta_{m,0} I$. Gauge transformations act via conjugation,
\be
X^a \mapsto gX^ag^\dagger = g \parens{[w^a,g^\dagger] + \sum_n X^a_n e(n) g^\dagger} + w^a \ ,
\ee
which is equivalent to
\be \sum_n X^a_n e(n) \mapsto g \parens{[w^a,g^\dagger] + \sum_n X^a_n e(n) g^\dagger} \ . \label{eq:tDualGaugeTrans1} \ee
Lie algebra elements take the form
\be h = \sum_n h_n e(n) \ , \ee
where $h_n^\dagger = h_{-n}$. As usual, this algebra acts via commutation:
\be \delta X^a = i\epsilon\brackets{\sum_n h_n e(n), w^a+\sum_m X^a_m e(m)} = i\epsilon \sum_n e(n)\brackets{n^a h_n + \sum_m [h_{n-m},X^a_m]} \ . \label{eq:smallTrans} \ee
This agrees with \eqref{eq:tDualGaugeTrans1} upon setting $g=e^{i\epsilon h}\approx 1+i\epsilon h$.

This is all suggestive of the following identifications:
\be e(n) \sim e^{in^a y_a} \ , \quad w^a \sim i\partial^a \ . \ee
Here, $y_a$ parametrizes the dual torus $\hat T^4=\RR^4/\Lambda^\vee$, where we define $\Lambda^\vee = \Hom(\Lambda, 2\pi \ZZ)$, and $\partial^a = \partial_{y_a}$. Indeed, this is a manifestation of T-duality! Under this duality, our D2-branes become D6-branes, and the fields $X_n$ which correspond to strings with $n$ units of winding map to Fourier modes of a gauge field with $n$ units of momentum. Now, a gauge field is not exactly in the adjoint representation of $U(k)$, but instead transforms as
\be \sum_n X^a_n e^{in^b y_b} \mapsto g(y)\parens{i\partial^a + \sum_n X^a_n e^{in^b y_b}}g(y)^\dagger \ . \label{eq:tDualGauge} \ee
\eqref{eq:tDualGaugeTrans1} shows that our infinite-dimensional matrices $X^a$ transform in the same way! In particular, this discussion makes it clear that the subgroup of $U(k\infty^4)$ that survives the orbifold projection is the group $\widehat{U(k)}$ of maps from $\hat T^4$ to $U(k)$. We call this the $U(k)$ four-loop group, or floop group for short. Similarly, the subalgebra of the $\mf{u}(k\infty^4)$ Lie algebra that survives the projection is the algebra $\widehat{\mf{u}(k)}$ of maps from $\hat T^4$ to $\mf{u}(k)$, which we call the $\mf{u}(k)$ floop algebra.

We now consider large gauge transformations. For concreteness, we focus here on the cases $k=1,2$. When $k=1$, there are gauge transformations of the form
\be \gamma_n = e^{in^a y_a} = e(n) \ . \ee
These comprise a disconnected $\ZZ^4$ factor in the $U(1)$ floop group, as follows from $\pi_1(U(1))=\ZZ$. They act on $X^a$ via
\be \gamma_n w^a \gamma_n^{-1} = w^a + n^a \ , \ee
which effects $X^a_0 \mapsto X^a_0 + n^a$. So, they compactify the zero-modes $X_0^a$ so that they parametrize $T^4$. This is T-dual to large gauge transformations compactifying the moduli space of Wilson lines in a D6-brane.

When $k=2$, we consider gauge transformations of the form
\be g = e^{in^a y_a (aI+c\, b\cdot \sigma)} = e^{in^ay_a a}(\cos(n^a y_a c) \, I + i \sin(n^a y_a c) \, b\cdot \sigma) \ , \label{eq:quasiLarge} \ee
where $|b|=1$. This is a single-valued function on $\hat T^4$ if $a,c\in \half\ZZ$ and $a\equiv c \modu 1$. Assuming these conditions, we can write \eqref{eq:quasiLarge} as
\begin{align}
g &= \half\brackets{\parens{e(n(a+c))+e(n(a-c))}I + \parens{e(n(a+c))-e(n(a-c))} b\cdot \sigma} \\
&= \half\brackets{e(n(a+c))\parens{I+ b\cdot\sigma} + e(n(a-c))\parens{I-b\cdot\sigma}} \ .
\end{align}
This is well-defined, since the arguments of $e(\cdot)$ are valued in $\Lambda$. It acts on $w^a$ via
\be g w^a g^{-1} = w^a + n^a(aI+c\, b\cdot\sigma) \ , \ee
which effects
\be X_0^a\mapsto X_0^a + n^a(aI+c\, b\cdot\sigma) \ . \label{eq:quasiLargeEffect} \ee
This will prove crucial in the next section. Unlike the $k=1$ case, we note that if $c\not=0$ then $g$ also acts non-trivially on $\sum_m X^a_m e(m)$, so \eqref{eq:quasiLargeEffect} does not give the complete change in $X^a$. Instead, we have 
\begin{align}
g X^a_m e(m) g^{-1} &= \half e(m) \parens{X^a_m +  b\cdot\sigma X^a_m  b\cdot\sigma} \nonumber \\
&+\frac{1}{4}\brackets{e(m+2c n)(I+ b\cdot \sigma) X^a_m (I- b\cdot\sigma) + e(m-2cn)(I- b\cdot \sigma)X^a_m (I+ b\cdot\sigma) } \ . \label{eq:quasiLargeEffect2}
\end{align}
That is, if we define
\be g\sum_m X^a_m e(m) g^{-1} \equiv \sum_m \tilde X_m^a e(m) \ , \ee
then we have
\begin{align}
\tilde X_m^a &= \half \parens{X^a_m+b\cdot\sigma X^a_m b\cdot\sigma} \nonumber \\
&+\frac{1}{4}\brackets{(I+b\cdot\sigma)X^a_{m-2cn}(I-b\cdot\sigma)+(I-b\cdot\sigma)X^a_{m+2cn}(I+b\cdot\sigma)} \\
&= \half \parens{X^a_m+b\cdot\sigma X^a_m b\cdot\sigma} + \frac{1}{4}\parens{X^a_{m-2cn}+X^a_{m+2cn}} \nonumber \\
&+ \frac{1}{4}\parens{[b\cdot\sigma,X^a_{m-2cn} - X^a_{m+2cn}] - b\cdot\sigma(X^a_{m-2cn}+X^a_{m+2cn})b\cdot\sigma } \ . \label{eq:quasiLargeEffect3}
\end{align}
We note that it is not strictly appropriate to call the transformations with $a=0$ large gauge transformations. For, the connected components of $\widehat{U(k)}$ are classified by $(\pi_1(U(k)))^4=\ZZ^4$; specifically, the connected components are comprised of those $g$ with the same discontinuities of $\frac{1}{2\pi i}\log\det g$ around all 1-cycles. For the transformations \eqref{eq:quasiLarge}, $\frac{1}{2\pi i}\log\det g=\frac{2a n^a y_a}{2\pi}$, and so the connected components containing them are labelled by $2an \in \Lambda\cong \ZZ^4$. So, if $a=0$ then $g$ is actually in the connected component of the identity. Nevertheless, these are not in the image of the exponential map, applied to the Lie algebra. We thus define $g$ to be `quasi-large' if it is not in the image of the exponential map; in particular, large gauge transformations (those which are not in the connected component of the identity) are quasi-large. This possibility of the Lie algebra not surjecting onto the connected component of the identity is, of course, dissimilar from the more familiar case of finite-dimensional compact Lie groups.

We have now specified the matter content and gauge group of our gauge theory. We next simply take the maximally supersymmetric $U(k\infty^4)$ Yang-Mills action and project all of the superfields as described above. In particular, it is now wise to package $X^a$ into complex fields
\be U = X^1 + iX^2 \ , \quad V = X^3 + iX^4 \ . \ee
We also define
\begin{align}
n^u &= n^1 + i n^2 \ , \quad n^{\bar u} = n^1 - i n^2 \ , \quad w^u = w^1 + iw^2 \ , \quad w^{\bar u} = w^1 - i w^2 \ , \\
U_n &= X^1_n + i X^2_n \ , \quad U^\dagger_n = X^1_n - i X^2_n \ ,
\end{align}
and introduce similar definitions for $V$. We then learn that $\Sym^k T^4$ may be obtained as the hyper-K\"ahler quotient of the quaternified $\mf{u}(k)$ floop algebra by its corresponding floop group. The moment maps are:\footnote{Note that, thanks to the orbifold projection, we can now add gauge-invariant FI terms of the form $\xi_n Ie(n)$ to the moment maps. However, as in \cite{kapustin:impurity}, these can be eliminated for all non-zero $n$ via a simple change of variables -- i.e., by adding a constant to some of $U_n,U_{-n},V_n,V_{-n}$. For $n=0$, this is the same FI term that we discarded in \S\ref{sec:hkStart}. \label{ft:fakeFI}}
\begin{align}
\mu_+ &= -2[U,V] \nonumber \\
&= -2\brackets{w^u+\sum_n U_n e(n)\, ,\  w^v+\sum_m V_m e(m)} \nonumber \\
&= -2\sum_n e(n)\brackets{U_n n^v - V_n n^u + \sum_m [U_{n-m}, V_m] } \ , \label{eq:muPT4} \\
\mu_\RR &= [U,U^\dagger] + [V,V^\dagger] \nonumber \\
&= \brackets{w^u+\sum_n U_n e(n) \, , \ w^{\bar u} + \sum_m U_m^\dagger e(-m)} + (U\mapsto V) \nonumber \\
&= \sum_n e(n)\brackets{ -n^u U^\dagger_{-n} + n^{\bar u} U_n + \sum_m [U_{n+m}, U^\dagger_m] + (U\mapsto V) } \ . \label{eq:muRT4}
\end{align}
Note that setting them equal to zero yields infinitely many equations. They, along with $\widehat{U(k)}$ gauge transformations, allow us to set to 0 all of the non-zero modes $U_n,V_n$. The zero mode moment map equations, plus gauge transformations, then leave us with $\Sym^k T^4$.

We comment briefly on the preserved supersymmetry of this theory. The projections imposed on $U$ and $V$ are holomorphic, so at least we preserve the $K$ complex structure. However, the projections on $U$ and $V^\dagger$ take different forms, so the R-symmetry \eqref{eq:rSymm} does not preserve the forms \eqref{eq:constSol} of $U$ and $V$. But, we can just regard $w^a$ as coupling constants which themselves transform under the R-symmetry. With this understood, the R-`symmetry' persists in the orbifold theory, and so we preserve the $\PP^1$ of complex structures.

$\Sym^k T^4$ is, of course, flat, but singular. It may be resolved to a smooth hyper-K\"ahler manifold, $\Hilb^k T^4$. However, the present formalism does not possess parameters that can implement this. (In contrast, in the next section we will be able to resolve $T^4/Z_2$ to smooth K3 surfaces.) We expect that introducing a D6-brane wrapping $T^4$, in addition to the $k$ D2-branes probing $T^4$, will introduce these parameters (similarly to \S9.3 of \cite{moore:ALEinst}). First, one can clearly obtain $\Sym^k T^4$ from this construction, since the $k$ D2-branes function as small instantons in the $U(1)$ gauge theory of the D6-brane. (There is also an extra $\hat T^4$ factor in the moduli space, coming from $U(1)$ Wilson lines of the D6-brane. But, this is easy to neglect, as the D6-brane $U(1)$ gauge field is free.) Then, turning on a self-dual B-field along the torus makes this a $U(1)$ gauge theory on a non-commutative torus, for which there exist finite-size instantons.\footnote{Alternatively, this B-field generalizes the objects of interest from holomorphic vector bundles -- as discussed at the end of this section -- to torsion free sheaves, as discussed in \S9.3 of \cite{moore:ALEinst}.} This should eliminate the singularities corresponding to coincident small instantons, as the instantons become fuzzy. This B-field also eliminates the Coulomb branch of our gauge theory, as it attracts the D2-branes to the D6-brane \cite{s:20DLCQ,nekrasov:noncomm,tong:kinky,sw:noncomm}. So, there are no small instanton singularities in the Higgs branch due to the existence of a Coulomb branch. Little string theories whose Coulomb branches are moduli spaces of non-commutative instantons on K3 and $T^4$ were studied in \cite{ganor:noncomm1,ganor:noncomm2}.

Lastly, we re-write the moment map equations in the T-dual language of differential equations. Defining
\be B^a = \sum_n X_n^a e^{in^a y_a} \ , \ee
we have
\be X^a \sim i(\partial^a - iB^a) \equiv i\nabla^a \ , \ee
where $\nabla^a$ is a covariant derivative associated to the $U(k)$ D6-brane gauge symmetry. Because $B^a$ is globally-defined (and periodic), there must be a global section of the D6-brane's principal $U(k)$-bundle, and so the latter is trivial. We repackage the coordinates $y_a$ into
\be \psi_1 = \frac{y_1-iy_2}{2} \ , \quad \psi_2 = \frac{y_3-iy_4}{2} \ , \ee
so that
\be \partial^{1'} = \partial^1+i\partial^2 \ , \quad \partial^{2'} = \partial^3+i\partial^4 \ . \ee
Primes on indices indicate that we are referring to the holomorphic coordinates $\psi_1,\psi_2$, as opposed to $y_a$. We thus have
\be U = X^1+iX^2 \sim i\nabla^{1'} \ , \quad V = X^3+iX^4 \sim i\nabla^{2'} \ , \quad U^\dagger \sim i\nabla^{\bar 1'} \ , \quad V^\dagger \sim i\nabla^{\bar 2'} \ . \ee
Using
\be [\nabla^a,\nabla^b]=-iF^{ab} \ , \quad F^{ab}=\partial^a B^b-\partial^b B^a - i[B^a, B^b] \ , \ee
we find that the moment maps are
\be \mu_+=-2[U,V]=-2iF^{1'2'} \ , \quad \mu_\RR = [U,U^\dagger] + [V,V^\dagger] = iF^{1'\bar 1'}+iF^{2'\bar 2'} \ . \ee
Written in terms of real coordinates, the moment map equations $\mu_+=\mu_\RR=0$ take the form
\be F^{13} = F^{24} \ ,\quad F^{14}=-F^{23} \ ,\quad F^{12}=-F^{34} \ , \ee
or equivalently
\be F=-*F \ , \ee
where the Hodge star is defined with respect to a metric which, in $y_a$ coordinates, is proportional to the identity. So, our moduli space is that of anti-self-dual $U(k)$ connections on a trivial bundle, up to $U(k)$ gauge equivalence. Triviality implies that $F=0$, since
\be \|F\|^2 \equiv \int \Tr F\wedge *F = - \int \Tr F\wedge F = 0 \ . \ee
So, we simply have the moduli space of $U(k)$ Wilson lines on $\hat T^4$, which is indeed $\Sym^k T^4$; the symmetric group quotient arises from the fact that the Weyl group of $U(k)$, which happens to be $S_k$, conjugates the maximal torus to itself, and so it is necessary to quotient by this discrete group of gauge symmetries.

It is worthwhile to reinterpret this moduli space from a holomorphic perspective. The equation $\mu_-=0$ implies that the antiholomorphic part of the curvature vanishes, or equivalently that the rank $k$ (complexified) vector bundle corresponding to the fundamental representation of $U(k)$ can canonically be given the structure of a holomorphic vector bundle. Furthermore, we have implicitly endowed our bundle with a Hermitian structure, in defining the adjoint. This gives an isomorphism (as complex vector bundles) between the holomorphic dual bundle and the antiholomorphic conjugate bundle. (The latter is, in turn, isomorphic as a complex vector bundle to the original bundle, as these are conjugate complexifications of a real vector bundle. The existence of an antiholomorphic structure on our bundle was guaranteed by $\mu_+=0$.) Conversely, any rank $k$ holomorphic Hermitian vector bundle possesses a canonical $U(k)$ connection, the Chern connection, whose curvature has no $(2,0)$ or $(0,2)$ part. So, ignoring stability, we are interested in the moduli space of holomorphic structures on a trivial rank $k$ Hermitian vector bundle on $\hat T^4$, up to isomorphism (i.e., $\widehat{GL(k,\CC)}$ equivalence). The Hermitian structure does not affect this moduli space; it simply provides the bridge between the present complex/algebraic geometry problem and that of the previous paragraph. Stability implies that our holomorphic vector bundles are isomorphic to the direct sum of $k$ topologically trivial holomorphic line bundles (since a flat $U(k)$ connection decomposes into $k$ flat $U(1)$ connections), and so our moduli space is $\Sym^k {\rm Pic}^\circ\, \hat T^4=\Sym^k T^4$.

\subsection{K3} \label{sec:k3HK}

Finally, we arrive at the promised land of K3. (We could easily study $\Sym^k K3$, but this introduces no new ideas, so we focus on the $k=1$ case. As for $\Sym^k T^4$, we expect to be able to resolve to $\Hilb^k K3$ only after adding a D6-brane.) Our approach to constructing K3 metrics via the hyper-K\"ahler quotient construction, which builds on results of \cite{ho:noncomm,waldram:lol,greene:lol}, combines those of the previous sections. For, we can obtain K3 by resolving the singularities of $T^4/Z_2$. We therefore orbifold $\RR^4$ by $\ZZ^4$, and then by $Z_2$; equivalently, we orbifold $\RR^4$ by the quadruply-infinite dihedral group $D_{\infty^4}=\ZZ^4 \rtimes Z_2$.

To study the D2-brane probe of this orbifold, we start with a D2-brane probe of $T^4$ and its $Z_2$ image and then impose $Z_2$ orbifold projections. That is, our starting point is the $\widehat{U(2)}$ gauge theory of the previous section. We then impose projections associated to the $Z_2$ action $x^a \mapsto \ell^a-x^a$, where $x^a$ are coordinates on $\RR^4$ and $\ell$ is a fixed element of $\Lambda$. These take the form
\begin{align}
X^a_{\ell-m;\ell-n} &= -\sigma_z X^a_{mn} \sigma_z + \ell^a I \delta_{mn} \ , \label{eq:z2X} \\
A^\mu_{\ell-m;\ell-n} &= \sigma_z A^\mu_{mn}\sigma_z \label{eq:z2A} \ ,
\end{align}
or equivalently
\begin{align}
X^a_{-n} &= - \sigma_z X^a_n \sigma_z \ , \label{eq:z2sol} \\
A^\mu_{-n} &= \sigma_z A^\mu_n \sigma_z \ .
\end{align}
One easily verifies that the $Z_2$ projections \eqref{eq:z2X} and \eqref{eq:z2A} are compatible with the $\ZZ^4$ projections of the previous section, in the sense that composing them gives no additional unwanted projections. Indeed, this is related to the shift matrices $e(n)$, together with the negation matrix
\be N_{im;jn} = \delta_{m,\ell-n} \,\sigma^z_{ij} \ , \ee
representing generators of $D_{\infty^4}$ in its regular representation. We have included $\ell$ in this discussion in order to highlight its irrelevance: it does not show up in \eqref{eq:z2sol}. This is as it should be, as any $\ell\in\Lambda$ yields the same orbifold of $\RR^4$.

The general solution to the constraints \eqref{eq:z2X} and \eqref{eq:z2sol} still takes the form
\be X^a = w^a + \sum_n X^a_n e(n) \ , \ee
but now these Fourier coefficients are further constrained to take the form
\begin{align}
X^a_n+X^a_{-n} &= X^a_{n,x}\sigma_x + X^a_{n,y}\sigma_y \ , \quad i(X^a_n - X^a_{-n}) = X^a_{n,0}I+X^a_{n,z}\sigma_z \quad (n\not=0) \nonumber \\
X^a_0 &= X^a_{0,x}\sigma_x+X^a_{0,y}\sigma_y \ . \label{eq:k3constr}
\end{align}
Note that these combinations are Hermitian. As usual, we introduce the complex combinations
\be U=X^1+iX^2=w^u + \sum_n U_n e(n) \ , \quad V = X^3+iX^4=w^v + \sum_n V_n e(n) \ . \ee
We also use notation as in \eqref{eq:k3constr}, but there is no need for $i$'s, since $U$ and $V$ need not be Hermitian:
\begin{align}
U_n + U_{-n} &= U_{n,x}\sigma_x+U_{n,y}\sigma_y \ , \quad U_n-U_{-n} = U_{n,0} I + U_{n,z}\sigma_z \quad (n\not=0) \nonumber \\
U_0 &= u_x \sigma_x + u_y \sigma_y \ , \label{eq:orbProjK3}
\end{align}
and similarly for $V$.

Gauge transformations take the form
\be g = \sum_n g_n e(n) \ , \ee
where $g^\dagger = g^{-1}$ and $g_{-n} = \sigma_z g_n \sigma_z$. In particular, this means that of the quasi-large gauge transformations of the previous section, we retain those with $a=b_z=0$. Since $a=0$, and $c$ is therefore an integer, we can absorb $c$ into $n$. So, quasi-large gauge transformations take the form
\be \gamma_{n,b} = e^{i n^a y_a b\cdot\sigma} \ , \ee
where $b=(b_x,b_y)$ has $|b|=1$. We note that $\gamma_{n,b}= \gamma_{-n,-b}$. Similarly, Lie algebra elements take the form
\be h = \sum_n h_n e(n) \ , \ee
with $h_n^\dagger = h_{-n}$ and $h_{-n} = \sigma_z h_n \sigma_z$. These conditions are solved by
\begin{align}
h_n+h_{-n} &= h_{n,0}I+h_{n,z}\sigma_z \ , \quad i(h_n-h_{-n}) = h_{n,x}\sigma_x+h_{n,y}\sigma_y \quad (n\not=0) \nonumber \\
h_0 &= h_{0,0} I + h_{0,z}\sigma_z \ . \label{eq:K3hProj}
\end{align}
As always, $g$ acts on $U$ and $V$ via conjugation and $h$ via commutation.

Lastly, we need the moment maps. We start with \eqref{eq:muPT4} and \eqref{eq:muRT4} and impose the orbifold projection. However, thanks to the orbifold projection we can now introduce 16 triplets of gauge-invariant FI parameters. We denote these by $\xi_{n,+}$, $\xi_{n,-}=\xi_{n,+}^*$, and $\xi_{n,\RR}$, where $n\in\Lambda$, but we stress that they only depend on the equivalence class $[n]$ of $n$ in $\Lambda/2\Lambda$. In particular, $[-n]=[n]$. The moment maps then take the form\footnote{The would-be FI parameters that we eliminated in footnote \ref{ft:fakeFI} via field redefinitions can again be eliminated here. For, the identity components of $U_n$ and $V_n$ are restricted to be odd under $n\mapsto -n$ by the orbifold projection, but they multiply components of $n$ in the moment maps, and these FI parameters are restricted to be even.}
\begin{align}
\mu_+ &= -2\sum_n e(n)\brackets{U_n n^v - V_n n^u + \sum_m [U_{n-m}, V_m] - \xi_{n,+} \sigma_z } \ , \label{eq:muPK3} \\
\mu_\RR &= \sum_n e(n)\brackets{ -n^u U^\dagger_{-n} + n^{\bar u} U_n + \sum_m [U_{n+m}, U^\dagger_m] + (U\mapsto V) - \xi_{n,\RR}\sigma_z } \ . \label{eq:muRK3}
\end{align}
Gauge invariance of these FI terms follows from
\begin{align}
g \sum_n e(n) \xi_{n,+}\sigma_z g^\dagger &= g \sum_n e(n) \xi_{n,+} \sigma_z \sum_m e(-m) g_m^\dagger \nonumber \\
&= g \sum_n e(n) \xi_{n,+} \sum_m e(-m) g_{-m}^\dagger \sigma_z \nonumber \\
&= g \sum_{m,n} e(n-m) \xi_{n,+} g_{-m}^\dagger \sigma_z \nonumber \\
&= g \sum_{m,n} e(n-m) \xi_{n-2m,+} g_m^\dagger \sigma_z \nonumber \\
&= g \sum_{m,n} e(n-m) \xi_{n,+} g_m^\dagger \sigma_z \nonumber \\
&= g g^\dagger \sum_n e(n) \xi_{n,+} \sigma_z \nonumber \\
&= \sum_n e(n) \xi_{n,+} \sigma_z \ ,
\end{align}
and the analogous computation for $\xi_{n,\RR}$. By adding and subtracting the coefficients of $e(n)$ and $e(-n)$ in the moment maps, one finds that the moment map equations $\mu_+=\mu_\RR=0$ are equivalent to
\begin{align}
2\xi_{n,+}\sigma_z &= n^v(U_n- U_{-n}) - n^u(V_n - V_{-n}) + \half \sum_{k+\ell=n} \parens{[U_k + U_{-k}, V_\ell + V_{-\ell}] + [U_k - U_{-k}, V_\ell - V_{-\ell}]} \label{eq:mm1} \\
0 &= n^v(U_n+ U_{-n}) - n^u(V_n + V_{-n}) + \half \sum_{k+\ell=n} \parens{[U_k + U_{-k}, V_\ell - V_{-\ell}] + [U_k - U_{-k}, V_\ell + V_{-\ell}]} \\
2\xi_{n,\RR}\sigma_z &= n^{\bar u}(U_n - U_{-n}) - n^u(U_{-n}^\dagger - U_n^\dagger) + \half \sum_{k+\ell=-n}\parens{[U_k+U_{-k}, U_\ell^\dagger + U_{-\ell}^\dagger]-[U_k - U_{-k}, U_\ell^\dagger - U_{-\ell}^\dagger]} \nonumber \\
&+ (U\mapsto V) \\
0 &= n^{\bar u}(U_n + U_{-n}) - n^u(U_{-n}^\dagger + U_n^\dagger) + \half \sum_{k+\ell=-n}\parens{[U_k+U_{-k}, U_\ell^\dagger - U_{-\ell}^\dagger]-[U_k - U_{-k}, U_\ell^\dagger + U_{-\ell}^\dagger]} \nonumber \\
&+ (U\mapsto V) \ . \label{eq:mm4}
\end{align}
Having written the moment map equations in this way, we can make explicit the orbifold projection by substituting in \eqref{eq:orbProjK3} (and the analogous equations for $V$). This completes our specification of the gauge theory.

We note that the moduli space
\be O(\Gamma^{3,19})\backslash O(3,19,\RR) / (O(3)\times O(19)) \times \RR^+ \ee
of Ricci-flat K3 metrics is 58-dimensional, and we similarly have 58 parameters: the 10 moduli of a metric on $T^4$, plus $3\times 16$ FI parameters.

T-duality maps the $Z_2$ involution of $T^4$ to the $Z_2$ involution of $\hat T^4$ (see, e.g., \cite{dabholkar:orientifolds}), and so the T-dual description of this theory now involves a D6-brane wrapping the orbifold $\hat T^4/Z_2$. We can describe $X^a$ as a $U(2)$ connection on a trivial bundle on $\hat T^4$ which satisfies the orbifold projections; similarly, we act on it using gauge transformations which are restricted by the orbifold projections. In this formulation, these projections take the form
\be g(-y) = \sigma_z g(y) \sigma_z \ , \quad h(-y) = \sigma_z h(y) \sigma_z \ , \quad B^a(-y) = -\sigma_z B^a(y) \sigma_z \ , \label{eq:TdualProj} \ee
since, e.g., $g(-y)=\sum_n g_n e(-n)=\sum_n g_{-n} e(n)$. The 16 real (resp., complex) independent FI parameters in the real (holomorphic) moment map multiply
\be \sum_m e(2m+n) \sigma_z \sim e^{in^ay_a} \sigma_z \sum_m e^{2im^a y_a} \propto e^{in^a y_a}\sigma_z \sum_{y'} \delta^4(y-y') = \sigma_z \sum_{y'} e^{in^a y'_a} \delta^4(y-y') \ , \label{eq:FITDual} \ee
where $n$ runs over representatives of $\Lambda/2\Lambda$ and $y'$ runs over the fixed points of the $Z_2$ action on $\hat T^4$. Note that $e^{in^ay'_a}=\pm 1$, and so \eqref{eq:FITDual} is Hermitian (and can therefore appear in the real moment map). The anti-self-duality equations are modified by these FI parameters to take the form
\be F = -*F + \sigma_z \sum_{y'} \eta_{y'} \delta^4(y-y') \ , \label{eq:modifiedEqns} \ee
where $\eta_{y'}$ is a constant real anti-symmetric 2-tensor which depends on $y'$. Application of the Hodge star to this equation shows that each $\eta_{y'}$ must be self-dual; this matches with the fact that we have 3 real FI parameters per $y'$. Gauge invariance of the FI parameters, in this language, follows from the fact that \eqref{eq:TdualProj} implies that $g(y')$ commutes with $\sigma_z$.

\eqref{eq:modifiedEqns} makes it clear that FI parameters cause our connections to be singular at the fixed points $y'$. We expect that this singular connection may alternatively be thought of as a smooth connection on a topologically non-trivial bundle on a resolution of $\hat T^4/Z_2$, in the same way that an orbifold CFT deformed by twist fields coincides with a K3 non-linear sigma model.\footnote{Such a correspondence is also known to mathematicians, at the level of (twisted) sheaves \cite{bridgeland:derivedMcKay,chen:twistedMcKay}. However, in this case it seems likely that (twisted) vector bundles suffice on the resolved space, as the moduli space is already a smooth K3 surface, so there are no small instanton singularities that need resolving.} Indeed, \cite{waldram:lol} explains that one has a D4-brane wrapping half the sum of the 16 collapsed 2-cycles at the fixed points (which is in the integral homology lattice), as well as a $1/4$-D2-brane at each fixed point, for a total of 4 D2-branes. So, after resolving $\hat T^4/Z_2$ one should obtain a rank 2 bundle with $c_1^2=-\half\cdot 16=-8$ and $ch_2=-4$ (i.e., $c_2=\half c_1^2-ch_2=0$) (twisted by half a unit of B-field on each resolved 2-cycle).\footnote{As a check, we note that the Mukai vector, i.e. the vector of (D6,D4,-D2) charges (which includes contributions to the D2 charge from D6-branes \cite{vafa:k3inst,moore:iBrane,Harvey:1996gc}), is $Q=(Q_6,Q_4,-Q_2)=(r,c_1,r+ch_2)=(2,c_1,-2)$ (where $r$ is the rank), and the inner product on the integral homology lattice gives $Q^2=8+c_1^2=0$. Similarly, in the D2 frame we have $Q=(0,0,-1)$ and $Q^2=0$. So, this duality invariant is the same in the two frames. Furthermore, the complex dimension of the moduli space is $Q^2+2=2$, which agrees with the fact that it is a K3 surface.}

Lastly, we note the curious fact that the FI terms are not in the Lie algebra, contrary to usual expectations, since the Lie algebra only involves smooth maps from $\hat T^4$ to $\mf{u}(2)$. And, accordingly, we cannot exponentiate the FI terms to yield central factors of the gauge group to which they correspond. To understand this, we recall that FI terms normally need to be in the Lie algebra so that we can evaluate the trace of their product with some adjoint superfield in the vector multiplet. But, this actually only requires FI parameters to be in the dual vector space to the Lie algebra, a.k.a. the Lie coalgebra! For finite-dimensional Lie algebras, this distinction is immaterial, but here we see the importance of the duality between function spaces and distribution spaces. More generally, the moment maps are valued in the Lie coalgebra / coadjoint representation.

We now turn to the study of the moduli space of our gauge theory. We begin by considering the case where all FI parameters vanish. The D6-brane point of view makes it clear that the non-zero modes -- that is, $U_n,V_n$ with $n\not=0$ -- can be taken to vanish, as they either have Kaluza-Klein masses or can be gauged away. To see this explicitly, note that we are studying the moduli space of flat $U(2)$ connections on $\hat T^4$ which satisfy \eqref{eq:TdualProj}, modulo $U(2)$ gauge transformations which satisfy \eqref{eq:TdualProj}. Flatness implies that $B=B^a dy_a$ is a closed 1-form valued in a Cartan subalgebra of $\mf{u}(2)$. \eqref{eq:TdualProj} shows that conjugation by $\sigma_z$ must preserve this Cartan subalgebra, so the latter is either of the form $\avg{I,\sigma_z}$ or $\avg{I,\alpha\sigma_x+\beta\sigma_y}$. Any exact piece $d\lambda$ of $B$ that satisfies \eqref{eq:TdualProj} can be gauged away, using transformations that satisfy \eqref{eq:TdualProj}. For, \eqref{eq:TdualProj} implies that $\partial^a(\lambda(-y))=\sigma_z\partial^a \lambda(y) \sigma_z$, so $\lambda(-y)=\sigma_z\lambda(y)\sigma_z+\kappa$ for all $y$, where $\kappa$ is a constant element of the Cartan subalgebra. Exchanging $y$ and $-y$ shows that $\kappa$ is in the span of $\sigma_x$ and $\sigma_y$. $\lambda(-y)-\kappa/2=\sigma_z\lambda(y)\sigma_z+\kappa/2=\sigma_z(\lambda(y)-\kappa/2)\sigma_z$ then implies that $g=\exp(-i(\lambda-\kappa/2))$ satisfies \eqref{eq:TdualProj} and can be used to cancel off $d\lambda$. The orbifold condition on $B^a$ does not mix the zero- and non-zero modes, so they must separately satisfy the condition. Since the non-zero modes make an exact contribution to $B$ (since all cohomology classes have a representative in the span of $dy^a$) which satisfies \eqref{eq:TdualProj}, we can gauge them away and assume that $B^a$ are constants.\footnote{Note that this same argument allows us to gauge away the non-zero modes of the $\mf{u}(1)$ components of $B^a$ even after we turn on the FI parameters. The orbifold projection further implies that the zero modes of these components vanish. So, we can always take $B$ to be in $\mf{su}(2)$ and $g$ to be in $SU^\pm(2)$, the group of unitary matrices with determinant $\pm1$. (We choose this group, rather than $SU(2)$, so that it contains the non-trivial element, $\sigma_z$, of the regular representation of $Z_2$. However, for the present purposes, this distinction is immaterial, since $\hat T^4$ is connected, so $\det g$ is constant, and transformations with $\det g=-1$ correspond to those with $\det g=1$ via multiplication by $i$, which has no effect on gauge transformations of the connection.) In contrast to this singular connection on a trivial bundle over $\hat T^4$, the corresponding smooth connection on a rank 2 bundle over a resolution of $\hat T^4/Z_2$ with non-trivial $c_1$ cannot be taken to be an $SU^\pm(2)$ conection, since a traceless connection has vanishing $c_1$.}

Substituting $U_n=V_n=0$ for $n\not=0$ into the moment map equations trivializes all of them except for the zero mode equations, which become identical to those of \S\ref{sec:douglasMoore}! Furthermore, the $U(1)^2$ subgroup of our gauge group consisting of elements of the form \eqref{eq:EHgauge}, i.e.
\be g=e^{i(\theta I + \alpha\sigma_z/2)} e(0) =e^{i\theta}\twoMatrix{e^{i\alpha/2}}{}{}{e^{-i\alpha/2}} e(0) \ , \ee
preserves $U_n=V_n=0$ for $n\not=0$. As in \S\ref{sec:douglasMoore}, $\theta$ acts trivially, but $\alpha$ allows us to set $u_y=v_y=0$, and in addition implements the quotient $(u_x,v_x)\sim (-u_x,-v_x)$. Because $u_x$ and $v_x$ play the privileged role of coordinates on our moduli space, we henceforth define
\be u\equiv u_x \ , \quad v \equiv v_x \ . \ee
Lastly, we consider the effects of the quasi-large gauge transformations. From \eqref{eq:quasiLargeEffect} and \eqref{eq:quasiLargeEffect3}, we see that if $b_y=0$ then $\gamma_{n,b}$ preserves $U_n=V_n=0$ for $n\not=0$ and $u_y=v_y=0$. Since $\gamma_{n,b}=\gamma_{-n,-b}$, we can set $b_x=1$. We are then left with the gauge transformations
\be e^{in^a y_a \sigma_x} \ , \ee
which are labelled by $n\in \Lambda$ and implement
\be (u,v)\sim (u+n^u,v+n^v) \ . \ee
We thus see that our moduli space is precisely $T^4/Z_2$!

We now perturb around this limit, working to first order in the FI parameters. Since $U_n,V_n$ $(n\not=0)$ and $u_y,v_y$ are of order $\xi$, the moment map equations and continuous gauge transformations decouple, $n$ by $n$! Explicitly, for $n\not=0$ we have the scalar equations
\begin{align}
0 &= n^v U_{n,0} - n^u V_{n,0} \label{eq:mm0P} \\
2\xi_{n,+} &= n^v U_{n,z} - n^u V_{n,z} + 2i(u V_{n,y}-v U_{n,y}) \\
0 &= n^v U_{n,x} - n^u V_{n,x} \label{eq:mmxP} \\
0 &= n^v U_{n,y} - n^u V_{n,y} + 2i(-V_{n,z}u + U_{n,z}v) \\
0 &= n^{\bar u} U_{n,0} + n^u U_{n,0}^* + n^{\bar v} V_{n,0} + n^v V_{n,0}^* \label{eq:mm0R} \\
2\xi_{n,\RR} &= n^{\bar u} U_{n,z} + n^u U_{n,z}^* + 2i(u U_{n,y}^* - U_{n,y}u^*) + (U\mapsto V) \\
0 &= n^{\bar u} U_{n,x} - n^u U_{n,x}^* + (U\mapsto V) \label{eq:mmxR} \\
0 &= n^{\bar u} U_{n,y} - n^u U^*_{n,y} + 2i(U^*_{n,z}u + U_{n,z}u^*) + (U\mapsto V) \ ,
\end{align}
while for $n=0$ we again have the equations from \S\ref{sec:douglasMoore}:
\begin{align}
\xi_+ &= 2i(u v_y-u_y v)  \\
\xi_\RR &= 2i(u u_y^* - u_y u^* + (U\mapsto V)) \ . 
\end{align}
We have used up most of the continuous gauge freedom in making $U_n,V_n,u_y,v_y$ order $\xi$, but we still have the freedom to perform order $\xi$ gauge transformations. \eqref{eq:smallTrans} shows that, to first order in $\xi$, the Hermitian generator $h_n e(n) + h_{-n} e(-n)$ (where $n\not=0$) only changes $X^a_{\pm n}$, and it does so via
\be \delta X_{\pm n}^a = i\epsilon([h_{\pm n},X_{0,x}^a\sigma_x] \pm n^a h_{\pm n}) \ . \ee
That is,
\be \delta X^a_n\pm \delta X^a_{-n} = i\epsilon\parens{[h_n\pm h_{-n}, X^a_{0,x}\sigma_x] + n^a(h_n\mp h_{-n})} \ . \ee
Substituting in \eqref{eq:K3hProj} gives
\begin{align}
\delta X^a_n + \delta X^a_{-n} &= \epsilon\parens{-2 h_{n,z} X^a_{0,x} \sigma_y + n^a(h_{n,x} \sigma_x + h_{n,y}\sigma_y)} \label{eq:nGaugeP} \\
\delta X^a_n - \delta X^a_{-n} &= i\epsilon\parens{ -2 h_{n,y} X^a_{0,x} \sigma_z + n^a(h_{n,0}I+h_{n,z}\sigma_z)} \ . \label{eq:nGaugeM}
\end{align}
Note that $h_{n,x}$ only affects $X^a_{n,x}$, and $h_{n,0}$ only affects $X^a_{n,0}$. And, conversely, only $h_{n,x}$ affects $X^a_{n,x}$, and only $h_{n,0}$ affects $X^a_{n,0}$. Furthermore, the moment map equations that involve $X^a_{n,x}$ and $X^a_{n,0}$, \eqref{eq:mm0P}, \eqref{eq:mmxP}, \eqref{eq:mm0R}, and \eqref{eq:mmxR}, involve no other components of $X^a$ and are not deformed by the FI parameters. So, $(X^a_{n,x},X^a_{n,0},h_{n,x},h_{n,0})$ are isolated from everything else, and as in the case with no FI parameters we can set $X^a_{n,x}=X^a_{n,0}=0$ via a combination of the moment map equations and gauge transformations. We then have 6 real linear equations in the 8 unknowns $Q_n=(U_{n,y},U^*_{n,y},U_{n,z},U_{n,z}^*,V_{n,y},V_{n,y}^*,V_{n,z},V_{n,z}^*)^T$. As in \S\ref{sec:douglasMoore}, we write these as
\be L_n(q)Q_n = \xi_n \ , \ee
where $q=(u,u^*,v,v^*)$, $L_n(q)$ is the $6\times 8$ matrix of coefficients, and $\xi_n=(2\xi_{n,+},2\xi_{n,-},0,0,2\xi_{n,\RR},0)^T$. Lastly, we deal with $n=0$. To first order, $h_0 e(0)$ only changes $X^a_0$, and it does so via
\be \delta X^a_0 = i\epsilon [h_0,X^a_{0,x}\sigma_x] = -2\epsilon h_{0,z}X^a_{0,x} \sigma_y \ . \label{eq:0Gauge} \ee
This is simply the $U(1)$ transformation from \S\ref{sec:douglasMoore}. We write the zero-mode moment map equations as
\be L_0(q) Q_0 = \xi_0 \ , \ee
where $L_0(q)$ is now a $3\times 4$ matrix, $Q_0=(u_y,u_y^*,v_y,v_y^*)$, and $\xi_0=(\xi_{0,+},\xi_{0,-},\xi_{0,\RR})^T$.

In terms of the complexified fields $U$ and $V$, the gauge transformations \eqref{eq:nGaugeP}, \eqref{eq:nGaugeM}, and \eqref{eq:0Gauge} take the form
\begin{align}
\delta U_{n,0} &= i\epsilon n^u h_{n,0} \\
\delta U_{n,x} &= \epsilon n^u h_{n,x} \\
\delta U_{n,y} &= \epsilon\parens{-2h_{n,z} u + n^u h_{n,y}} \\
\delta U_{n,z} &= i\epsilon\parens{-2h_{n,y} u + n^u h_{n,z}} \\
\delta u_y &= -2\epsilon h_{0,z} u \ ,
\end{align}
and similarly for $V$. One may explicitly verify that the moment map equations are invariant under these transformations. Before solving the moment map equations, we need to pick a gauge. There are gauge fixing conditions analogous to the $\Real \frac{u_y}{u_x}=0$ gauge that we studied in \S\ref{sec:douglasMoore}. In addition, for generic $n$ and $u$ we can make choices such as $U_{n,y}=0$. However, we find that many results are particularly nice in the `least norm gauge' introduced in \S\ref{sec:douglasMoore}:
\be Q_n = L_n^\dagger(L_n L_n^\dagger)^{-1}\xi_n \ . \label{eq:embedding} \ee
Other gauges obtain by adding a (possibly $q$-dependent) order $\xi$ element of the nullspace of $L_n$ to $Q_n$. Henceforth adopting least norm gauge, and defining
\be N^u_\pm = n^u \pm 2 u \ , \quad N^v_\pm = n^v \pm 2 v \ , \quad D_\pm = |N^u_\pm|^2 + |N^v_\pm|^2 \ , \label{eq:PSdef} \ee
we have
\begin{align}
U_{n,y} &= \frac{i(2\xi_{n,+}\bar N^v_+ + \xi_{n,\RR} N^u_+)}{2 D_+} - \frac{i(2\xi_{n,+}\bar N^v_- + \xi_{n,\RR} N^u_-)}{2 D_-} \\
U_{n,z} &= \frac{ 2\xi_{n,+}\bar N^v_+ + \xi_{n,\RR} N^u_+}{2 D_+} + \frac{2\xi_{n,+}\bar N^v_- + \xi_{n,\RR} N^u_-}{2 D_-} \\
V_{n,y} &= \frac{i(-2\xi_{n,+}\bar N_+^u+\xi_{n,\RR} N^v_+)}{2 D_+} - \frac{i(-2\xi_{n,+}\bar N_-^u+\xi_{n,\RR} N^v_-)}{2 D_-} \\
V_{n,z} &= \frac{-2\xi_{n,+}\bar N^u_+ + \xi_{n,\RR} N^v_+}{2 D_+} + \frac{-2\xi_{n,+}\bar N^u_- + \xi_{n,\RR} N^v_-}{2 D_-} \\
u_y &= \frac{i(2\xi_{0,+} \bar v + \xi_{0,\RR} u)}{4(|u|^2+|v|^2)} \\
v_y &= \frac{i(-2\xi_{0,+}\bar u + \xi_{0,\RR} v)}{4(|u|^2+|v|^2)} \ .
\end{align}
Up to global issues, this defines embeddings (to first order in the FI parameters) of K3 surfaces into an infinite-dimensional flat space!

We do not concern ourselves with such global issues -- namely, the effects of both quasi-large gauge transformations and the zero-mode $Z_2\subset U(1)$, as well as choosing good coordinates near the fixed points of $T^4/Z_2$. (Looking at $L_0$ near $u=v=0$ makes it clear that $u_y$ and $v_y$ cease to be small near the origin.) These are, of course, important, e.g. for topology, but do not matter for our purposes. For, having fixed the continuous gauge symmetry, we are able to determine the hyper-K\"ahler structure of K3 near a generic point $(u,v)$.

To do so, we first start with the K\"ahler forms on $\CC^{2\cdot (2\infty^4)^2}$, \eqref{eq:flatI}, \eqref{eq:flatJ}, and \eqref{eq:flatK}, and impose the $T^4$ orbifold projections to get K\"ahler forms on $\CC^{2\cdot 2^2\infty^4}$:\footnote{Here, we define $\Tr e(n)=\delta_{n,0}$. The definition \eqref{eq:eDef} would instead suggest $\Tr e(n)=\delta_{n,0}\sum_m 1\propto \delta_{n,0}\delta^4(0)$, where $\delta^4(0)$ is a delta function on $\hat T^4$. This discrepancy can formally be addressed by rescaling the Yang-Mills coupling of the pre-orbifold-projection $U(2\infty^4)$ gauge theory by the order of the $D_{\infty^4}$ orbifold group (up to an irrelevant factor of 2 that can be absorbed into $g^2$). It is of no concern, since we only care about the orbifold projected theory \cite{wati:dBraneT}.}
\begin{align}
\tilde \omega_I &= i \sum_n \Tr \parens{-dU_n\wedge dV_{-n} + dU_n^\dagger \wedge dV_{-n}^\dagger} \\
\tilde \omega_J &= -\sum_n \Tr \parens{dU_n\wedge dV_{-n} + dU_n^\dagger \wedge dV_{-n}^\dagger} \\
\tilde \omega_K &= i \sum_n \Tr \parens{dU_n \wedge dU^\dagger_n + dV_n\wedge dV^\dagger_n } \ .
\end{align}
We then further impose the $Z_2$ orbifold projection, to get K\"ahler forms on $\CC^{2\cdot 2\infty^4}$. Using
\be U_n = \frac{U_n+U_{-n}}{2}+\frac{U_n-U_{-n}}{2}=\half\parens{U_{n,0}I+U_{n,x}\sigma_x+U_{n,y}\sigma_y+U_{n,z}\sigma_z} \ , \ee
as well as $U_{-n,x}=U_{n,x}$, $U_{-n,y}=U_{n,y}$, $U_{-n,z}=-U_{n,z}$, and $U_{-n,0}=-U_{n,0}$, and similar equations for $V$, we have
\begin{align}
\tilde \omega_I &= \frac{i}{2} \sum_{n\not=0} \parens{-dU_{n,x}\wedge dV_{n,x}-dU_{n,y}\wedge dV_{n,y}+dU_{n,z}\wedge dV_{n,z}+dU_{n,0}\wedge dV_{n,0} - {\rm c.c.}} \nonumber \\
&+2i\parens{-du\wedge dv-du_y\wedge dv_y+du^*\wedge dv^*+du_y^*\wedge dv_y^*} \\
\tilde \omega_J &= - \frac{1}{2} \sum_{n\not=0} \parens{dU_{n,x}\wedge dV_{n,x}+dU_{n,y}\wedge dV_{n,y}-dU_{n,z}\wedge dV_{n,z}-dU_{n,0}\wedge dV_{n,0}+{\rm c.c.}} \nonumber \\
&-2\parens{du\wedge dv+du_y\wedge dv_y+du^*\wedge dv^*+du_y^*\wedge dv_y^*} \\
\tilde \omega_K &= \frac{i}{2} \sum_{n\not=0} \parens{dU_{n,x}\wedge dU^*_{n,x}+dU_{n,y}\wedge dU^*_{n,y}+dU_{n,z}\wedge dU^*_{n,z}+dU_{n,0}\wedge dU^*_{n,0} + (U\mapsto V)} \nonumber \\
&+2i\parens{ du\wedge du^*+du_y\wedge du_y^* + dv\wedge dv^* + dv_y\wedge dv_y^* } \ .
\end{align}
We finally pull back these K\"ahler forms to K3, using \eqref{eq:embedding}. That is, we substitute
\be dU_{n,y} = \frac{\partial U_{n,y}}{\partial u} du+\frac{\partial U_{n,y}}{\partial u^*} d u^*+\frac{\partial U_{n,y}}{\partial v} dv+\frac{\partial U_{n,y}}{\partial v^*} d v^* \ , \ee
etc. When all FI parameters vanish, this yields
\be \omega_+^{\rm orb} = -4idu\wedge dv \ , \quad \omega_K^{\rm orb}=2i(du\wedge d u^*+dv\wedge d v^*) \label{eq:HiggsSF} \ .\ee
The `orb' superscripts stand for `orbifold.'

We now state the leading corrections due to FI parameters. We write them as
\begin{align}
\varpi(\zeta) &= \varpi^{\rm orb}(\zeta) + \varpi^{\rm pert}(\zeta) \\
\varpi^{\rm pert}(\zeta) &= -\frac{i}{2\zeta} \omega_+^{\rm pert} + \omega_K^{\rm pert} - \frac{i\zeta}{2}\omega_-^{\rm pert} \\
&= \sum_{n} \parens{-\frac{i}{2\zeta}\omega'_{n+}+\omega'_{nK}-\frac{i\zeta}{2}\omega'_{n-} } \ .
\end{align}
Here, `pert' stands for `perturbation.' For all $n$, we have
\begin{align}
\omega'_{n+\, u\bar u} &= \frac{2i(2\xi_{n,+} \bar N^v_+ + \xi_{n,\RR} N^u_+)(2\xi_{n,+} \bar N^u_+ - \xi_{n,\RR} N^v_+)}{D_+^3} + \frac{2i(2\xi_{n,+} \bar N^v_- + \xi_{n,\RR} N^u_-)(2\xi_{n,+} \bar N^u_- - \xi_{n,\RR} N^v_-)}{D_-^3} \nonumber \\
\omega'_{n+\, uv} &= 0 \nonumber \\
\omega'_{n+\, u\bar v} &= - \frac{2i(2\xi_{n,+}\bar N^u_+ - \xi_{n,\RR} N^v_+)^2}{D_+^3} - \frac{2i(2\xi_{n,+}\bar N^u_- - \xi_{n,\RR} N^v_-)^2}{D_-^3} \nonumber \\
\omega'_{n+\, \bar u v} &= - \frac{2i(2\xi_{n,+} \bar N^v_+ + \xi_{n,\RR} N^u_+)^2}{D_+^3} - \frac{2i(2\xi_{n,+} \bar N^v_- + \xi_{n,\RR} N^u_-)^2}{D_-^3} \nonumber \\
\omega'_{n+\, \bar u \bar v} &= 0 \nonumber \\
\omega'_{n+\, v \bar v} &= - \omega'_{n+\, u\bar u} \nonumber \\
\omega'_{nK\, u\bar u} &= - \frac{i\parens{(-4|\xi_{n,+}|^2+\xi_{n,\RR}^2)(|N^u_+|^2-|N^v_+|^2)+4\xi_{n,\RR}(\xi_{n,+}\bar N^u_+ \bar N^v_+ + \xi_- N^u_+ N^v_+)}}{D_+^3} \nonumber \\
&\quad - \frac{i\parens{(-4|\xi_{n,+}|^2+\xi_{n,\RR}^2)(|N^u_-|^2-|N^v_-|^2)+4\xi_{n,\RR}(\xi_{n,+}\bar N^u_- \bar N^v_- + \xi_- N^u_- N^v_-)}}{D_-^3} \nonumber \\
\omega'_{nK\, uv} &= 0 \nonumber \\
\omega'_{nK\, u\bar v} &= \frac{2i(2\xi_{n,-} N^v_+ + \xi_{n,\RR}\bar N^u_+)(2\xi_{n,+}\bar N^u_+ - \xi_{n,\RR} N^v_+)}{D_+^3} + \frac{2i(2\xi_{n,-} N^v_- + \xi_{n,\RR}\bar N^u_-)(2\xi_{n,+}\bar N^u_- - \xi_{n,\RR} N^v_-)}{D_-^3} \nonumber \\
\omega'_{nK\, \bar u v} &= \frac{2i(-2\xi_{n,-} N^u_+ + \xi_{n,\RR}\bar N^v_+)(2\xi_{n,+}\bar N^v_+ + \xi_{n,\RR} N^u_+)}{D_+^3} + \frac{2i(-2\xi_{n,-} N^u_- + \xi_{n,\RR}\bar N^v_-)(2\xi_{n,+}\bar N^v_- + \xi_{n,\RR} N^u_-)}{D_-^3} \nonumber \\
\omega'_{nK\, \bar u \bar v} &= 0 \nonumber \\
\omega'_{nK\, v \bar v} &= -\omega'_{nK\, u\bar u} \ .
\end{align}
For later convenience, we combine terms from the $n$ term and the $-n$ term to obtain
\begin{align}
\varpi^{\rm pert}(\zeta) &= \sum_{n} \parens{-\frac{i}{2\zeta}\omega_{n+}+\omega_{nK}-\frac{i\zeta}{2}\omega_{n-} } \label{eq:higgsTower} \\
\omega_{n+\, u\bar u} &= \frac{4i(2\xi_{n,+} \bar N^v_+ + \xi_{n,\RR} N^u_+)(2\xi_{n,+} \bar N^u_+ - \xi_{n,\RR} N^v_+)}{D_+^3} \nonumber \\
\omega_{n+\, uv} &= 0 \nonumber \\
\omega_{n+\, u\bar v} &= - \frac{4i(2\xi_{n,+}\bar N^u_+ - \xi_{n,\RR} N^v_+)^2}{D_+^3} \nonumber \\
\omega_{n+\, \bar u v} &= - \frac{4i(2\xi_{n,+} \bar N^v_+ + \xi_{n,\RR} N^u_+)^2}{D_+^3} \nonumber \\
\omega_{n+\, \bar u \bar v} &= 0 \nonumber \\
\omega_{n+\, v \bar v} &= - \omega_{n+\, u\bar u} \nonumber \\
\omega_{nK\, u\bar u} &= - \frac{2i\parens{(-4|\xi_{n,+}|^2+\xi_{n,\RR}^2)(|N^u_+|^2-|N^v_+|^2)+4\xi_{n,\RR}(\xi_{n,+}\bar N^u_+ \bar N^v_+ + \xi_- N^u_+ N^v_+)}}{D_+^3} \nonumber \\
\omega_{nK\, uv} &= 0 \nonumber \\
\omega_{nK\, u\bar v} &= \frac{4i(2\xi_{n,-} N^v_+ + \xi_{n,\RR}\bar N^u_+)(2\xi_{n,+}\bar N^u_+ - \xi_{n,\RR} N^v_+)}{D_+^3} \nonumber \\
\omega_{nK\, \bar u v} &= \frac{4i(-2\xi_{n,-} N^u_+ + \xi_{n,\RR}\bar N^v_+)(2\xi_{n,+}\bar N^v_+ + \xi_{n,\RR} N^u_+)}{D_+^3} \nonumber \\
\omega_{nK\, \bar u \bar v} &= 0 \nonumber \\
\omega_{nK\, v \bar v} &= -\omega_{nK\, u\bar u} \ . \label{eq:hkForms}
\end{align}
We have thus determined the hyper-K\"ahler structure of K3 (to first order in $\xi$)!

Using \eqref{eq:fromW}, we can now determine the metric and complex structures. These take the form
\be g = g^{\rm orb}+\sum_n g_n \ , \quad J_\sigma = J^{\rm orb}_\sigma + \sum_n J_{n\sigma} \ , \ee
where
\begin{align}
g_n &= - \omega_I^{\rm orb}(\omega_J^{\rm orb})^{-1} \omega_{nK} + \omega_I^{\rm orb}(\omega_J^{\rm orb})^{-1} \omega_{nJ}(\omega_J^{\rm orb})^{-1}\omega_K^{\rm orb} - \omega_{nI} (\omega_J^{\rm orb})^{-1} \omega_K^{\rm orb} \\
J_{nI} &= -(\omega_J^{\rm orb})^{-1}\omega_{nK} + (\omega_J^{\rm orb})^{-1}\omega_{nJ} (\omega_J^{\rm orb})^{-1} \omega_K^{\rm orb} \\
J_{nJ} &= -(\omega_K^{\rm orb})^{-1}\omega_{nI} + (\omega_K^{\rm orb})^{-1}\omega_{nK} (\omega_K^{\rm orb})^{-1} \omega_I^{\rm orb} \\
J_{nK} &= -(\omega_I^{\rm orb})^{-1}\omega_{nJ} + (\omega_I^{\rm orb})^{-1}\omega_{nI} (\omega_I^{\rm orb})^{-1} \omega_J^{\rm orb} \ . 
\end{align}
Explicitly, the orbifold expressions, written in matrix form with the coordinates ordered $u,\bar u,v,\bar v$, are
\begin{align}
(J_I^{\rm orb})^\bullet{}_\bullet &= \begin{pmatrix} &&&-i\\ &&i&\\ &i&&\\ -i&&& \end{pmatrix} \ , \quad (J_J^{\rm orb})^\bullet{}_\bullet = \begin{pmatrix} &&&1\\&&1&\\&-1&&\\-1&&&\end{pmatrix} \ , \quad (J_K^{\rm orb})^\bullet{}_\bullet = \begin{pmatrix} i & & & \\ & -i &&\\ &&i&\\ &&&-i \end{pmatrix} \nonumber \\
g^{\rm orb}{}_{\bullet\bullet} &= \begin{pmatrix} &2&&\\ 2&&& \\ &&&2 \\ &&2& \end{pmatrix} \ , \label{eq:orb}
\end{align}
while the corrections take the form
\begin{align}
J_{n\sigma}{}^v{}_{\bar u} &= J_{n\sigma}{}^u{}_{\bar v} \ , \quad 
J_{n\sigma}{}^v{}_v = - J_{n\sigma}{}^u{}_u \ ,\quad 
J_{n\sigma}{}^{\bar u}{}_u =(J_{n\sigma}{}^u{}_{\bar u})^* \ , \nonumber \\
J_{n\sigma}{}^{\bar u}{}_{\bar u} &=  (J_{n\sigma}{}^u{}_u)^* \ , \quad
J_{n\sigma}{}^{\bar u}{}_{\bar v} = (J_{n\sigma}{}^u{}_v)^* \ , \quad
J_{n\sigma}{}^{\bar u}{}_v = J_{n\sigma}{}^{\bar v}{}_u =  (J_{n\sigma}{}^u{}_{\bar v})^* \ , \nonumber \\
J_{n\sigma}{}^{\bar v}{}_{\bar u} &= (J_{n\sigma}{}^v{}_u)^* \ , \quad
J_{n\sigma}{}^{\bar v}{}_v = (J_{n\sigma}{}^v{}_{\bar v})^* \ ,\quad
J_{n\sigma}{}^{\bar v}{}_{\bar v} = (J_{n\sigma}{}^v{}_v)^* \nonumber \\
J_{nI}{}^u{}_u&= \frac{i(4\xi_{n,-}^2 N^u_+ N^v_+ - 4 \xi_{n,+}^2 \bar N^u_+ \bar N^v_+ + 2 (\xi_{n,-}-\xi_{n,+}) \xi_{n,\RR} (|N^u_+|^2-|N^v_+|^2) + \xi_{n,\RR}^2(N^u_+ N^v_+ - \bar N^u_+ \bar N^v_+))}{D_+^3} \nonumber \\
J_{nI}{}^u{}_{\bar u} &= \frac{2i(-2\xi_{n,-} N^u_+ + \xi_{n,\RR}\bar N^v_+)(2\xi_{n,+} \bar N^v_+ + \xi_{n,\RR}N^u_+)}{D_+^3} \nonumber \\
J_{nI}{}^u{}_v &= - \frac{i(4\xi_{n,-}^2 (N^u_+)^2 + 4 \xi_{n,+}^2 (\bar N^v_+)^2 + 4\xi_{n,\RR}(\xi_{n,+}-\xi_{n,-})N^u_+ \bar N^v_+ + \xi_{n,\RR}^2((\bar N^v_+)^2 + (N^u_+)^2))}{D_+^3} \nonumber \\
J_{nI}{}^u{}_{\bar v} &= - \frac{i(4\xi_{n,\RR}(\xi_{n,-}N^u_+ N^v_+ + \xi_{n,+}\bar N^u_+ \bar N^v_+) + (\xi_{n,\RR}^2 - 4|\xi_{n,+}|^2)(|N^u_+|^2 - |N^v_+|^2))}{D_+^3} \nonumber \\
J_{nI}{}^v{}_u &= \frac{i(4\xi_{n,-}^2 (N^v_+)^2 + 4\xi_{n,+}^2 (\bar N^u_+)^2 + 4\xi_{n,\RR}(\xi_{n,-}-\xi_{n,+})N^v_+ \bar N^u_+ + \xi_{n,\RR}^2((N^v_+)^2 + (\bar N^u_+)^2))}{D_+^3} \nonumber \\
J_{nI}{}^v{}_{\bar v} &= (J_{nI}{}^u{}_{\bar u})^* \nonumber \\
J_{nJ}{}^u{}_u&= \frac{-4(\xi_{n,-}^2 N^u_+ N^v_+ + \xi_{n,+}^2 \bar N^u_+ \bar N^v_+) - 2 (\xi_{n,-}+\xi_{n,+}) \xi_{n,\RR} (|N^u_+|^2-|N^v_+|^2) + \xi_{n,\RR}^2(N^u_+ N^v_+ + \bar N^u_+ \bar N^v_+)}{D_+^3} \nonumber \\
J_{nJ}{}^u{}_{\bar u} &= \frac{2(2\xi_{n,-} N^u_+ - \xi_{n,\RR}\bar N^v_+)(2\xi_{n,+} \bar N^v_+ + \xi_{n,\RR}N^u_+)}{D_+^3} \nonumber \\
J_{nJ}{}^u{}_v &= \frac{(2\xi_{n,-}N^u_+ + 2\xi_{n,+}\bar N^v_+ - \xi_{n,\RR}(\bar N^v_+-N^u_+))(2\xi_{n,-}N^u_+ - 2\xi_{n,+}\bar N^v_+ - \xi_{n,\RR}(\bar N^v_++N^u_+))}{D_+^3} \nonumber \\
J_{nJ}{}^u{}_{\bar v} &= \frac{4\xi_{n,\RR}(\xi_{n,-}N^u_+ N^v_+ + \xi_{n,+} \bar N^u_+ \bar N^v_+) + (-4|\xi_{n,+}|^2+\xi_{n,\RR}^2)(|N^u_+|^2-|N^v_+|^2)}{D_+^3} \nonumber \\
J_{nJ}{}^v{}_u &= - \frac{(2\xi_{n,-} N^v_+ - 2\xi_{n,+} \bar N^u_+ + \xi_{n,\RR}(N^v_+ + \bar N^u_+))(2\xi_{n,-} N^v_+ + 2 \xi_{n,+} \bar N^u_+ - \xi_{n,\RR}(N^v_+ - \bar N^u_+))}{D_+^3} \nonumber \\
J_{nJ}{}^v{}_{\bar v} &= - (J_{nJ}{}^u{}_{\bar u})^* \nonumber \\
J_{nK}{}^u{}_u &= J_{nK}{}^u{}_v = J_{nK}{}^v{}_u = 0 \nonumber \\
J_{nK}{}^u{}_{\bar u} &= \frac{2i(2\xi_{n,+}\bar N^v_+ + \xi_{n,\RR} N^u_+)^2}{D_+^3} \nonumber \\
J_{nK}{}^u{}_{\bar v} &= - \frac{2i(2\xi_{n,+} \bar N^v_+ + \xi_{n,\RR} N^u_+)(2\xi_{n,+}\bar N^u_+ - \xi_{n,\RR} N^v_+)}{D_+^3} \nonumber \\
J_{nK}{}^v{}_{\bar v} &= \frac{2i(-2\xi_{n,+} \bar N^u_+ + \xi_{n,\RR} N^v_+)^2}{D_+^3} \nonumber \\
g_{n\, uu} &= (g_{n\, \bar u\bar u})^* = \frac{4(2\xi_{n,-}N^v_+ + \xi_{n,\RR}\bar N^u_+)^2}{D_+^3} \nonumber \\
g_{n\, u\bar u} &= -g_{n\, v\bar v} = - \frac{2(4\xi_{n,\RR}(\xi_{n,-}N^u_+ N^v_+ + \xi_{n,+}\bar N^u_+ \bar N^v_+) + (-4|\xi_{n,+}|^2+\xi_{n,\RR}^2)(|N^u_+|^2-|N^v_+|^2))}{D_+^3} \nonumber \\
g_{n\, uv} &= (g_{n\, \bar u \bar v})^* = - \frac{4(2\xi_{n,-}N^v_+ + \xi_{n,\RR}\bar N^u_+)(2\xi_{n,-} N^u_+ - \xi_{n,\RR} \bar N^v_+)}{D_+^3} \nonumber \\
g_{n\, u\bar v} &= (g_{n\, \bar u v})^* = - \frac{4(2\xi_{n,-}N^v_+ + \xi_{n,\RR} \bar N^u_+)(-2\xi_{n,+}\bar N^u_+ + \xi_{n,\RR}N^v_+)}{D_+^3} \nonumber \\
g_{n\, vv} &= (g_{n\, \bar v\bar v})^* = \frac{4(-2\xi_{n,-} N^u_+ + \xi_{n,\RR}\bar N^v_+)^2}{D_+^3} \ .
\end{align}
(The equalities in the first 3 rows hold for all $\sigma=I,J,K$.) Having worked out these formulae, we may verify properties such as Ricci-flatness and \eqref{eq:quats}. We provide a couple examples of such calculations. Working to leading (i.e., quadratic) order in $\xi$, we have
\begin{align}
R_{ik\ell m}&=\half\parens{g_{im,k\ell}+g_{k\ell,im}-g_{i\ell,km}-g_{km,i\ell}}+g_{np}\parens{\Gamma^n_{k\ell}\Gamma^p_{im}-\Gamma^n_{km}\Gamma^p_{i\ell}} \nonumber \\
&\approx \half\sum_n\parens{g_{n\, im,k\ell}+g_{n\, k\ell,im}-g_{n\, i\ell,km}-g_{n\, km,i\ell}} \nonumber \\
R_{km} &= R^\ell{}_{k\ell m} \approx (g^{\rm orb})^{\ell i} R_{ik\ell m} \nonumber \\
&\approx \half \sum_n (g^{\rm orb})^{\ell i}\parens{g_{n\, im,k\ell}+g_{n\, k\ell,im}-g_{n\, i\ell,km}-g_{n\, km,i\ell}}\ ,
\end{align}
and one may verify that the summand vanishes for each $n$. Similarly,
\be J_\sigma^2 \approx (J^{\rm orb}_\sigma)^2 + \sum_n \{J^{\rm orb}_\sigma, J_{n\sigma}\} \ , \ee
and these anticommutators vanish for all $n$ and $\sigma$.

We now explain how this perturbation theory may be carried beyond first order. One may fear that we lose the crucial simplifying feature that everything decouples, $n$ by $n$, and so instead of a finite-dimensional linear algebra problem we face an infinite-dimensional non-linear algebra problem. Fortunately, this is not the case. Suppose that one has solved for $Q_n$ to order $\xi^{\nu-1}$. We now explain how to improve this approximation to order $\xi^\nu$. Write $Q_n=Q_n^{(\nu-1)}+\delta Q_n^{(\nu)}$, where the perturbations are of order $\xi^{\nu}$. Then, the infinite sum in \eqref{eq:mm1} takes the form
\be f(\{U_k^{(\nu-1)}\},\{V_\ell^{(\nu-1)}\}) + \parens{u[\sigma_x,\delta V_n^{(\nu)}+\delta V_{-n}^{(\nu)}]+v[\delta U_n^{(\nu)}+\delta U_{-n}^{(\nu)}, \sigma_x]} + \Oo(\xi^{\nu+1}) \ . \ee
The other infinite sums in the moment maps behave similarly. So, at each order $\nu$, we again have a simple linear algebra problem -- indeed, nearly the same one we solved above (i.e., the $L_n$ are unchanged)!\footnote{By this, we really mean the $9\times 12$ matrices that include the equations that determine $U_{n,x}$ and $V_{n,x}$. For, these variables (and their conjugates) now must be included in $Q_n$.} The only change is in the vector $\xi_n$, whose entries involve infinite sums over $\{U_k^{(\nu-1)}\},\{V_\ell^{(\nu-1)}\}$. That is, we iteratively compute
\be \delta Q_n^{(\nu)} = L_n^\dagger(L_n L_n^\dagger)^{-1} \xi_n^{(\nu)} \ , \ee
where $\xi_n^{(\nu)}$ may be easily determined by looking at the moment map equations.

Lastly, we note that while we have focused on (resolutions of) the $T^4/Z_2$ orbifold, our results may be immediately applied to $(\RR^{4-N}\times T^N)/Z_2$, for $N=0,\ldots,4$ (where we take $T^0$ to be a point). We simply specialize the 4-dimensional lattice $\Lambda\subset \RR^4$ to be of the form $\Lambda'\times \Lambda''$, where the former factor is embedded in $\RR^{4-N}$ and the latter in $\RR^N$, and wherever we used to sum over $n\in\Lambda$, we now sum only over $n\in\Lambda''$ (i.e., $(0,n)\in \Lambda$). In particular, we studied the $N=0$ case in \S\ref{sec:douglasMoore}; this explains the similarities between the zero-mode moment map equations and gauge transformations in this section and those of \S\ref{sec:douglasMoore}. In the sense of \S\ref{sec:3dmirror}, for $N\ge1$ these hyper-K\"ahler quotients are 3d mirror to a $(N+2)$-dimensional theory compactified on $T^{N-1}$. Specifically, the $N=3$ case is dual to 5d $\N=1$ $SU(2)$ $N_f=8$ on $T^2$, the $N=2$ case to 4d $\N=2$ $SU(2)$ $N_f=4$ on $S^1$, the $N=1$ case to 3d $\N=4$ $SU(2)$ $N_f=2$ at finite coupling, and the $N=0$ case to 3d $\N=4$ $U(1)$ $N_f=2$ at infinite coupling. We will sometimes adopt a binary notation to label elements of $\Lambda''/2\Lambda''$. That is, we denote the FI parameters by $\xi_{\lambda,+/-/\RR}$, where $\lambda\in\{0,1\}^N$.

\section{BPS states and metrics: there and back again} \label{sec:coulomb}

We turn now to the Coulomb branch construction. As we explained in \cite{mz:k3} (to which we refer the reader for more details), following the investigations of 4d $\N=2$ field theories in \cite{GMN:walls,GMN:framed}, the heterotic 5-brane on $T^3$ point of view gives another useful way of thinking about the moduli space in the limit where the $T^3$ degenerates to $T^2\times S^1_R$, where $S^1_R$ is a circle of radius $R$ and $R$ becomes much larger than the other length scales in the problem. For, in this limit there is an intermediate energy scale at which the physics in a small patch of the moduli space is well-approximated by a 4d abelian gauge theory compactified on $S^1_R$, and the expectation values of supersymmetric Wilson-'t Hooft lines wrapping this circle provide canonical local holomorphic coordinates on the moduli space. Wall crossing phenomena (and the absence of interesting globally-defined holomorphic functions on compact spaces) force these coordinates to discontinuously jump at certain loci. Indeed, they are characterized by their asymptotics and these jumps -- i.e., by a so-called `Riemann-Hilbert problem.' The data specifying these jumps is a BPS index of the little string theory compactified on $T^2$. Given these piecewise-constant integer invariants, the solution to the Riemann-Hilbert problem given in \cite{GMN:walls} determines the metric. (Via the magic of hyper-K\"ahler geometry, specifying these canonical coordinates suffices to determine the metric, as well as all of the complex structures and K\"ahler forms.) Intuitively, the BPS states of the 4d theory yield instantons upon compactification on $S^1_R$, and the solution to the Riemann-Hilbert problem accounts for all of their effects.

Geometrically, large $R$ means that the K3 surface is nearly elliptically fibered\footnote{We will slightly modify this statement in section \S\ref{sec:lst}.} and its fibers are vanishingly small. This limit also goes by the names `large complex structure' and `semi-flat'; the latter name refers to the fact that in this limit, first studied in \cite{greene:cosmicString}, one has a Ricci-flat metric which is flat on the torus fibers. We note that this metric is singular at singular fibers.

We therefore begin with the following semi-flat orbifold geometry:
\be \omega_+ = da\wedge dz \ , \quad \omega_K = \frac{i}{2}\parens{R\tau_{F,2} \, da\wedge d\bar a + \frac{1}{R\tau_{F,2}} dz\wedge d\bar z} \ . \label{eq:CoulombSF} \ee
Here, $z$ and $a$ are coordinates on the product $(T^2_F\times T^2_B)/Z_2$, where $F$ stands for fiber and $B$ for base; these names refer to the fact that we may think of this manifold as an elliptic fibration. That is, we consider the `isotrivial' (all non-singular fibers are the same) fibration of $T^2_F$ over $T^2_B/Z_2$. These coordinates are well-defined up to
\be z \sim z+1 \sim z+\tau_F \ , \quad a \sim a+1\sim a+\tau_B \ , \quad (z,a)\sim (-z,-a) \ . \ee
We sometimes write $\tau_F=\tau_{F,1}+i\tau_{F,2}$ and $\tau_B=\tau_{B,1}+i\tau_{B,2}$.

The canonical coordinates mentioned above are denoted by
\be \Y_\gamma(\zeta) = \log \X_\gamma(\zeta) \ , \ee
where as usual $\zeta\in\CC\cup\{\infty\}$ (we will actually restrict to $\zeta\in\CC^\times$) denotes a complex structure, and where $\gamma$ labels a conserved charge (which is unbroken in the infrared) in the little string theory compactified on $T^2$. The $\X_\gamma$ are holomorphic (in complex structure $\zeta$) functions of $a,\bar a,z,\bar z$. In addition, they are piecewise holomorphic in $\zeta$, i.e., holomorphic away from certain rays, where they are discontinuous. They also depend on the parameters of the little string theory. The latter come in two varieties: there are complex mass parameters, which affect the semi-flat limit, and real mass parameters, which do not.

In contrast, the BPS invariants which appear in our smooth metrics depend only on $a$ and the complex mass parameters. We will therefore turn on only the real mass parameters, as we then have only to solve a single BPS state counting problem (as a function of $a$) in order to determine a large family of smooth K3 metrics. (Plus, complex masses cause $\tau_F$ to vary non-trivially with $a$, which complicates things.) If one wishes, one can nevertheless employ our approach with complex mass parameters. Since there are 20 real masses, plus two real parameters each from $\tau_F$ and $\tau_B$, and the real parameter $R$ (measured in units of the little string length), the solution of the counting problem on which we focus determines a 25-dimensional family of unit volume K3 metrics. (We will use `complex mass' and `flavor central charge' interchangeably, but what we mean by `taking all complex masses to vanish' is that we take all flavor central charges to be tuned in accordance with the semi-flat orbifold labelled by $\tau_B$ and $\tau_F$.)

To find the FI parameters of the previous section that we expect to match to real masses, we study the semi-flat limit. Matching the semi-flat hyper-K\"ahler structures \eqref{eq:HiggsSF} and \eqref{eq:CoulombSF} yields
\be u = \frac{i\rho}{2} a \ , \quad v = \frac{1}{2\rho} z \ , \label{eq:coordChange0} \ee
where we have defined
\be \rho = \sqrt{R\tau_{F,2}} \ . \ee
Since $z\sim z+1\sim z+\tau_F$, we have $v\sim v+\frac{1}{2\rho} \sim v+\frac{\tau_F}{2\rho}$. Similarly, we have $u\sim u+\frac{i\rho}{2}\sim u+\frac{i\rho\tau_B}{2}$. Since gauge transformations identify $v\sim v+n^v$ and $u\sim u+n^u$, we parametrize $n^u$ and $n^v$ via
\be n^u = \frac{i\rho}{2}(\tilde n^1+\tau_B \tilde n^2) \ , \quad n^v = \frac{1}{2\rho}(\tilde n^3+\tau_F \tilde n^4) \ , \label{eq:Nchange0} \ee
where $\tilde n^a\in \ZZ$. Now, since the complex masses transform like $\omega_+$ under the $U(1)$ R-`symmetry' that mixes the $I$ and $J$ complex structures while fixing the $K$ complex structure, and the same can be said for the holomorphic FI parameters $\xi_{n,+}$, one might reasonably expect that the real FI parameters match to real masses, at least at first order in perturbation theory about the semi-flat limit. To confirm this, we can examine the corrections away from the orbifold limit that were studied in the previous section. Even after turning on FI parameters, a semi-flat limit should still obtain as we take $R\to\infty$. Unfortunately, this limit is difficult to study using the form of these corrections presented in the last section, since no matter how large $R$ is we can always make $n$ larger. However, later it will become clear that in the large $R$ limit all contributions from the real FI parameters disappear. Indeed, we will show that the real FI parameters correspond to the real mass parameters.

The $\X_\gamma$ satisfy
\be \X_{\gamma} \X_{\gamma'} = (-1)^{\avg{\gamma,\gamma'}} \X_{\gamma+\gamma'} \ , \ee
where $\avg{,}$ denotes the integral symplectic pairing on the charge lattice whose existence is guaranteed by the Dirac quantization condition. So, all of these $\X_\gamma$ are determined by $\X_e\equiv\X_{\gamma^e}\,, \, \X_m\equiv \X_{\gamma^m} \, , \, \X_{\gamma^f}$, where $\{\gamma^e,\gamma^m\}$ represent a basis for the gauge charge lattice $\Gamma_a\cong \ZZ^2$ ($e$ for electric, $m$ for magnetic, referring to charges under the IR $U(1)$ gauge group), with $\avg{\gamma^m,\gamma^e}=1$, and $\gamma^f\in \Gamma_{\rm{flavor}}\cong \ZZ^{20}$ is a general flavor charge, which satisfies $\avg{\gamma^f,\gamma}=0$ for all charges $\gamma$. The $a$ subscript on $\Gamma$ refers to the fact that charges undergo monodromies around singular points of the base \cite{sw:theory1,sw}, and so the charge lattice is fibered over the base.\footnote{\label{ft:local} More precisely, it is the fiber of a local system -- that is, there is a notion of parallel transport for charges which only depends on the homotopy class of a path. We also note that the `flavor part' of a charge is not well-defined, as one can always add gauge charge to a charge without affecting the physical transformation implemented by symmetries. So, instead we have the exact sequence of local systems
\be 0\to \Gamma_{\rm{flavor}} \to \hat\Gamma \to \Gamma \to 0 \ , \ee
where $\hat\Gamma$ is locally the direct sum of the gauge and flavor lattices and $\Gamma_{\rm{flavor}}$ is a trivial local system. In particular, it makes sense to say a charge is `pure flavor,' i.e. that its gauge part vanishes, but not that it is `pure gauge.' This also means that monodromies can add linear combinations of gauge charges to global charges and, dually, that they can add complex masses to periods \cite{sw}.} Here, a gauge charge simply undergoes the monodromy $\gamma\mapsto -\gamma$ as it winds around a singular fiber (but, as described in footnote \ref{ft:local}, its monodromy when regarded as a gauge-plus-flavor charge is somewhat more complicated). 

In the semi-flat limit, the $\X_\gamma$ take the form
\be \X_\gamma^{\rm sf} = \exp\brackets{\frac{\pi R}{\zeta} Z_\gamma + i\theta_\gamma + \pi R \zeta \overline{Z_\gamma}} \ . \ee
When $\gamma$ is pure flavor, this is actually always exact. In this case, $Z:\Gamma_{\rm flavor}\to\CC$ and $\theta:\Gamma_{\rm flavor}\to \RR/2\pi\ZZ$ are homomorphisms; the former defines the flavor central charges $Z_{\gamma^f}$, which are functions (that we will specify below) of $\tau_F$ and $\tau_B$, while the latter defines the real mass parameters $\theta_{\gamma^f}$. For more general charges, $Z:\hat\Gamma_a\to \CC$ remains a homomorphism from the full gauge-plus-flavor charge lattice, while $\theta:\hat\Gamma_a\to \RR/2\pi\ZZ$ defines a twisted unitary character, i.e.,
\be e^{i(\theta_\gamma + \theta_{\gamma'})} = (-1)^{\avg{\gamma,\gamma'}} e^{i\theta_{\gamma+\gamma'}} \ . \ee
Of course, when restricted to $\Gamma_{\rm flavor}$, these functions agree with the parameters mentioned earlier in this paragraph; hence the overloaded notation. As with the $\X_\gamma$, $\theta$ and $Z$ are determined by their actions on a basis for $\hat\Gamma_a$. We denote the results of applying these functions to $\gamma^e$ and $\gamma^m$ by $Z_e\equiv Z_{\gamma^e}\equiv a$, $Z_m\equiv Z_{\gamma^m}\equiv \tau_F \, a$, $\theta_{\gamma^e}\equiv \theta_e$, and $\theta_{\gamma^m}\equiv \theta_m$. The latter two are related to the coordinate $z$ on fibers via
\be z = \frac{\theta_m-\tau_F\, \theta_e}{2\pi} \ , \quad \theta_e = \frac{i\pi}{\tau_{F,2}}(z-\bar z) \ , \quad \theta_m = \frac{i\pi}{\tau_{F,2}}(z\bar\tau_F-\bar z \tau_F) \ . \label{eq:zTheta} \ee
Note that $\X_\gamma^{\rm sf}$ is not invariant under the monodromy phenomena described in footnote \ref{ft:local} that affect $Z_\gamma$. In contrast, it is invariant under shifting $\theta_e$ or $\theta_m$ by $2\pi$.

For general charges, $\X_\gamma$ are now determined by the following integral equation:
\be \Y_\gamma(\zeta) = \Y_\gamma^{\rm sf}(\zeta) - \frac{1}{4\pi i} \sum_{\gamma'} \Omega(\gamma';a) \avg{\gamma,\gamma'} \int_{\ell_{\gamma'}(a)} \frac{d\zeta'}{\zeta'} \frac{\zeta'+\zeta}{\zeta'-\zeta} \log(1-\X_{\gamma'}(\zeta')) \ . \label{eq:intEqn} \ee
Here,
\be \ell_\gamma(a) = \{\zeta\in\CC^\times \, | \, Z_\gamma(a) / \zeta \in \RR_- \} \label{eq:ray} \ee
is a ray in the complex plane running from the origin to infinity, while $\Omega(\cdot;a) : \hat\Gamma_a\to \ZZ$ are piecewise-constant BPS invariants. Physically, they are the flavored second helicity supertrace of the little string theory on $T^2$. The factor $\frac{1}{\zeta'-\zeta}$ in the integral kernel introduces a discontinuity in $\Y_\gamma$ along $\ell_{\gamma'}(a)$. For sufficiently large $R$, \cite{GMN:walls,mz:k3} show that this equation may be solved by iteration. That is, one plugs $\X^{(0)}=\X^{\rm sf}$ into the right hand side and calls the left hand side $\Y^{(1)}$, then one plugs this in to get $\Y^{(2)}$, etc. At large $R$, $\Y^{(1)}$ is already a fantastic approximation -- that is, it yields an approximation to the hyper-K\"ahler structure that is exponentially close to the true one. (Technically, this is only true near the singular fibers if the latter are generic -- i.e., of type $I_1$ (or more generally $I_N$) -- since in this case near the singular fibers there is only a single species (or $N$ mutually local species) of light BPS particle(s) and one recovers the results of \cite{vafa:spacetimeInsts,seiberg:mirrorT}, plus exponential corrections.)

Finally, we state the relationship between these canonical coordinates $\Y_\gamma$ and the hyper-K\"ahler structure of K3:
\be \varpi(\zeta) = \frac{1}{4\pi^2 R} d\Y_m(\zeta) \wedge d\Y_e(\zeta) \ . \label{eq:darboux} \ee
(Differentials will always treat $\zeta$ as a constant.) This is the holomorphic symplectic form, as in \eqref{eq:holoSymp}. As we explained earlier, from this one may easily extract the entire hyper-K\"ahler structure. We will be particularly interested in the first approximation to this answer \cite{vafa:spacetimeInsts,seiberg:mirrorT,GMN:walls}:
\begin{align}
\varpi^{(1)}(\zeta) &= \varpi^{\rm sf}(\zeta) + \frac{1}{4\pi^2R} \parens{d(\Y^{(1)}_m(\zeta)-\Y^{{\rm sf}}_m(\zeta))\wedge d\Y_e^{\rm sf}(\zeta) + d\Y^{\rm sf}_m(\zeta)\wedge d(\Y_e^{(1)}(\zeta)-\Y_e^{\rm sf}(\zeta)) } \nonumber \\
&\equiv \varpi^{\rm sf}(\zeta) + \sum_{\gamma\in\hat\Gamma_a} \Omega(\gamma;a) \varpi^{\rm inst}_\gamma(\zeta) \nonumber \\
&\equiv \varpi^{\rm sf}(\zeta) + \varpi^{\rm inst}(\zeta) \label{eq:firstApprox} \\
\varpi^{\rm inst}_\gamma(\zeta) &= \frac{i}{8\pi^2} d\Y^{\rm sf}_\gamma(\zeta) \wedge \brackets{ |Z_\gamma| A^{\rm inst}_\gamma d\log (Z_\gamma/\overline{Z_\gamma}) - V^{\rm inst}_\gamma\parens{\frac{1}{\zeta} dZ_\gamma - \zeta d\overline{Z_\gamma} } } \\
A^{\rm inst}_\gamma &= \sum_{n>0} e^{in\theta_\gamma} K_1(2\pi R n |Z_\gamma|) \ , \quad V^{\rm inst}_\gamma = \sum_{n>0} e^{in\theta_\gamma} K_0(2\pi R n |Z_\gamma|) \ .
\end{align}
Here, $K_\nu$ are Bessel functions, and the asymptotics $K_\nu(x) \sim \sqrt{\frac{\pi}{2x}} e^{-x}$ as $x\to\infty$ show that these corrections to the semi-flat limit are exponentially suppressed away from singular fibers, where $Z_\gamma=0$ for some $\gamma$ with $\Omega(\gamma)\not=0$.

In fact, although this is not manifest, these formulae are also fairly well-behaved near singular fibers. To see this, we exploit the fact that $\Omega(-\gamma;a)=\Omega(\gamma;a)$ in order to see that it is natural to study the combination
\begin{align}
\varpi^{\rm inst}_\gamma(\zeta) &+ \varpi^{\rm inst}_{-\gamma}(\zeta) = \frac{i}{8\pi^2} d\Y^{\rm sf}_\gamma(\zeta) \wedge \brackets{ |Z_\gamma| \tilde A^{\rm inst}_\gamma d\log(Z_\gamma/\overline{Z_\gamma}) - \tilde V^{\rm inst}_\gamma \parens{\frac{1}{\zeta} dZ_\gamma - \zeta d\overline{Z_\gamma}} } \ , \\
\tilde A^{\rm inst}_\gamma &= \sum_{n\not=0} (\sgn n) e^{in\theta_\gamma} K_1(2\pi R |Z_\gamma| |n|) \ , \quad \tilde V^{\rm inst}_\gamma = \sum_{n\not=0} e^{in\theta_\gamma} K_0(2\pi R |Z_\gamma| |n|) \ .
\end{align}
Here, we used $\Y_{-\gamma}=-\Y_\gamma$. By re-writing these functions as\footnote{The functions multiplying $\frac{\sin(\pi x)}{\pi x}$ are present in order to cancel out the contributions from the Bessel functions as $x\to 0$. The factor of $\frac{\sin(\pi x)}{\pi x}$ ensures that the $n\not=0$ terms are unaffected.}
\begin{align}
\tilde A^{\rm inst}_\gamma &= \sum_{n\in\ZZ} \lim_{x\to n} e^{ix\theta_\gamma} \brackets{(\sgn x) K_1(2\pi R |Z_\gamma| |x|) - \frac{\sin(\pi x)}{2\pi^2 R |Z_\gamma| x^2} } \ , \\
\tilde V^{\rm inst}_\gamma &= \sum_{n\in\ZZ} \lim_{x\to n} \brackets{e^{ix\theta_\gamma} K_0(2\pi R |Z_\gamma| |x|) + \frac{\sin(\pi x)}{\pi x}\parens{ \log(\pi R |Z_\gamma| |x|) + \gamma_{EM} } } \ ,
\end{align}
where $\gamma_{EM}$ is the Euler-Mascheroni constant, we can Poisson resum them \cite{vafa:spacetimeInsts,seiberg:mirrorT}\footnote{That is, we multiply both sides of
\be \sum_n \delta(x-n) = \sum_k e^{2\pi i k x} \ee
by a function $f(x)$ and integrate, in order to find
\be \sum_n \lim_{x\to n} f(x) = \sum_k \F[f](k) \ . \ee
Since $(\sgn x)K_1(C |x|)=-\frac{1}{C} \partial_x K_0(C |x|)$, we need only evaluate $\F[K_0(C|x|)]$, with $C>0$. To do so, we use the integral representation
\be K_0(z) = \half\int_{-\infty}^\infty du\, e^{iz\sinh u} \ , \ee
which is valid for $z>0$. Note that if $z<0$ then
\be K_0(-z) = \half\int_{-\infty}^\infty du\, e^{iz\sinh(-u)} = \half \int_{-\infty}^\infty du\, e^{iz\sinh u} \ . \ee
We then have
\begin{align}
\F[K_0(C|x|)](k) &= \half\int_{-\infty}^\infty dxdu\, e^{ix(C\sinh u+2\pi k)} 
= \pi \int du\, \delta(C\sinh u+2\pi k) \\
&= \pi \int du\, \frac{\delta(u+\sinh^{-1} \frac{2\pi k}{C})}{C\cosh u} 
= \frac{\pi}{C\cosh \sinh^{-1} \frac{2\pi k}{C}} \\
&= \frac{\pi}{\sqrt{C^2+(2\pi k)^2}} \ .
\end{align}} to
\begin{align}
\tilde A^{\rm inst}_\gamma &= \frac{i}{2R|Z_\gamma|} \sum_{k\in \ZZ} \parens{\frac{k+\theta_\gamma/2\pi}{\sqrt{R^2|Z_\gamma|^2+(k+\theta_\gamma/2\pi)^2}} - \lambda_k } \ , \\
\lambda_k &= 
\left\{\begin{array}{rl}
1 &: k+\frac{\theta_\gamma-\pi}{2\pi} \ge 0 \\
2\parens{k+\frac{\theta_\gamma}{2\pi}} &: k+\frac{\theta_\gamma-\pi}{2\pi}<0<k+\frac{\theta_\gamma+\pi}{2\pi} \\
-1 &: k+\frac{\theta_\gamma+\pi}{2\pi} \le 0
\end{array}\right.\quad ,\\
\tilde V^{\rm inst}_\gamma &= \log R |Z_\gamma| + \half\sum_{k\in\ZZ} \parens{\frac{1}{\sqrt{R^2|Z_\gamma|^2 + (k+\theta_\gamma/2\pi)^2}} - \kappa_k} \ , \\
\kappa_k &= \piecewise{\log \frac{2|k|+1}{2|k|-1}}{k\not=0}{0}{k=0} \ .
\end{align}
These expressions make manifest their $|Z_\gamma|\to 0$ behavior, but now the exponential decay at large $R$ is hidden and relies on miraculous cancellations!

In what follows, we will discuss a different Poisson resummation that is available near orbifold limits. This will take us between `Coulomb branch' expressions of the sort we have just described and the perturbative `Higgs branch' results that we obtained from the hyper-K\"ahler quotient. We will thus gain insight into 3d mirror symmetry for compactified higher-dimensional theories, as we will be able to relate the effects of winding (or, T-dually, momentum) modes on the Higgs branch side to those of BPS states on the Coulomb branch side. As above, Poisson resummation of the Coulomb branch formulae will make their behavior near (but not necessarily arbitrarily so) singular fibers more transparent, at the cost of obscuring the exponential suppression in $R$.

\subsection{$SU(2)$ $N_f=4$} \label{sec:fieldTheory}

We begin with a theory whose BPS spectrum is well-known \cite{sw,ferrari:finite}: 4d $\N=2$ $SU(2)$ gauge theory with $N_f=4$ hypermultiplets. The moduli space of this field theory, when compactified on a circle and with vanishing complex masses, coincides with ours near one of the 4 singular fibers. (Real masses allow the moduli space to no longer be elliptically fibered, so when we refer to fibers we have in mind the semi-flat geometry.) In particular, the semi-flat moduli space is $(\CC\times T^2_F)/Z_2$.

We claim that this geometry is uncorrected at finite $R$ if the real masses are tuned appropriately. That is, all instanton corrections exactly cancel each other out. In this case, the infrared physics enjoys two $U(1)$ global symmetries which are spontaneously broken by the vacuum expectation values of $\theta_e$ and $\theta_m$.\footnote{Depending on one's duality frame these may be 0-form global symmetries or 1-form global symmetries \cite{s:genGlob}. For, when the infrared physics is thought of as a compactified 4d $U(1)$ gauge theory then these are electric and magnetic 1-form symmetries under which Wilson or 't Hooft lines are charged \cite{w:ramify,w:surface}, but if we dualize the photon and regard this as a 3d non-linear sigma model then these are familiar 0-form symmetries.} This can likely be explained using instanton calculus, as in \cite{chen:instMat,tong:AHmatter}. Here, we will instead simply explain it by embedding this field theory into string theory, as in \cite{sen:FOrientifolds,s:fBranes,s:K3,sw:3d}. Namely, the 4d field theory of interest arises on the worldvolume of a D3-brane probing an O7-plane plus 4 D7-branes in the T-dual of type I on $T^2$, or equivalently in F-theory on a $T^4/Z_2$ orbifold. In order for the moduli space to be flat, we assume that we have broken $SO(32)$ to $SO(8)^4$ -- i.e., we have placed 4 D7-branes on top of each of the 4 O7-planes and the F-theory base is $T^2/Z_2$. We now compactify on an additional circle and T-dualize. The D3-brane becomes a D2-brane, and its moduli space is the geometry seen by the M2-brane to which this D2-brane lifts in M-theory. This is roughly the same as the orbifold upon which we compactified F-theory, except now the fibers are part of spacetime.

The only remaining question is the values of the real masses -- or $SO(8)$ flavor holonomies in the 4d gauge theory on $S^1_R$ -- that tune us to the orbifold limit. We can address this from the type IIA point of view. In order for the moduli space to remain exactly flat, at the one-loop level, we need all RR charges to exactly cancel at each O6-plane, and so we must position two D6-branes on top of each O6-plane. That is, each O7-plane divides into two O6-planes at opposite ends of the circle, and we divide the 4 D7-branes positioned at the O7-plane between the two O6-planes. This corresponds to choosing real masses $\theta_1=\theta_2=0$, $\theta_3=\theta_4=\pi$. Near each O6-plane, the low-energy physics is 3d $\N=4$ $SU(2)$ $N_f=2$, whose quantum-corrected moduli space \cite{sw:3d} is the $D_2\cong A_1\times A_1$ ALF manifold $(\RR^3\times S^1)/Z_2$ with two $A_1$ singularities at opposite ends of the circle (which is parametrized by $\theta_m$, which in this context is regarded as the dual photon). So, including all of the effects of the 4 D6-branes and 2 O6-planes that correspond to the 7-branes that are included in the 4d gauge theory, we find the 4 $A_1$ singularities of $(\CC\times T^2)/Z_2$.

The above reasoning was a bit heuristic, but we can shore up our confidence by examining the integral equation \eqref{eq:intEqn}. The BPS index we supply it is as follows. It is independent of $a$, since there is no dimensionful scale in the 4d SCFT with vanishing complex masses to which $|a|$ may be compared, and there is a $U(1)_R$ symmetry that makes physics independent of the phase of $a$. For all relatively prime $p,q\in \ZZ$, there is a vector multiplet ($\Omega=-2$) in the trivial representation of the $\Spin(8)$ flavor symmetry with gauge charges $(2p,2q)$ (i.e., $\pm 2(p\gamma^m+q\gamma^e)$) and a half-hypermultiplet ($\Omega=1$) in one of the 8-dimensional representations of $\Spin(8)$ with gauge charges $(p,q)$. Which of the three such flavor representations the half-hypermultiplet is in depends on whether $p$ and/or $q$ are even or odd (they cannot both be even). When $p$ is even and $q$ is odd, one has the $\mathbf{8_v}$ representation, whose weights (decomposition into irreducible representations of the Cartan subgroup $U(1)^4\subset \Spin(8)$) are $(\pm 1,0,0,0)$, $(0,\pm 1,0,0)$, $(0,0,\pm 1,0)$, and $(0,0,0,\pm 1)$. When $p$ and $q$ are both odd, one has the $\mathbf{8_s}$ spinor representation, with weights $\half(\pm 1,\pm 1,\pm 1,\pm 1)$ with an even number of $+$'s. Lastly, when $p$ is odd and $q$ is even, one has the $\mathbf{8_c}$ conjugate spinor representation, with weights $\half(\pm 1,\pm 1,\pm 1,\pm 1)$ with an odd number of $+$'s. The integral equation, with vanishing flavor central charges and real masses $\theta_j$, $j=1,\ldots,4$, then takes the form
\be \Y_\gamma = \Y_\gamma^{\rm sf} - \frac{1}{4\pi i} \sum_{p,q} \avg{\gamma,(p,q)} \int_{\ell_{p,q}} \frac{d\zeta'}{\zeta'} \frac{\zeta'+\zeta}{\zeta'-\zeta} \log\frac{\prod_{\gamma^f} (1-e^{i\theta_{\gamma^f}}\X_{p,q}(\zeta'))}{\parens{1-\X^2_{p,q}(\zeta')}^4} \ . \label{eq:su2intEqn} \ee
Here, we have used the fact that the complex masses vanish in order to rename $\ell_\gamma$ to $\ell_{p,q}$; note that $\ell_{2p,2q}=\ell_{p,q}$. We similarly have
\be \X_{p,q}=(-1)^{pq}\X_m^p \X_e^q \ ,\quad Z_{p,q} = (p\tau_F+q)Z_e \ .\ee
Explicitly, the product over flavor charges takes the following form:\footnote{$\wedge$ reads `and' and $\vee$ reads `or.'}
\be
\prod_{\gamma^f} (1-e^{i\theta_{\gamma^f}} \X_{p,q}) = \left\{\begin{array}{rl}
\prod_{j=1,\ldots,4, s=\pm 1} (1-e^{is\theta_j}\X_{p,q}) &: 2|p\wedge 2\nmid q \\
\prod_{\{s_j\}} (1-e^{\half i \sum_j s_j \theta_j}\X_{p,q}) &: {\rm else}
\end{array}\right. \ , \label{eq:flavorProd}
\ee
where in the second line $\{s_j\}$ runs over collections of four signs, where there is either an even or odd number of total $+$ signs, depending on $p$ and $q$. When $\theta_1=\theta_2=0$ and $\theta_3=\theta_4=\pi$,
\be \prod_{\gamma^f}(1-e^{i\theta_{\gamma^f}} \X_{p,q}) = \left\{\begin{array}{rl}
\prod_j (1-e^{i\theta_j} \X_{p,q})^2 &: 2|p\wedge 2\nmid q \\
\prod_{\{s_j\}/\sim} (1-\X_{p,q}^2) &: {\rm else}
\end{array}\right.
= (1-\X_{p,q}^2)^4 \ , \label{eq:firstVanishes}
\ee
where $\sim$ identifies collections that differ by flipping both the first and last signs, and so the logarithm vanishes and the semi-flat answer is exact.

We now study corrections away from this limit, writing $\theta_j=\theta_j^{(0)}+\delta\theta_j$. We begin by proving that
\be \Y_\gamma(\zeta) = \Y^{(\nu)}_\gamma(\zeta) + \Oo(\delta\theta^{2(\nu+1)}) \ ,\label{eq:orbExpand} \ee
and so the iterative scheme is a systematic expansion in $\delta\theta$ (where we take all $\delta\theta_j$ to be of order $\delta\theta$). (This is, a priori, surprising, since as we explained above the iterative approximation is essentially a large $R$ expansion. We still rely on large $R$ in order to guarantee that the iterative scheme converges -- otherwise, there is no guarantee that the coefficients of $\delta\theta^{2\nu}$ do not grow rapidly with $\nu$. However, even without sufficiently large $R$ for convergence, \eqref{eq:induct} holds as long as $\Y^{(\nu+1)}$ exists.) We do so by induction, by proving that
\be \Y^{(\nu+1)}_\gamma(\zeta) = \Y^{(\nu)}_\gamma(\zeta) + \Oo(\delta\theta^{2(\nu+1)}) \ . \label{eq:induct} \ee
First, we prove the base case:
\be \Y^{(1)}_\gamma(\zeta) = \Y^{(0)}_\gamma(\zeta) + \Oo(\delta\theta^2) \ . \ee
By definition, the left hand side is given by the right hand side of \eqref{eq:su2intEqn} with $\X_{p,q}$ replaced by $\X_{p,q}^{\rm sf}$, which is independent of the real mass parameters. We showed that the logarithm vanishes when $\theta_j=\theta_j^{(0)}$ for all $j$, so we just need to show that it still vanishes at order $\delta\theta$. We prove this by noting the stronger result that the inside of the logarithm is an even function of $\{\delta\theta_j\}$, in the sense that it is invariant under simultaneously negating all $\delta\theta_j$.\footnote{Actually, the same reasoning that we are about to use implies that it is invariant under negation of the pairs $(\delta\theta_1,\delta\theta_2)$ and $(\delta\theta_3,\delta\theta_4)$.} This follows from \eqref{eq:flavorProd}: in the first case, negating $\delta\theta_{j'}$ can be undone by negating $s$ in the factors with $j=j'$, and $e^{is\theta_{j'}^{(0)}}$ is independent of $s$; in the second case, negating $\delta\theta_j$ for all $j$ can be undone by negating all $s_j$, which does not change whether $s_3=s_4$, which is all that matters in $e^{\half i\sum_j s_j\theta_j^{(0)}}$. Next, we prove \eqref{eq:induct}, assuming that it holds with $\nu$ replaced by $\nu-1$. The definition of $\Y^{(\nu+1)}_\gamma$ takes the form $\Y^{(\nu+1)}_\gamma=\Y^{\rm sf}_\gamma+f_\gamma(\Y^{(\nu)},\{\theta_j\})$, where $f_\gamma(\cdot,\{\theta_j^{(0)}\})$ vanishes for any value of its first argument (such that the integrals defining $f_\gamma$ converge), as a consequence of \eqref{eq:firstVanishes}. So,
\begin{align}
\Y^{(\nu+1)}_\gamma-\Y^{(\nu)}_\gamma &= f_\gamma(\Y^{(\nu)},\{\theta_j\}) - f_\gamma(\Y^{(\nu-1)},\{\theta_j\}) \\
&= \int_{\CC} |d\zeta'|^2\,\sum_{\gamma'=\gamma_e,\gamma_m} \left.\frac{\delta}{\delta\Y_{\gamma'}(\zeta')} f_\gamma(\Y,\{\theta_j\}) \right|_{\Y=\Y^{(\nu-1)}} \cdot\Oo(\delta\theta^{2\nu}) \\
&= \half \int_{\CC} |d\zeta'|^2\,\sum_{\gamma',i,j}\left.\frac{\delta}{\delta\Y_{\gamma'}(\zeta')}\partial_{\theta_i}\partial_{\theta_j} f_\gamma(\Y,\{\theta_j\})\right|_{\Y=\Y^{(\nu-1)},\theta_j=\theta_j^{(0)}}\cdot\Oo(\delta\theta^{2(\nu+1)}) \ .
\end{align}
This completes the proof.

\eqref{eq:orbExpand}, with $\nu=1$, implies that \eqref{eq:firstApprox} should exactly agree with our perturbative hyper-K\"ahler quotient results, up to a change of variables! We now demonstrate this explicitly. By summing up the contributions to $\varpi^{\rm inst}$ from half-hypermultiplets with gauge charge $\gamma=(p,q)$ and the vector multiplet with charge $(2p,2q)$, we obtain
\begin{align}
\varpi^{\rm{eff}}_\gamma(\zeta) &= -\frac{i}{8\pi^2} d\Y^{\rm sf}_\gamma(\zeta) \wedge \brackets{\sum_{n>0} e^{in\theta_\gamma}\parens{\piecewise{2 \sum_j \cos(n\theta_j)}{2|p\wedge 2\nmid q}{\sum_{\{s_j\}} \prod_j e^{\frac{in}{2} s_j \theta_j}}{\rm else} + \piecewise{-8}{n\, \rm{even}}{0}{n\, \rm{odd}}}  \right. \nonumber \\
&\left.\times\parens{-|Z_\gamma| K_1(2\pi R n |Z_\gamma|) \,d\log(Z_\gamma/\bar Z_\gamma) + K_0(2\pi R n |Z_\gamma|) \parens{\frac{1}{\zeta} dZ_\gamma - \zeta d\bar Z_\gamma} } } \nonumber \\
&\approx \frac{i}{16\pi^2} d\Y^{\rm sf}_\gamma(\zeta) \wedge \brackets{\sum_{n>0} n^2 e^{in\theta_\gamma} \left\{\begin{array}{rl}
2\parens{\delta\theta_1^2+\delta\theta_2^2+(-1)^n(\delta\theta_3^2+\delta\theta_4^2)} &: 2|p\wedge 2\nmid q \nonumber \\
\sum_{s=\pm 1} s^n \brackets{ (\delta\theta_1-s\delta\theta_2)^2+(\delta\theta_3-s\delta\theta_4)^2 } &: 2\nmid p\wedge 2\nmid q \nonumber \\
\sum_{s=\pm 1} s^n \brackets{ (\delta\theta_1+s\delta\theta_2)^2+(\delta\theta_3-s\delta\theta_4)^2 } &: 2\nmid p\wedge 2|q
\end{array}\right. \right. \nonumber \\
&\left.\times\parens{-|Z_\gamma| K_1(2\pi R n |Z_\gamma|) \,d\log(Z_\gamma/\bar Z_\gamma) + K_0(2\pi R n |Z_\gamma|) \parens{\frac{1}{\zeta} dZ_\gamma - \zeta d\bar Z_\gamma} } } \nonumber \\
&= \frac{i}{16\pi^2} \sum_{n>0} e^{i\theta_{n\gamma}} \left\{\begin{array}{rl}
2\parens{\delta\theta_1^2+\delta\theta_2^2+(-1)^n(\delta\theta_3^2+\delta\theta_4^2)} &: 2|p\wedge 2\nmid q \nonumber \\
\sum_{s=\pm 1} s^n \brackets{ (\delta\theta_1-s\delta\theta_2)^2+(\delta\theta_3-s\delta\theta_4)^2 } &: 2\nmid p\wedge 2\nmid q \nonumber \\
\sum_{s=\pm 1} s^n \brackets{ (\delta\theta_1+s\delta\theta_2)^2+(\delta\theta_3-s\delta\theta_4)^2 } &: 2\nmid p\wedge 2|q
\end{array}\right. \\
&\times d\Y^{\rm sf}_{n\gamma}(\zeta) \wedge \parens{-|Z_{n\gamma}| K_1(2\pi R |Z_{n\gamma}|) \,d\log(Z_{n\gamma}/\bar Z_{n\gamma}) + K_0(2\pi R |Z_{n\gamma}|) \parens{\frac{1}{\zeta} dZ_{n\gamma} - \zeta d\bar Z_{n\gamma}} } \ . \label{eq:wEff}
\end{align}
This sum over $n>0$ then combines nicely with the sum over relatively prime $p,q\in\ZZ$ to yield
\begin{align}
\varpi^{\rm inst}(\zeta) &\approx \frac{i}{16\pi^2} \sum_{\gamma\in \ZZ^2\backslash (0,0)} e^{i\theta_{\gamma}} \Xi(\gamma) \nonumber \\
&\times d\Y^{\rm sf}_{\gamma}(\zeta) \wedge \parens{-|Z_{\gamma}| K_1(2\pi R |Z_{\gamma}|) \,d\log(Z_{\gamma}/\bar Z_{\gamma}) + K_0(2\pi R |Z_{\gamma}|) \parens{\frac{1}{\zeta} dZ_{\gamma} - \zeta d\bar Z_{\gamma}} } \ , \label{eq:combinedSum}
\end{align}
where if $\gamma=n(p,q)=(p',q')$ then
\begin{align}
\Xi(\gamma) &= \left\{\begin{array}{rl}
2\parens{\delta\theta_1^2+\delta\theta_2^2+(-1)^n(\delta\theta_3^2+\delta\theta_4^2)} &: 2|p\wedge 2\nmid q \\
\sum_{s=\pm 1} s^n \brackets{ (\delta\theta_1-s\delta\theta_2)^2+(\delta\theta_3-s\delta\theta_4)^2 } &: 2\nmid p\wedge 2\nmid q \\
\sum_{s=\pm 1} s^n \brackets{ (\delta\theta_1+s\delta\theta_2)^2+(\delta\theta_3-s\delta\theta_4)^2 } &: 2\nmid p\wedge 2|q
\end{array}\right. \\
&= \piecewise{2\parens{(\delta\theta_1^2+\delta\theta_2^2)+(-1)^n(\delta\theta_3^2+\delta\theta_4^2)}}{2|p'}{4\parens{(-1)^{q'}\delta\theta_1\delta\theta_2-\delta\theta_3\delta\theta_4}}{2\nmid p'} \\
&= (1+(-1)^{p'})\parens{(\delta\theta_1^2+\delta\theta_2^2)+(-1)^n(\delta\theta_3^2+\delta\theta_4^2)} + 2(1-(-1)^{p'})\parens{(-1)^{q'}\delta\theta_1\delta\theta_2-\delta\theta_3\delta\theta_4} \ . \label{eq:xi}
\end{align}
We can make the dependence on $\gamma$ very explicit, using
\begin{align}
Z_\gamma &= (p'\tau_F + q')Z_e\ , \quad dZ_\gamma = (p'\tau_F + q') dZ_e \quad \Rightarrow\quad d\log Z_\gamma = d\log Z_e \\
|Z_\gamma| &= |Z_e| \sqrt{(p'^2 |\tau_F|^2 + q'^2)+2p'q' \tau_{F,1}} \\
d\Y_\gamma &= p'\,d\Y_m + q'\,d\Y_e \\
e^{i\theta_\gamma} &= (-1)^{p'q'}e^{i(p'\theta_m+q'\theta_e)} \ . \label{eq:thetaPart}
\end{align}
We now use
\begin{align}
(-1)^{p'q'+n} &= \piecewise{1}{(2|p'\wedge 2|q')\vee(2\nmid p'\wedge 2\nmid q')}{-1}{\rm else} = (-1)^{p'+q'} \\
(-1)^{p'q'} &= \piecewise{1}{2|p'\vee 2|q'}{-1}{\rm else} = 1 - \frac{(1-(-1)^{p'})(1-(-1)^{q'})}{2}
\end{align}
to re-write $(-1)^{p'q'}\Xi(\gamma)$ as an analytic function of $p'$ and $q'$:
\begin{align}
(-1)^{p'q'}\Xi(\gamma) &= (1+(-1)^{p'})\parens{(\delta\theta_1^2+\delta\theta_2^2)+(-1)^{q'}(\delta\theta_3^2+\delta\theta_4^2)}+2(1-(-1)^{p'})(\delta\theta_1\delta\theta_2-(-1)^{q'}\delta\theta_3\delta\theta_4) \nonumber \\
&= \parens{\delta\theta_1+\delta\theta_2}^2+e^{i\pi p'}\parens{\delta\theta_1-\delta\theta_2}^2+e^{i\pi q'}\parens{\delta\theta_3-\delta\theta_4}^2+e^{i\pi(p'+q')}\parens{\delta\theta_3+\delta\theta_4}^2 \ . \label{eq:thetaFI}
\end{align}
(Of course, we also have $(-1)^{p'q'}=e^{i\pi p'q'}$. But the above equation will prove more convenient to Fourier transform.) The various terms in this equation will soon be matched to the four FI parameters of the Higgs branch approach.

Having done this, we can now implement a two-dimensional Poisson resummation. (There is no danger in including the $\gamma=(0,0)$ term in the sum, as the functions we are Fourier transforming tend to zero as $(p',q')\to (0,0)$.) We denote the Fourier duals of $p'$ and $q'$ by $k_p$ and $k_q$. Let
\be A = \twoMatrix{|\tau_F|}{\frac{\tau_{F,1}}{|\tau_F|}}{0}{\frac{\tau_{F,2}}{|\tau_F|}} \ . \ee
$A$ is invertible, since $\tau_{F,2}>0$.
Defining
\be \column{x}{y} = A \column{p'}{q'} \ , \ee
we find that
\be |Z_\gamma| = |Z_e| \sqrt{x^2 + y^2} \ , \quad p'\tau_F + q' = \frac{\tau_F}{|\tau_F|}(x-iy) \ . \ee
We also define
\be \column{k_x}{k_y} = (A^T)^{-1} \column{k_p}{k_q} \ , \ee
so that $k_p p'+k_q q'=k_x x + k_y y$. This makes Fourier transformation nicer; we just have to account for the Jacobian of this transformation:
\be dp' dq' = \frac{dx dy}{\tau_{F,2}} \ . \ee
The first line of \eqref{eq:combinedSum} just shifts the Fourier duals $k_x$ and $k_y$ of $x$ and $y$, so we can ignore it for now. Defining
\be C = 2\pi R |Z_e| \ , \ee
the first term in the second line of \eqref{eq:combinedSum} boils down to the Fourier transform of
\be x \sqrt{x^2+y^2} K_1(C \sqrt{x^2+y^2}) = - \frac{1}{C} (x^2+y^2) \partial_x K_0(C \sqrt{x^2+y^2}) \ , \label{eq:toTrans1} \ee
(as well as the analogous expression with the roles of $x$ and $y$ exchanged), and the second to the Fourier transforms of
\be x^2 K_0(C \sqrt{x^2+y^2}) \ , \quad  xy K_0(C \sqrt{x^2+y^2}) \ . \label{eq:toTrans2} \ee
All of these, in turn, follow straightforwardly (by multiplying by and differentiating with respect to $k_x$ and $k_y$) from the Fourier transform
\begin{align}
\F[K_0(C \sqrt{x^2+y^2})](k) &= \int dxdy\, e^{2\pi i(k_x x+k_y y)} K_0(C\sqrt{x^2+y^2}) \\
&= \int_0^\infty dr\, \int_0^{2\pi} d\theta\, r e^{2\pi i r |k| \cos\theta} K_0(Cr) \\
&= 2\pi \int_0^\infty dr\, r J_0(2\pi r |k|) K_0(Cr) \\
&= \frac{2\pi}{(2\pi |k|)^2 + C^2} \ .
\end{align}
$J_0$ is another Bessel function. This last integral,
\be I(B,C)\equiv \int_0^\infty dr\, r J_0(Br) K_0(Cr) \ , \quad B,C>0 \ , \ee
may be evaluated by using the defining differential equation
\be r J_0(r) = - r \partial_r^2 J_0(r) - \partial_r J_0(r) \ , \ee
integrating by parts, and then using
\be r\partial_r^2 K_0(r) + \partial_r K_0(r) = r K_0(r) \ . \ee
This gives an algebraic equation,
\be I = \frac{1}{B^2} - \frac{C^2}{B^2} I \ , \ee
whose solution is as above.

Lastly, we acccount for the first line of \eqref{eq:combinedSum}. Let $F(k_x,k_y)$ denote the Fourier transform of one of the expressions in the second line of \eqref{eq:combinedSum}. For example, we might let $F$ be the Fourier transform of the coefficient of $d\Y_m^{\rm sf}\wedge \frac{dZ_e}{\zeta}=\frac{i}{\zeta} d\theta_m \wedge dZ_e + \pi R \bar\tau_F d\bar Z_e\wedge dZ_e$, namely
\be \frac{\tau}{\tau_2|\tau|^2}\parens{\F[x^2 K_0(C\sqrt{x^2+y^2})] + \parens{-i-\frac{\tau_1}{\tau_2}}\F[xy K_0(C\sqrt{x^2+y^2})]+i\frac{\tau_1}{\tau_2}\F[y^2 K_0(C\sqrt{x^2+y^2})]} \ . \ee
We account for the factor of $e^{i(p'\theta_m+q'\theta_e)}$ in \eqref{eq:thetaPart}, as well as the change of variables from $k_x,k_y$ to $k_p,k_q$, by defining
\begin{align}
\tilde F(k_p,k_q,\theta_m,\theta_e) &= F\parens{k_x + \frac{\theta_m}{2\pi |\tau|},\, k_y + \frac{1}{2\pi \tau_2}\parens{-\frac{\tau_1}{|\tau|}\theta_m + |\tau|\theta_e}} \\
&= F\parens{\frac{1}{|\tau|}(k_p+\theta_m/2\pi),\, \frac{1}{\tau_2}\parens{-\frac{\tau_1}{|\tau|}(k_p + \theta_m/2\pi) + |\tau|(k_q + \theta_e/2\pi)}} \ .
\end{align}
Accounting for the rest of the first line of \eqref{eq:combinedSum} then gives (up to the factor of $i/16\pi^2$)
\begin{align}
&(\delta\theta_1+\delta\theta_2)^2 \tilde F(k_p,k_q,\theta_m,\theta_e) + (\delta\theta_1-\delta\theta_2)^2 \tilde F(k_p,k_q,\theta_m+\pi,\theta_e) \nonumber \\
&+(\delta\theta_3-\delta\theta_4)^2 \tilde F(k_p,k_q,\theta_m,\theta_e+\pi)+(\delta\theta_3+\delta\theta_4)^2 \tilde F(k_p,k_q,\theta_m+\pi,\theta_e+\pi) \\
&=(\delta\theta_1+\delta\theta_2)^2 \tilde F(k_p,k_q,\theta_m,\theta_e) + (\delta\theta_1-\delta\theta_2)^2 \tilde F(k_p+\half,k_q,\theta_m,\theta_e) \nonumber \\
&+(\delta\theta_3-\delta\theta_4)^2 \tilde F(k_p,k_q+\half,\theta_m,\theta_e)+(\delta\theta_3+\delta\theta_4)^2 \tilde F(k_p+\half,k_q+\half,\theta_m,\theta_e) \ . \label{eq:FIcoulomb}
\end{align}
Note that, while originally we were summing over $k_p,k_q\in\ZZ$, we now effectively have $k_p,k_q\in\half\ZZ$.

We can now begin to compare the Coulomb and Higgs branch results. We begin by noting that while there may be an $\Oo(\delta\theta^2)$ correction to \eqref{eq:coordChange0} and \eqref{eq:Nchange0} -- i.e., we might need to slightly modify the torus that is being orbifolded on the Higgs branch side as we blow up the orbifold singularities in order to match the Coulomb and Higgs branch expressions -- this correction will not matter at the order in perturbation theory to which we work. Comparing the form of \eqref{eq:FIcoulomb} with the structure of the Higgs branch results causes us to identify
\be \tilde n_3+\tau_F \tilde n_4 = 2(k_p-\tau_F k_q) \ , \ee
since both sides change by units of 1 and $\tau_F$, and the blow-up parameters are labelled by $(\tilde n^3,\tilde n^4) \!\! \mod{2\ZZ^2}$. (Recall, from the end of \S\ref{sec:hk}, that there is no $n^u$.)

After a bit of algebra, using some foresight to identify
\begin{align}
\xi_{00,\RR}^2&=\frac{1}{16\pi^4 R^2} (\delta\theta_1+\delta\theta_2)^2 \ , \quad \xi_{10,\RR}^2=\frac{1}{16\pi^4 R^2} (\delta\theta_1-\delta\theta_2)^2 \ , \nonumber \\
\xi_{01,\RR}^2&=\frac{1}{16\pi^4 R^2} (\delta\theta_3-\delta\theta_4)^2 \ , \quad \xi_{11,\RR}^2=\frac{1}{16\pi^4 R^2} (\delta\theta_3+\delta\theta_4)^2 \ , \label{eq:thetaFIfinal}
\end{align}
employing the felicitous result
\be k_p-\tau_F k_q = -\frac{i\tau_{F,2}\tau_F}{|\tau_F|}(k_x-i k_y) \Rightarrow |k_p-\tau_F k_q|=\tau_{F,2}|k_x-ik_y| \ , \ee
and recalling \eqref{eq:PSdef}, which when combined with \eqref{eq:coordChange0} yields
\be N^u_\pm=\pm i\rho a + \Oo(\delta\theta^2) \ , \quad N^v_\pm=n^v\pm \frac{1}{\rho}z + \Oo(\delta\theta^2) \ , \quad D_\pm = |N^u_\pm|^2+|N^v_\pm|^2 \ , \ee
we arrive at the following:\footnote{The vanishing of $\omega_{+\, \alpha\beta}$, where $\alpha,\beta=\bar a,z,\bar z$ -- that is, neither $\alpha$ nor $\beta$ is $a$ -- follows from the fact, demonstrated in \cite{GMN:walls} (see, e.g., (5.47) and (5.56) therein), that there is a change of variables $(z,a)\mapsto (\Upsilon(z,\bar z,a,\bar a),a)$ such that $\omega_+=da\wedge d\Upsilon$. \label{ft:holoCoords}}
\begin{align}
\varpi^{\rm inst}(\zeta) &= -\frac{i}{2\zeta} \omega_+^{\rm inst} + \omega_K^{\rm inst} - \frac{i\zeta}{2}\omega_-^{\rm inst} \\
&= \sum_{n} \parens{-\frac{i}{2\zeta}\omega_{n+}+\omega_{nK}-\frac{i\zeta}{2}\omega_{n-} } \label{eq:coulombTower} \\
\omega_{n+\, a\bar a} &= - \xi_{n,\RR}^2 \cdot \frac{2i\rho^2 N^u_+N^v_+}{D_+^3} \\
\omega_{n+\, az} &= \xi_{n,\RR}^2 \cdot \frac{|N^v_+|^2-|N^u_+|^2}{D_+^3} \\
\omega_{n+\, a\bar z} &= \xi_{n,\RR}^2 \cdot \frac{2 (N^v_+)^2}{D_+^3} \\
\omega_{n+\, \bar a z} &= 0 \\
\omega_{n+\, \bar a \bar z} &= 0 \\
\omega_{n+\, z \bar z} &= 0 \\
\omega_{nK\, a\bar a} &= i\rho^2\xi_{n,\RR}^2 \cdot\frac{|N^v_+|^2-|N^u_+|^2}{D_+^3} \\
\omega_{nK\, az} &= \xi_{n,\RR}^2 \cdot \frac{\bar N^v_+ \bar N^u_+}{D_+^3} \\
\omega_{nK\, a\bar z} &= \xi_{n,\RR}^2 \cdot \frac{N^v_+ \bar N^u_+}{D_+^3} \\
\omega_{nK\, \bar a z} &= \xi_{n,\RR}^2 \cdot \frac{\bar N^v_+ N^u_+}{D_+^3} \\
\omega_{nK\, \bar a \bar z} &= \xi_{n,\RR}^2 \cdot \frac{N^v_+ N^u_+}{D_+^3} \\
\omega_{nK\, z \bar z} &= 0 \ .
\end{align}
(`Inst' stands for `instanton.') To complete the identification of the Coulomb and Higgs branch results, the last step is to identify the first subleading corrections to \eqref{eq:coordChange0}, which take the form
\be u = u^{\rm sf}+\sum_n u_n \ , \quad v = v^{\rm sf} + \sum_n v_n \ , \ee
with $u_n$ and $v_n$ of order $\delta\theta^2$.

For this purpose, it is useful to first match the vanishing components of $\omega_+$. Working to order $\delta\theta^2$, we have
\begin{align}
0 &= \omega_{+\, z\bar a} \nonumber \\
&= \partial_z v^{\rm sf} \partial_{\bar a}\bar u^{\rm sf} \sum_n \omega_{n+\, v\bar u} + \omega^{\rm sf}_{+\, vu}\partial_z v^{\rm sf} \sum_n \partial_{\bar a} u_n   \nonumber \\
&= - \frac{i}{4} \sum_n \omega_{n+\, v\bar u} + \frac{2i}{\rho} \sum_n \partial_{\bar a} u_n \ .
\end{align}
We are thus led to integrate
\be \partial_{\bar a} u_n = \frac{\rho}{8} \omega_{n+\, v\bar u} \ , \ee
(where on the right side, we substitute the zeroth order expressions \eqref{eq:coordChange0} into $\omega_{n+\, v\bar u}$ from \eqref{eq:hkForms} with $\xi_{n,+}=0$), which yields
\be u_n = \xi_{n,\RR}^2 \parens{ \frac{N^u_+}{4D_+^2} + f(a,z,\bar z) } \ . \label{eq:unFirst} \ee
Here, $f$ is a constant of integration (which is an arbitrary function of $a,z,\bar z$). Similarly, we have
\begin{align}
0 &= \omega_{+\, z\bar z} \nonumber \\
&= \partial_z v^{\rm sf}\partial_{\bar z} \bar v^{\rm sf}\sum_n \omega_{n+\, v\bar v} + \omega_{+\, vu}^{\rm sf} \partial_z v^{\rm sf} \sum_n \partial_{\bar z}u_n \nonumber \\
&= \frac{1}{4\rho^2} \sum_n \omega_{n+\, v\bar v} + \frac{2i}{\rho} \sum_n \partial_{\bar z} u_n \ ,
\end{align}
which leads to
\be \partial_{\bar z} u_n = \frac{i}{8\rho} \omega_{n+\, v\bar v} \ . \ee
Plugging in \eqref{eq:unFirst} implies that $f$ is a function only of $a$ and $z$. The final vanishing component simply yields
\be 0 = \omega_{+\, \bar a\bar z} = \partial_{\bar a}\bar u^{\rm sf}\partial_{\bar z}\bar v^{\rm sf} \sum_n \omega_{n+\,\bar u\bar v} \ , \ee
which does not help with determining the change of variables; but, it does agree with $\omega_{n+\, \bar u\bar v}=0$.

Proceeding similarly with $\omega_{+\, a\bar a}$ and $\omega_{+\, a\bar z}$ yields, respectively,
\begin{align}
\partial_{\bar a} v_n &= -\frac{i\rho \xi_{n,\RR}^2}{2} \frac{N^u_+ N^v_+}{D_+^3} \\
\partial_{\bar z} v_n &= \frac{\xi_{n,\RR}^2}{2\rho} \frac{(N^v_+)^2}{D_+^3} \ ,
\end{align}
whose solution is
\be v_n = \xi_{n,\RR}^2 \parens{-\frac{N^v_+}{4D_+^2} + g(a,z)} \ . \ee
Sparing the reader further details, one may verify that the Coulomb and Higgs formalisms exactly agree if we set $f=g=0$. This is quite satisfying, as the problem was overdetermined -- not only do we have multiple formulae that constrain partial derivatives of the two functions $u_n$ and $v_n$, but we even have some partial derivatives which are themselves determined by different equations. For example,
\begin{align}
\omega_{K\, \bar a\bar z} = \sum_n \parens{\omega^{\rm sf}_{K\, \bar uu}\partial_{\bar a}\bar u^{\rm sf} \partial_{\bar z} u_n + \omega^{\rm sf}_{K\, v\bar v}\partial_{\bar a} v_n \partial_{\bar z} \bar v^{\rm sf}}
\end{align}
involves only derivatives which we already determined above.

In summary, we have the change of variables
\be u_n = \xi_{n,\RR}^2 \cdot \frac{N^u_+}{4D_+^2} \ , \quad v_n = -\xi_{n,\RR}^2 \cdot  \frac{N^v_+}{4D_+^2} \ . \label{eq:changeVars} \ee
Note that $u_0$ and $v_0$ are singular near the origin. This agrees with the observation in \S\ref{sec:k3HK} that the coordinates $u,v$ break down there (since $a,z$ can be used to parametrize the entire moduli space). Along similar lines, we note that while Poisson resummation has made manifest the behavior of $\varpi(\zeta)$ near singular fibers at order $\xi^2$, higher order corrections could nevertheless become important sufficiently close to the singular fibers. The Poisson resummed formulae are therefore most illuminating when $R|Z_\gamma|$ is small for some $\gamma$ with $\Omega(\gamma)\not=0$, but not so small as to overwhelm the smallness of $\xi$.

This completes the demonstration that the Coulomb and Higgs branch approaches agree. To recap, we noted that certain properties of the BPS spectrum a) led to intense cancellations at the orbifold point, and b) enabled a 2-dimensional Poisson resummation at first order in $\delta\theta^2$. The infinite BPS spectrum, with contributions from all relatively prime $p,q$, was essential for property (b), as the sum over $p,q$ combined with the sum over $n>0$. These properties are likely characteristic of theories with orbifold limits. In the next section, we will proceed in the opposite direction: we will start with the hyper-K\"ahler structure of K3 and derive part of a BPS spectrum.

\subsection{Little string theory} \label{sec:lst}

\subsubsection{Semi-flat limit} \label{sec:genus1}

We now transition from the 4d field theory, and its associated moduli space, to the heterotic little string theory and K3. Our strategy will be to Poisson resum the K3 metric of section \ref{sec:k3HK} and extract the BPS index from the resulting expressions. Amusingly, the result will be found to be moduli-independent, and it will furthermore only detect four infinite towers of copies of the 4d $SU(2)$ $N_f=4$ spectrum. However, as we explain in \S\ref{sec:moreStates}, there are other contributions to the index which we have not yet found.

The main novelty turns out to be the existence of 4 real mass parameters in the little string theory which are not related to FI parameters on the Higgs branch side. Instead, they correspond to deforming the torus which we orbifold away from the $T^2\times T^2$ locus. (The moduli of $T^4$ are the 10 independent components of the metric -- a symmetric $4\times 4$ matrix; in contrast, $T^2\times T^2$ has 6 moduli. In our notation, these are $\tau_F$, $\tau_B$, $R$ (the relative size of $T^2_F$ and $T^2_B$), and the overall volume, which we always neglect.) They have no analogues in the 4d field theory, and lead to interesting new phenomena. From the Coulomb side, the 16 FI parameters also play a distinct role from the other 4, as the former are associated to a non-abelian $SO(8)^4$ global symmetry, whereas the other 4 are associated to abelian global symmetries. We note that turning on these 4 special real masses is not necessary to determine the fully flavored BPS index, since we will be able to freely vary the 16 other real masses plus 4 flavor central charges, which together will function as the 20 desired `chemical potentials.' The main point of turning on these 4 extra parameters is to demonstrate that they require a small, but interesting modification of the results of \cite{mz:k3}.

To generalize the torus being orbifolded on this Higgs side away from $T^2_F\times T^2_B$, we deform \eqref{eq:Nchange0} to
\begin{align}
n^u &= \frac{i\rho}{2} n_B + \frac{\epsilon^z}{\rho^3}(n_F+\bar n_F) \ , \nonumber \\
n^v &= \frac{1}{2\rho} n_F + \delta n^v  \ , \nonumber \\
n_B &= \tilde n^1+\tau_B \tilde n^2 \ , \quad n_F = \tilde n^3+\tau_F \tilde n^4 \ , \quad \delta n^v = \frac{\epsilon^a}{\rho}n_B  \ , \label{eq:NchangeEps}
\end{align}
Similarly, we generalize \eqref{eq:coordChange0} to
\begin{align}
u &= \frac{i\rho}{2}a+\frac{\epsilon^z}{\rho^3}(z+\bar z) + \Oo(\xi^2) \ , \nonumber \\
v &= \frac{1}{2\rho} z + \frac{\epsilon^a}{\rho} a + \Oo(\xi^2) \ . \label{eq:UVchangeEps}
\end{align}
For convenience, we define
\be E = 1+\frac{16|\epsilon^z|^2}{\rho^4} \ . \ee
Lastly, \eqref{eq:CoulombSF} now takes the form
\begin{align}
\omega_+^{\rm orb} &= -4i du\wedge dv = da\wedge dz+\frac{4i\epsilon^a\epsilon^z}{\rho^4} da\wedge (dz+d\bar z) + \frac{2i\epsilon^z}{\rho^4} dz\wedge d\bar z \nonumber \\
\omega_K^{\rm orb} &= 2i(du\wedge d\bar u+dv\wedge d\bar v) \nonumber \\
&= \frac{i\rho^2}{2}\parens{1+\frac{4|\epsilon^a|^2}{\rho^4}}da\wedge d\bar a+\frac{i}{2\rho^2}dz\wedge d\bar z  \nonumber \\
&\quad+\frac{1}{\rho^2}\brackets{- \bar\epsilon^z da\wedge dz - \parens{\bar\epsilon^z-i\epsilon^a}da\wedge d\bar z - \parens{\epsilon^z+i\bar\epsilon^a}d\bar a\wedge dz - \epsilon^z d\bar a\wedge d\bar z } \ . \label{eq:orbW}
\end{align}
Note that when $\epsilon^z=0$, both $T^2_F$ and $T^2_B$ remain holomorphic submanifolds of $T^4$ in complex structure $K$. However, we will see shortly that after orbifolding by $Z_2$ and resolving using real FI parameters we obtain a K3 surface which is (away from singular fibers) a perturbation of a semi-flat genus 1 fibration, but not necessarily an elliptic fibration -- that is, there need not be a section. Indeed, this is suggested by the identifications in \eqref{eq:NchangeEps}.

We note that there is significant freedom in how we describe deformations away from $T^2_F\times T^2_B$, since any lattice embedding $\Lambda\hookrightarrow \RR^4$ related by $SO(4)$ transformations defines the same torus. However, by looking at the $SO(4)$-invariant Gram matrix associated to the basis $(\tilde n^1,\tilde n^2,\tilde n^3,\tilde n^4)=\{(1,0,0,0),(0,1,0,0),(0,0,1,0),(0,0,0,1)\}$ of $\Lambda$ (using the metric \eqref{eq:orb}), namely
\begin{align}
\hspace*{-2.5em}\frac{1}{\rho^2}\begin{pmatrix}
\rho^4 + 4 |\epsilon^a|^2 & \tau_{B,1}\parens{\rho^4+4|\epsilon^a|^2} & 2i(\bar\epsilon^z-\epsilon^z)+\epsilon^a+\bar\epsilon^a & 2i\tau_{F,1}(\bar\epsilon^z-\epsilon^z)+\tau_F \bar\epsilon^a + \bar\tau_F \epsilon^a \\
 & |\tau_B|^2\parens{\rho^4 + 4 |\epsilon^a|^2} & 2i(\tau_B \bar\epsilon^z-\bar\tau_B \epsilon^z)+\tau_B\epsilon^a+\bar\tau_B\bar\epsilon^a & 2i\tau_{F,1}(\tau_B \bar\epsilon^z - \bar\tau_B \epsilon^z) + \tau_F \bar\tau_B \bar\epsilon^a + \bar\tau_F \tau_B \epsilon^a \\
 &  & E & \tau_{F,1}E \\
& & & \tau_{F,1}^2E+\tau_{F,2}^2
\end{pmatrix}
\end{align}
(where only half of the matrix is shown, to save space), we find that \eqref{eq:NchangeEps} is sufficiently general so as to describe the 4 new moduli of $T^4$. That is, the complex parameters $\epsilon^z$ and $\epsilon^a$ encode 4 real moduli which cannot be eliminated by $SO(4)$ transformations. (The top $2\times 2$ block of the Gram matrix involves 4 linearly independent combinations of $\epsilon^a,\bar\epsilon^a,\epsilon^z$, and $\bar\epsilon^z$. And, furthermore, looking at the Gram matrix makes it clear that these are the 4 moduli of interest, as opposed to the overall volume modulus. For instance, after rescaling it so that its first entry is 1, it still has 9 independent moduli.)

We now turn to the geometric interpretation of these parameters. We begin by considering the case with $\epsilon^z=0$ but $\epsilon^a\not=0$. Then, \eqref{eq:orbW} arises from the general prescription \eqref{eq:darboux} if we set
\begin{align}
\Y^{\rm orb}_m(\zeta) &= \Y^{\rm sf}_m(\zeta) + \frac{2\pi}{\tau_{F,2}} \parens{-\bar\tau_F\epsilon^a a + \tau_F\bar\epsilon^a \bar a}  \nonumber \\
\Y^{\rm orb}_e(\zeta) &= \Y^{\rm sf}_e(\zeta) + \frac{2\pi}{\tau_{F,2}}\parens{-\epsilon^a a + \bar\epsilon^a \bar a} \ .
\end{align}
This suggests that we replace $\theta_e$ and $\theta_m$ by fiber coordinates which differ from them by $a$-dependent translations:
\begin{align}
\frac{\theta'_m}{2\pi} &= \frac{\theta_m}{2\pi} + \frac{a \epsilon^a \bar\tau_F - \bar a \bar\epsilon^a \tau_F}{-i\tau_{F,2}} \\
\frac{\theta'_e}{2\pi} &= \frac{\theta_e}{2\pi} + \frac{a \epsilon^a - \bar a \bar\epsilon^a}{-i\tau_{F,2}} \\
z' &= \frac{\theta'_m - \tau_F \theta'_e}{2\pi} = z + 2\epsilon^a a \ .
\end{align}
The functions $\Y^{\rm orb}_\gamma$, as opposed to $\Y^{\rm sf}_\gamma$, are actually the correct functions to appear in the integral equation. This is surprising, since real mass parameters do not affect 4d physics, as they correspond to flavor holonomies along $S^1_R$, and so their effects should na\"ively disappear in the large $R$ limit.

To understand this, consider an M2-brane probing a genus 1 fibration in the limit where its fibers shrink to zero size. As stressed in \cite{morrison:triples,morrison:genus1,wati:genus1}, the usual adiabatic argument that relates M-theory on an elliptic fibration with vanishing fiber volume to F-theory applies equally well to genus 1 fibrations with vanishing fiber volume. The multi-valued function $\tau_F(a)$ defining the F-theory vacuum coincides with that of the Jacobian elliptic fibration of the genus 1 fibration -- i.e., the fibration obtained by replacing each fiber by its Jacobian. As long as the fibers of a genus 1 fibration have finite size, there is no submanifold that can be identified as the base, but in the zero size limit such a manifold emerges, and indeed it coincides with the base of the Jacobian fibration.

The limiting metric on the genus 1 fibration has every right to be called `semi-flat,' since the fibers are still flat. And, again thanks to the validity of the adiabatic argument in this limit, the Strominger-Yau-Zaslow description of mirror symmetry in terms of fiberwise T-dualities \cite{strominger:mirrorT}, which generally governs the geometry of nearly semi-flat manifolds (see \S\ref{sec:stringWebs}), applies. Proceeding in the reverse direction, from F-theory to M-theory, we can now see how real mass parameters can affect the semi-flat geometry. 4d physics determines the multi-valued function $\tau_F(a)$, and the metric on the base of the associated elliptic fibration, but it is otherwise indifferent to the semi-flat geometry of the total space that is obtained by fattening up the fibers. So, we can take this geometry to be any semi-flat genus 1 fibration with the correct $\tau_F(a)$. It describes how the fibers over the 4d Coulomb branch fatten up, up to exponentially suppressed instanton corrections.

Lastly, we turn on $\epsilon^z$. The guesses
\begin{align}
\tilde\Y^{\rm orb}_m(\zeta) &= \Y^{\rm sf}_m + \frac{2i\epsilon^z\tau_F}{\zeta \rho^2\tau_{F,2}} \parens{\tau_{F,1} \theta_e - \theta_m } + \frac{2\pi}{\tau_{F,2}} \parens{-\bar\tau_F\epsilon^a a + \tau_F\bar\epsilon^a \bar a} - \frac{2i\bar\epsilon^z \bar\tau_F \zeta}{\rho^2\tau_{F,2}}(\tau_{F,1}\theta_e-\theta_m) \nonumber \\
\tilde\Y^{\rm orb}_e(\zeta) &= \Y^{\rm sf}_e + \frac{2i\epsilon^z}{\zeta \rho^2\tau_{F,2}}\parens{\tau_{F,1}\theta_e - \theta_m} + \frac{2\pi}{\tau_{F,2}}\parens{-\epsilon^a a + \bar\epsilon^a \bar a} - \frac{2i\bar\epsilon^z \zeta}{\rho^2\tau_{F,2}}(\tau_{F,1}\theta_e-\theta_m) \nonumber
\end{align}
yield \eqref{eq:orbW}, but they are not correct. To see this, note that $\X_\gamma=e^{\Y_\gamma}$ should be a periodic function of $\theta_\gamma$. Motivated by this observation, we search for a hyper-K\"ahler rotation
\be \twoMatrix{\omega'_K}{\omega'_-}{\omega'_+}{-\omega'_K} = \twoMatrix{P}{Q}{-Q^*}{P^*}\twoMatrix{\omega_K}{\omega_-}{\omega_+}{-\omega_K}\twoMatrix{P^*}{-Q}{Q^*}{P} \ , \quad |P|^2+|Q|^2=1 \ , \ee
with $\omega'_{+\, z\bar z}=0$. We also require that $P\to 1$ and $Q\to 0$ as $\epsilon^z\to 0$. Up to a phase, these requirements determine
\be
P = \sqrt{\frac{1}{2}\parens{1+\frac{1}{\sqrt{E}}}} \ ,\quad
Q = \frac{2\sqrt{2}\bar\epsilon^z }{\rho^2 \sqrt{E+\sqrt{E}}} \ .
\ee
If we now define
\be
\varpi'(\zeta) = -\frac{i}{2\zeta} \omega'_+ + \omega'_K - \frac{i\zeta}{2} \omega'_- \ ,
\ee
then we have
\be \varpi'^{\rm orb}(\zeta) = \frac{1}{4\pi^2 R'} d\Y_m^{\rm orb}(\zeta)\wedge d\Y_e^{\rm orb}(\zeta) \ , \label{eq:wRot} \ee
with
\begin{align}
\Y^{\rm orb}_\gamma(\zeta) &= \frac{\pi R'}{\zeta} Z'_\gamma + i \theta'_\gamma + \pi R' \zeta \overline{Z'_\gamma} \\
Z_e' &= a' = \half\parens{\sqrt{E}+\frac{\rho^4+8i\epsilon^a\epsilon^z}{\rho^4}}a + \frac{\epsilon^z}{2\bar\epsilon^z}\parens{\sqrt{E}-\frac{\rho^4-8i\bar\epsilon^a\bar\epsilon^z}{\rho^4}} \bar a \\
\tau'_F &= \half\parens{\tau_F\parens{1+\frac{1}{\sqrt{E}}} + \bar\tau_F\parens{1-\frac{1}{\sqrt{E}}}} 
= \tau_{F,1} + \frac{i\tau_{F,2}}{\sqrt{E}} \\
Z_m' &= \tau'_F Z_e' \ , \quad R' = E^{-1/2}R \ , \quad \rho' = \sqrt{R'\tau'_{F,2}} \\
\frac{\theta_m'}{2\pi} &= \frac{\theta_m}{2\pi} + \frac{a\brackets{\epsilon^a(\bar\tau_F + 16\rho^{-4}|\epsilon^z|^2 \tau_{F,1})+2\tau_{F,2}\bar\epsilon^z}-{\rm c.c.}}{-i\tau_{F,2}E} \\
\frac{\theta_e'}{2\pi} &= \frac{\theta_e}{2\pi} + \frac{a\epsilon^a-\bar a \bar\epsilon^a}{-i\tau_{F,2}} \\
z' &= \frac{\theta'_m - \tau'_F \theta'_e}{2\pi} \ .
\end{align}
If one desires, the translational identifications on $a'$ (which also affect $\theta'_e$ and $\theta'_m$) can be made to take the form $a'\sim a'+1\sim a'+\tau'_B$ by combining the remaining hyper-K\"ahler rotation freedom by a matrix of the form $\twoMatrix{e^{i\phi/2}}{}{}{e^{-i\phi/2}}$, i.e. $\zeta\mapsto e^{i\phi}\zeta$, with a rescaling of the form $a'\mapsto e^{i\phi} \kappa a'$, $R'\mapsto R'/\kappa$, and $\varpi'\mapsto \kappa\varpi'$ with $\kappa>0$. (Note that rescaling $\varpi'$ rescales the overall volume. That we should need to implement such a rescaling between the Higgs and Coulomb branch formalisms is unsurprising, since in the former the overall volume is arbitrary, while in the latter it is determined by the other moduli of the K3 surface.) In contrast, we already have $z'\sim z'+1\sim z'+\tau'_F$. Similarly, we have $(a',z')\sim (-a',-z')$.

The geometric interpretation of all of this is simply that after turning on $\epsilon^z$, we are still describing an orbifold semi-flat genus 1 fibration, but the manifold has this structure in a complex structure different from the $K$ complex structure of the Higgs branch formalism. Our argument at the beginning of this section now implies that the FI parameters that match up with real mass parameters are not the real FI parameters, but rather those which are real in this other complex structure. Lastly, $T^2_F$ and $T^2_B$ have complex structures $\tau'_F$ and $\tau'_B$, respectively, as opposed to $\tau_F$ and $\tau_B$. (By $T^2_B$, we mean the double cover of the base of the Jacobian fibration, a.k.a. the double cover of the 4d Coulomb branch.)

In summary, \eqref{eq:intEqn} is now replaced by
\be \Y_\gamma(\zeta) = \Y_\gamma^{\rm orb}(\zeta) - \frac{1}{4\pi i} \sum_{\gamma'} \Omega(\gamma';a') \avg{\gamma,\gamma'} \int_{\ell'_{\gamma'}(a')} \frac{d\zeta'}{\zeta'} \frac{\zeta'+\zeta}{\zeta'-\zeta} \log(1-\X_{\gamma'}(\zeta')) \ , \label{eq:intEqn2} \ee
with
\be \ell'_\gamma(a') = \{\zeta\in\CC^\times \, | \, Z'_\gamma(a') / \zeta \in \RR_- \} \ . \label{eq:ray2} \ee
(We will explicitly state the values of the real and complex mass parameters below. We still define $\theta'_\gamma$ and $Z'_\gamma$ for all $\gamma\in\hat\Gamma$, given their values for a basis of $\hat\Gamma$, by
\be e^{i\theta'_{\gamma+\gamma'}} = (-1)^{\avg{\gamma,\gamma'}}e^{i(\theta'_\gamma+\theta'_{\gamma'})} \ , \quad Z'_{\gamma+\gamma'} = Z'_\gamma + Z'_{\gamma'} \ .) \ee
The derivation in \cite{GMN:walls} of the approximation \eqref{eq:firstApprox} is unchanged, with $\Y^{\rm sf}$ replaced by $\Y^{\rm orb}$ and unprimed quantities replaced by primed ones. As we will now see, the key fact of the previous section, namely that the Higgs side first order perturbation theory about the orbifold point coincided with the Coulomb side first iteration of the integral formula, starting with the semi-flat expression, also holds here, as long as we employ \eqref{eq:intEqn2} and start our iteration with $\Y^{\rm orb}$.

\subsubsection{Poisson resummation}

Without further ado, we again commence Poisson resummation. We will do this separately for each equivalence class in $\Lambda/2\Lambda$, so that we can treat the FI parameters as constants. Motivated by the SYZ picture of mirror symmetry, as well as the successes of the previous subsection, we still only Poisson resum over $\tilde n^3$ and $\tilde n^4$. We thus fix $\lambda^3,\lambda^4\in \{0,1\}$, as well as $\tilde n^1$ and $\tilde n^2$, and sum over $\tilde n^3$ and $\tilde n^4$ such that
\be \tilde x^3 = (\tilde n^3+\lambda^3)/2 \ , \quad \tilde x^4 = (\tilde n^4+\lambda^4)/2 \ee
are integers. Denoting the function being resummed as $F(\tilde x-\lambda/2)$, we have
\be \int d^2\tilde x\, e^{2\pi i k\cdot \tilde x} F(\tilde x-\lambda/2) = e^{\pi i k\cdot \lambda}\int d^2x\, e^{2\pi i k\cdot x} F(x) \ , \ee
where
\be x^3 = \tilde n^3/2 \ , \quad x^4 = \tilde n^4/2 \ . \ee
So, we can assume that $\lambda^3=\lambda^4=0$, since otherwise we just multiply by $e^{\pi i k\cdot\lambda}$ in the end.

Next, we note that $F$ takes the form $N/D_+^3$, where $N$ is a polynomial in $x^3,x^4$ and $D_+$ is a quadratic polynomial in $x^3,x^4$. By an affine change of variables, $D_+$ can be put into the form
\be D_+ = x'^2 + y'^2 + C \ , \label{eq:niceForm} \ee
with $C>0$. The effects of $N$ may be incorporated in the end by differentiating the Fourier transform of $1/D_+^3$ with respect to the entries of $k$, so the main computation is
\begin{align}
\int d^2x'\, \frac{e^{2\pi i k'\cdot x'}}{D_+^3} 
&= \int rdrd\theta\, \frac{e^{2\pi i |k'| r\cos\theta}}{(r^2+C)^3} \nonumber \\
&= 2\pi \int_0^\infty dr\, r\frac{J_0(2\pi |k'|r)}{(r^2+C)^3} \nonumber \\
&= \frac{\pi^3}{C}|k'|^2 K_2(2\pi |k'|\sqrt{C}) \ . \label{eq:N2fourier}
\end{align}
We note that $D_+$ is actually quadratic in all four components of $n$, and so the same reasoning can be employed in Poisson resumming over 1, 3, or 4 of its entries. In all cases, $D_+$ can be put in the form $r^2+C$, with $C>0$ and $r$ the radial coordinate, and the relevant integrals take the form
\begin{align}
\int dx'\, \frac{e^{2\pi ik'x'}}{D_+^3} &= \frac{e^{-2\pi \sqrt{C} |k'|} \pi(3+6\pi \sqrt{C} |k'| + 4\pi^2 C k'^2)}{8C^{5/2}} \ ,  \label{eq:N1fourier} \\
\int d^3x'\, \frac{e^{2\pi i k'\cdot x'}}{D_+^3} &= \int_0^\pi \sin\theta\, d\theta \int_0^{2\pi} d\phi \int_0^\infty r^2 dr\, \frac{e^{2\pi i |k'| r \cos\theta}}{(r^2+C)^3} \nonumber \\
&=  \int dr\, \frac{2r\sin(2\pi |k'| r)}{|k'|(r^2+C)^3} \nonumber \\
&= \frac{\pi^2 e^{-2\pi \sqrt{C} |k'|} (1+2\pi \sqrt{C} |k'|)}{4C^{3/2}} \ , \label{eq:N3fourier}  \\
\int d^4x'\, \frac{e^{2\pi i k'\cdot x'}}{D_+^3} &= \int_0^\pi \sin^2\theta_1 d\theta_1 \int_0^\pi \sin\theta_2 d\theta_2 \int_0^{2\pi} d\theta_3 \int_0^\infty r^3 dr\, \frac{e^{2\pi i |k'| r \cos\theta_1}}{(r^2+C)^3} \nonumber \\
&= \int dr\, \frac{2\pi r^2 J_1(2\pi r |k'|)}{|k'|(r^2+C)^3} \nonumber \\
&= \frac{\pi^3 |k'|}{\sqrt{C}} K_1(2\pi \sqrt{C} |k'|) \ . \label{eq:N4fourier}
\end{align}
These results are likely of interest in limits of K3 moduli space other than the semi-flat limit. However, we will henceforth stick with \eqref{eq:N2fourier}. (\eqref{eq:N4fourier} is deceptively similar to the sort of formulae we are trying to match: the inside of the Bessel function is not of the correct form.)

Returning to the $d=2$ case of interest, we will now be a bit more explicit. $D_+$ takes the form
\be D_+ = c_1 (x^3)^2 + c_2 x^3x^4 + c_3 (x^4)^2 + c_4 x^3 + c_5 x^4 + c_6 \ , \ee
with
\begin{align}
c_1 &= \frac{E}{\rho^2} \nonumber \\
c_2 &= \frac{2\tau_{F,1}E}{\rho^2} \nonumber \\
c_3 &= \frac{\tau_{F,1}^2 E + \tau_{F,2}^2 }{\rho^2} \nonumber \\
c_4 &= \frac{4\bar\epsilon^z((i\rho/2) n_B + 2u) + \rho^2(\delta n^v +2v)}{\rho^3} + {\rm c.c.} \nonumber \\
c_5 &= \frac{4\tau_{F,1}\bar\epsilon^z((i\rho/2) n_B + 2u) + \rho^2\bar\tau_F (\delta n^v +2v) }{\rho^3 } +{\rm c.c.} \nonumber \\
c_6 &= \abs{\frac{i\rho}{2}n_B +2u}^2 + \abs{\delta n^v +2v}^2 \ .
\end{align}
We also note the value of the following commonly-occuring quantity:
\be 4c_1c_3-c_2^2 = \frac{4\tau_{F,2}^2 E}{\rho^4} = \frac{4}{R'^2} \ . \ee
Defining $x'$ and $y'$ via
\be x^3 = \frac{1}{\sqrt{c_1}} x' - \frac{c_2}{\sqrt{c_1(4c_1c_3-c_2^2)}} y' + \frac{c_2c_5-2c_3c_4}{4c_1c_3-c_2^2} \ , \quad x^4 = 2\sqrt{\frac{c_1}{4c_1c_3-c_2^2}} \,y' + \frac{c_2c_4-2c_1c_5}{4c_1c_3-c_2^2} \label{eq:primeVars} \ee
yields \eqref{eq:niceForm}, with
\begin{align}
C &= \frac{c_2c_4c_5-c_3c_4^2-c_1c_5^2}{4c_1c_3-c_2^2} + c_6 \nonumber \\
&= \frac{4}{E\rho^4}\abs{(-\frac{i\rho}{2}\bar n_B +2\bar u) \epsilon^z-(\frac{i\rho}{2}n_B +2u) \bar\epsilon^z}^2
 + \frac{1}{E} \abs{(i\rho/2)n_B +2u-\frac{2\epsilon^z}{\rho^2}(\delta n^v  +2v+\delta \bar n^v + 2\bar v)}^2 \nonumber \\
&= \frac{1}{E\rho^2}\abs{\epsilon^z(\bar n_B+2\bar a)+\bar\epsilon^z(n_B+2a)}^2+\frac{\rho^2}{4E}\abs{n_B+2a+\frac{4i\epsilon^z}{\rho^4}\parens{\epsilon^a(n_B+2a)+\bar\epsilon^a(\bar n_B+2\bar a)}}^2 \ ;
\end{align}
the last line holds at leading order in $\xi$. The Jacobian of the transformation \eqref{eq:primeVars} gives
\be dx^3 dx^4 = \frac{2}{\sqrt{4c_1c_3-c_2^2}} \, dx' dy' \ . \ee
We define $k'$ so that $k'\cdot x'=k\cdot x-\frac{c_2c_5-2c_3c_4}{4c_1c_3-c_2^2} k_3 - \frac{c_2c_4-2c_1c_5}{4c_1c_3-c_2^2} k_4$; that is,
\be k'_x = \frac{k_3}{\sqrt{c_1}} \ , \quad k'_y = \frac{1}{\sqrt{c_1(4c_1c_3-c_2^2)}}\parens{ - c_2 k_3 + 2 c_1 k_4 } \ , \ee
and in particular
\be |k'|^2 = \frac{\rho^2|\tau_F k_3-k_4|^2+16\rho^{-2}|\epsilon^z|^2(\tau_{F,1}k_3-k_4)^2}{\tau_{F,2}^2 E} \ . \ee
Then,
\begin{align}
\int d^2x\, \frac{e^{2\pi i k\cdot x}}{D_+^3}
&= \frac{2}{\sqrt{4c_1c_3-c_2^2}} e^{2\pi i\parens{ C_3 k_3 + C_4 k_4 }} \cdot \int d^2x'\, \frac{e^{2\pi i k'\cdot x'}}{D_+^3} \nonumber\\
&= \frac{2}{\sqrt{4c_1c_3-c_2^2}} e^{2\pi i\parens{ C_3 k_3 + C_4 k_4 }} \cdot \frac{\pi^3}{C} |k'|^2 K_2(2\pi|k'|\sqrt{C}) \ ,
\end{align}
where
\begin{align}
C_3 &= \frac{c_2c_5-2c_3c_4}{4c_1c_3-c_2^2} \\
&= - \frac{\theta_m}{2\pi} - \frac{(n_B+2a)\brackets{\epsilon^a(\bar\tau_F+16\rho^{-4} |\epsilon^z|^2\tau_{F,1})+2\tau_{F,2} \bar\epsilon^z}-{\rm c.c.}}{-2i\tau_{F,2} E} \\
C_4 &= \frac{c_2c_4-2c_1c_5}{4c_1c_3-c_2^2} \\
&= \frac{\theta_e}{2\pi} + \frac{\epsilon^a(n_B+2a)-\bar\epsilon^a(\bar n_B+2\bar a)}{-2i\tau_{F,2}} \ .
\end{align}
Note that all dependence on the coordinates $(\theta_e,\theta_m,a,\bar a)$ is in $C_3,C_4$, and $C$. All of the desired Fourier transforms now follow from derivatives of this result. The identities
\begin{align}
K_{n+1}(x)&=\frac{2n}{x} K_n(x) + K_{n-1}(x) \ , \label{eq:Krecur} \\
K'_n(x) &= - \frac{n}{x} K_n(x) - K_{n-1}(x) \ , \quad K_{-n}(x) = K_n(x) \ .
\end{align}
prove useful for simplifying expressions. The recurrence \eqref{eq:Krecur} can be solved:
\be K_n(x) = \alpha_n(x) K_0(x) + \beta_n(x) K_1(x) \ , \ee
where
\be \alpha_n(x) = x\parens{I_n(-x)K_1(x)+I_1(x)K_n(x)} \ , \quad \beta_n(x) = x\parens{-I_n(-x)K_0(x)+I_0(x)K_n(x)} \ . \ee
The $I_n$ are another class of Bessel function. Remarkably, the functions $\alpha_n,\beta_n$ are always polynomials in $1/x$.

In order to interpret these expressions, we rename $(-k_3,k_4)$ to $\gamma^g=(p',q')$ and define
\begin{align}
\frac{\theta_{f,m}}{2\pi} &= \frac{n_B\brackets{\epsilon^a(\bar\tau_F+16\rho^{-4} |\epsilon^z|^2\tau_{F,1})+2\tau_{F,2} \bar\epsilon^z}-{\rm c.c.}}{-2i\tau_{F,2} E} \\
\frac{\theta_{f,e}}{2\pi} &= \frac{\epsilon^a n_B-\bar\epsilon^a\bar n_B}{-2i\tau_{F,2}} \\
\theta'_{\gamma^f} &= p' \theta_{f,m} + q' \theta_{f,e} \\
e^{i\theta'_\gamma} &= (-1)^{p' q'} e^{i(p' \theta'_m + q' \theta'_e + \theta'_{\gamma^f})} \\
n'_B &=  \half\parens{\sqrt{E}+\frac{\rho^4+8i\epsilon^a\epsilon^z}{\rho^4}}n_B + \frac{\epsilon^z}{2\bar\epsilon^z}\parens{\sqrt{E}-\frac{\rho^4-8i\bar\epsilon^a\bar\epsilon^z}{\rho^4}} \bar n_B  \\
Z'_{\gamma^g} &= (p' \tau'_F + q') a' \\
Z'_{\gamma^f} &= \half (p'\tau'_F+q') n'_B \\
Z'_\gamma &= Z'_{\gamma^g} + Z'_{\gamma^f} = (p'\tau'_F + q')(a'+\half n'_B) \\
\gamma^f &= (p' \tilde n^1, p' \tilde n^2, q' \tilde n^1, q' \tilde n^2) \ .
\end{align}
In particular, we identify $Z'_e = Z'_{\gamma_e}$ and $Z'_m = Z'_{\gamma_m}$. We also identify
\begin{align}
C_3 &= - \frac{\theta'_m+\theta_{f,m}}{2\pi} \ , \quad C_4 = \frac{\theta'_e+\theta_{f,e}}{2\pi} \ , \quad C = \rho'^2 |a'+\half n'_B|^2 \ , \quad \sqrt{C} |k'| = R' |Z'_\gamma| \ .
\end{align}
So,
\be \int d^2x\, \frac{e^{2\pi i k\cdot x}}{D_+^3} = R' e^{i(p'\theta'_m+q'\theta'_e+\theta'_{\gamma^f})} \cdot \frac{\pi^3 |Z'_\gamma|^2}{\tau_{F,2}'^2 |a'+\half n'_B|^4} K_2(2\pi R' |Z'_\gamma|) \ . \ee
This is starting to look familiar. In particular, we observe that the real masses appearing in $\theta_{\gamma^f}$ (of which there are 4 -- namely, the coefficients of $p'\tilde n^1$, $p'\tilde n^2$, $q' \tilde n^1$, and $q' \tilde n^2$) are associated to the same charges that play a role in $Z'_{\gamma^f}$, and so we lose no information about the BPS spectrum by setting $\epsilon^a=\epsilon^z=0$. We henceforth set $\epsilon^z=0$, in order to avoid pointless suffering; in contrast, keeping $\epsilon^a$ around entails minimal unpleasantness. Correspondingly, we turn on only the real FI parameters. We can also drop the primes on a number of symbols: $R'$, $a'$, $Z'_\gamma$, $\tau_F'$, $n_B'$, etc.

We denote the order $\xi^2$ corrections to the semi-flat limit by
\begin{align}
\varpi(\zeta) &= \varpi^{\rm orb}(\zeta)+\varpi^{\rm inst}(\zeta) \nonumber \\
\varpi^{\rm inst}(\zeta) &= -\frac{i}{2\zeta}\omega_+^{\rm inst} + \omega_K^{\rm inst} - \frac{i\zeta}{2}\omega_-^{\rm inst} \nonumber \\
&= \sum_{(\tilde n^1,\tilde n^2)\in \ZZ^2} \sum_{\gamma^g\in\ZZ^2} \sum_{\lambda\in\{0,1\}^2} e^{i\pi \gamma^g\cdot\lambda} \varpi^{\rm inst}_{n_B\gamma^g\lambda} \nonumber \\
&= \sum_{(\tilde n^1,\tilde n^2)\in \ZZ^2} \sum_{\gamma^g\in\ZZ^2} \sum_{\lambda\in\{0,1\}^2} e^{i\pi \gamma^g\cdot\lambda} \parens{-\frac{i}{2\zeta}\omega_{n_B  \gamma^g \lambda +} + \omega_{n_B  \gamma^g \lambda K} - \frac{i\zeta}{2}\omega_{n_B \gamma^g \lambda -}} \ .
\end{align}
Similarly, the change of variables, to quadratic order in the FI parameters, takes the form
\begin{align}
u &= \frac{i\rho}{2} a + \frac{\epsilon^z}{\rho^3}(z+\bar z) + \sum_{n_B,\gamma^g,\lambda} e^{i\pi \gamma^g\cdot\lambda} u_{n_B\gamma^g\lambda} \nonumber \\
&= \frac{i\rho}{2} a + \frac{\epsilon^z}{\rho^3}(z+\bar z) + \sum_n u_n  \ , \nonumber \\
v &= \frac{z}{2\rho} + \frac{\epsilon^a}{\rho} a + \sum_{n_B,\gamma^g,\lambda} e^{i\pi \gamma^g\cdot\lambda} v_{n_B\gamma^g\lambda}  \nonumber \\
&= \frac{z}{2\rho} + \frac{\epsilon^a}{\rho} a + \sum_n v_n  \ . 
\end{align}
Following \eqref{eq:changeVars}, we make the ansatz
\be u_n = \xi_{n,\RR}^2 \cdot \frac{N^u_+}{4D_+^2} \ , \quad v_n = -\xi_{n,\RR}^2 \cdot  \frac{N^v_+}{4D_+^2} \ , \label{eq:changeVars2} \ee
where we substitute \eqref{eq:NchangeEps} and \eqref{eq:UVchangeEps} in these definitions. These expressions may be Poisson resummed, using
\begin{align}
\int d^2x\, \frac{e^{2\pi i k\cdot x}}{D_+^2} &= \frac{2}{\sqrt{4c_1c_3-c_2^2}} e^{2\pi i(C_3 k_3+C_4 k_4)} \cdot \int d^2x'\, \frac{e^{2\pi i k'\cdot x'}}{D_+^2} \nonumber \\
&= \frac{2}{\sqrt{4c_1c_3-c_2^2}} e^{2\pi i(C_3 k_3+C_4 k_4)} \cdot \frac{2\pi^2}{\sqrt{C}} |k'| K_1(2\pi |k'| \sqrt{C}) \ .
\end{align}
Alternatively, one can make the change of variables and then Poisson resum. After Poisson resumming, we denote the FI parameters by $\xi_{n_B\lambda,\RR}$; of course, they only depend on $n_B$ via $(\tilde n^1,\tilde n^2)$ mod 2.

After some algebra, we finally arrive at the Poisson resummed $\varpi^{\rm inst}$:
\begin{align}
\varpi^{\rm inst}_{n_B\gamma^g\lambda} &= i\pi^2 R^2 \xi_{n_B\lambda,\RR}^2 e^{i(p'\theta'_m + q' \theta'_e+\theta'_{\gamma^f})} \nonumber \\
&\times d\Y^{\rm orb}_\gamma(\zeta) \wedge \parens{-|Z_\gamma| K_1(2\pi R |Z_\gamma|) d\log(Z_\gamma/\bar Z_\gamma) + K_0(2\pi R |Z_\gamma|)\parens{\frac{1}{\zeta} dZ_\gamma - \zeta d\bar Z_\gamma}} \ .
\end{align}
We now follow \S\ref{sec:fieldTheory} in reverse. First, following \eqref{eq:thetaFIfinal}, we define
\begin{align}
\delta\theta_{n_B,1} &= 2\pi^2 R(\xi_{n_B,00,\RR} + \xi_{n_B,10,\RR}) \ , \quad \delta\theta_{n_B,2} = 2\pi^2 R(\xi_{n_B,00,\RR} - \xi_{n_B,10,\RR}) \nonumber \\
\delta\theta_{n_B,3} &= 2\pi^2 R(\xi_{n_B,11,\RR} + \xi_{n_B,01,\RR}) \ , \quad \delta\theta_{n_B,4} = 2\pi^2 R(\xi_{n_B,11,\RR} - \xi_{n_B,01,\RR})
\end{align}
(which only depend on $(\tilde n^1,\tilde n^2)$ mod 2). We then perform the sum over $\lambda$:
\begin{align}
\varpi^{\rm inst}(\zeta) &= \frac{i}{16\pi^2} \sum_{(\tilde n^1,\tilde n^2)\in \ZZ^2} \sum_{\gamma^g\in\ZZ^2\backslash (0,0)} e^{i\theta'_\gamma} \Xi_{n_B}(\gamma) \nonumber \\
&\times d\Y^{\rm orb}_\gamma(\zeta) \wedge \parens{-|Z_\gamma| K_1(2\pi R |Z_\gamma|) d\log(Z_\gamma/\bar Z_\gamma) + K_0(2\pi R |Z_\gamma|)\parens{\frac{1}{\zeta} dZ_\gamma - \zeta d\bar Z_\gamma}} \ ,
\end{align}
where $\Xi_{n_B}$ is defined similarly to \eqref{eq:thetaFI}:
\be (-1)^{p' q'}\Xi_{n_B} = (\delta\theta_{n_B,1}+\delta\theta_{n_B,2})^2 + e^{i\pi p'}(\delta\theta_{n_B,1}-\delta\theta_{n_B,2})^2 + e^{i\pi q'}(\delta\theta_{n_B,3}-\delta\theta_{n_B,4})^2 + e^{i\pi(p'+q')} (\delta\theta_{n_B,3}+\delta\theta_{n_B,4})^2 \ . \ee
Next, we write $(p',q')=n(p,q)$, where $p,q$ are coprime and $n>0$. Redefining $\gamma^g=(p,q)$, $\gamma^f=(p\tilde n^1,p \tilde n^2, q \tilde n^1, q \tilde n^2)$, and $\gamma=\gamma^g\oplus \gamma^f$, we have
\begin{align}
\varpi^{\rm inst}(\zeta) &= \sum_{p,q,\tilde n^1,\tilde n^2} \varpi^{\rm eff}_{\gamma} \nonumber \\
\varpi^{\rm eff}_\gamma &= -\frac{i}{8\pi^2} d\Y^{\rm orb}_\gamma(\zeta) \wedge \brackets{\sum_{n>0} e^{in\theta'_\gamma} \parens{\piecewise{2 \sum_j \cos(n\theta_{n_B,j})}{2|p\wedge 2\nmid q}{\sum_{\{s_j\}} \prod_j e^{\frac{in}{2} s_j \theta_{n_B,j}}}{\rm else} + \piecewise{-8}{n\, \rm{even}}{0}{n\, \rm{odd}}}  \right. \nonumber \\
&\left.\times\parens{-|Z_\gamma| K_1(2\pi R n |Z_\gamma|) \,d\log(Z_\gamma/\bar Z_\gamma) + K_0(2\pi R n |Z_\gamma|) \parens{\frac{1}{\zeta} dZ_\gamma - \zeta d\bar Z_\gamma} } } \ .
\end{align}
Here, we have introduced
\be \theta_{n_B,1}=\delta\theta_{n_B,1} \ , \quad \theta_{n_B,2}=\delta\theta_{n_B,2} \ , \quad \theta_{n_B,3}=\pi+\delta\theta_{n_B,3} \ , \quad \theta_{n_B,4}=\pi+\delta\theta_{n_B,4} \ . \ee
At this point, we can easily read off the BPS spectrum of the theory -- or at least that portion which contributes to the metric at order $\xi^2$.

We begin by specifying the flavor charge lattice. It takes the form $(\ZZ^4)^{\oplus 4}\oplus \ZZ^4$, where the first four $\ZZ^4$ lattices (whose associated central charges vanish) correspond to a $\Spin(8)^4$ global symmetry and the last to a $U(1)^4$ global symmetry. (This lattice has a natural inner product that turns it into $\Gamma^{2,18}$, the even unimodular lattice with signature $(2,18)$, but this will play no role here.) Next, the gauge charge lattice is $\ZZ^2$. Now, we can state the spectrum. For each relatively prime $p,q$ and $\tilde n^1,\tilde n^2\in \ZZ^2$, we have a vector multiplet with gauge charge $(2p,2q)$ and a half-hypermultiplet with gauge charge $(p,q)$. The half-hypermultiplet has charges $(p\tilde n^1,p\tilde n^2,q\tilde n^1,q\tilde n^2)$ under the $U(1)^4$ symmetries, while the vector multiplet has charges $(2p\tilde n^1,2p\tilde n^2,2q\tilde n^1,2q\tilde n^2)$. Lastly, the half-hypermultiplet transforms under the $\Spin(8)$ group labelled by $(\tilde n^1,\tilde n^2)$ mod 2, and it does so in one of the three 8-dimensional representations; the latter is determined by $(p,q)$ mod 2, as in \S\ref{sec:fieldTheory}. In short, up to the $U(1)^4$ charges, we have four doubly infinite sets of copies of the 4d $SU(2)$ $N_f=4$ spectrum, each with its own $\Spin(8)$ flavor symmetry.

We emphasize that this spectrum does not wall cross as we vary the 4d Coulomb modulus! In the case of the 4d $SU(2)$ $N_f=4$ field theory, our explanation of the absence of wall crossing relied on superconformal invariance, which the little string theory on $T^2$ does not enjoy. Another strange feature is the strongly sub-Hagedorn growth of the BPS index. (The sum of the absolute value of the BPS index over all charges with $|p|,|q|,|\tilde n^1|,|\tilde n^2|<n$ grows only as a power of $n$.) One possible explanation for these observations could have been that severe cancellations in the BPS index -- likely due to fermi zero modes (beyond those Goldstinos required by broken supersymmetry) -- are at work. However, in \S\ref{sec:moreStates}, we will show that, in fact, there are contributions to the index which we have not yet found because they affect the metric only at subleading orders in the $\xi$ expansion.

\section{Dual counting problems} \label{sec:counting}

In this section, we explain a number of counting problems which are dual to the one we have studied above.

\subsection{$A_1$ $\N=(1,1)$ little string theory} \label{sec:11}

In this section, we explain a relationship between the $\N=(1,0)$ little string theory we have studied thus far and the simplest maximally supersymmetric little string theory. It involves the moduli spaces, BPS spectra, and canonical coordinates $\Y_\gamma$ of the two theories.

We begin by reviewing the observation of \cite{ganor:noncomm1,ganor:noncomm2} that certain K3 surfaces -- which happen to be special cases of the ones we have already studied, near the $T^4/Z_2$ locus -- arise as the moduli space of the $A_1$ $\N=(1,1)$ little string theory compactified on $T^3$ and deformed by background R-symmetry holonomies which preserve 8 supercharges. At low energies, this little string theory behaves like a 6d $U(2)$ gauge theory and these holonomies behave like masses for the adjoint hypermultiplet. So, we call this the $\N=(1,0)^*$ little string theory. Note that we are privileging the latter factor in the $\Spin(4)=SU(2)_L\times SU(2)_R$ R-symmetry group, which we treat as an $\N=2$ global symmetry; the former is still thought of as an R-symmetry.

One way to think about the compactified $A_1$ $\N=(1,1)$ little string theory involves two parallel type IIB NS5-branes wrapping $T^3$. S-duality and three T-dualities takes this configuration to two D2-branes probing $T^3$, or two M2-branes probing $T^4$. The moduli space is thus seen to be $\Sym^2(T^4)$. The masses then smooth this out to $\Hilb^2(T^4)$. The latter manifold happens to be an `isotrivial' K3 fibration over $T^4$ -- that is, all fibers are the same. Furthermore, these fibers all have $Z_2^4$ symplectic automorphisms. The mild non-triviality of the K3 fibration is due to such $Z_2$ actions as we traverse around a 1-cycle of the $T^4$ base. But, this is still sufficiently simple that the metric on $\Hilb^2(T^4)$ restricts to the exact hyper-K\"ahler metric on all K3 fibers (and the metric on the base is simply the flat metric). One way to see this is that quotienting by the fiberwise $Z_2^4$ action gives a trivial $K3/Z_2^4$ fibration over $T^4$. Another way to see this is that our isotrivial fibration is itself the quotient of a trivial K3 fibration over a bigger 4-torus by a $Z_2^4$ action which simultaneously translates by a half-period of the base and acts by a symplectomorphism of the fiber.

To explain these claims, we will work with the symmetric product, as the resolution does not change this story meaningfully. So, we have two unordered points $(z, w)$ on $T^4$. This is an isotrivial $T^4/Z_2$ fibration over $T^4$; we can use $z$ as a coordinate on a fiber and $z+w$ on the base. One can roughly think of the $T^4$ base as being the center of mass, but this is more properly thought of as ``$\frac{z+w}{2}$"; the problem is that on $T^4$, there are 16 different points that this could refer to, and so only $z+w$ is well-defined. The $Z_2$ quotient in a fiber with $z+w=c$ arises from the identification $z\sim w=c-z$. The $Z_2^4$ action simply involves adding the same half-period to both $z$ and $w$; this does not affect $z+w$, but it does change the fiber coordinate. The quotient manifold described near the end of the previous paragraph can be parametrized by $z-w=2z-c$ in a fiber and $z+w$ on the base.

Mathematically, what we have just sketched is the fact that K3 arises as a `generalized Kummer variety,' i.e. as the fiber over 0 of the natural fibration $\Hilb^n(T^4)\to T^4$ whose projection map is the addition map.

If the $T^3$ wrapped by the NS5-branes splits as $T^2\times S^1_R$ with $R$ large, then the Coulomb branch formalism applies. Therefore, BPS state counts in this little string theory on $T^2$ determine a family of K3 metrics. As usual, we will only turn on real mass parameters, which here means that the only non-trivial R-symmetry holonomy is on $S^1_R$. The relevant BPS state counting problem then involves the simplest maximally supersymmetric little string theory!

However, in many respects it still has the flavor of a BPS state counting problem with 8 supercharges, rather than 16. For, in order to turn on the R-symmetry holonomy on $S^1_R$, we need a flavored index which keeps track of R-charges. This is the index $B_2(z)$ of \cite{sen:N4refine},
\be B_2(z,\gamma;u) = \sum_r \Omega(\gamma+r \gamma_R; u) z^r \ , \ee
where $\gamma_R$ denotes a single unit of R-charge, whose wall crossing behavior is that of a 4d $\N=2$ index, rather than of the typical $\N=4$ index $B_6$. Furthermore, while the moduli space of the theory on $T^2$ is $\Sym^2(T^2\times \RR^4)$, we can only compute $B_2(z)$ on the `Coulomb branch' sublocus $\Sym^2(T^2)$ where the R-symmetry is unbroken. On the one hand, this is precisely where we want to be in order to get K3 metrics. But, on the other hand, it means that (unlike in the field theory examples studied in \cite{sen:N4refine}) there is no weak coupling limit that we can study in order to compute $B_2(z)$. However, one can make predictions about the weak coupling spectrum by first specializing $B_2(z)$ to $B_6$ and then using the wall crossing formula for $B_6$ to determine its weak coupling limit. For $1/2$-BPS states, the situation is even better, as one may similarly extract the index $B_4$ from $B_2(z)$ and, in contrast to $B_2(z)$ and $B_6$, this index does not wall cross, and so again it yields predictions about the weak coupling spectrum.

As discussed in \cite{sen:N4refine}, there are two types of massive multiplets which can contribute to $B_2(z)$. First, there are $1/2$-BPS, or short, multiplets. And second is the class of $1/4$-BPS, or intermediate, multiplets whose unbroken supercharges transform trivially under $SU(2)_R$. However, we will soon find that the BPS states that we found in the $\N=(1,0)$ little string theory all correspond to $1/2$-BPS vector multiplets. These each make a contribution of 1 to $B_4$. One way to see this correspondence is that the spectrum we found in the $\N=(1,0)$ theory did not wall cross, and neither do counts of these $1/2$-BPS states. We will provide more intuition for this in the next section.

However, in the meantime we first follow the example of the previous section and study the Coulomb branch geometry associated to this spectrum. This will give us a physical explanation for the $Z_2^4$ symmetries of our K3 surfaces and allow us to find the relationships between the parameters and canonical coordinates of the two theories.

We warm up with a field theory example. At low energies, near the origin of moduli space, our little string theory reduces to the 4d $U(2)$ $\N=4$ gauge theory. The correspondence we have just explained between the $\N=(1,1)$ and $\N=(1,0)$ little string theories descends to one between the 4d $U(2)$ $\N=4$ and $SU(2)$ $N_f=4$ gauge theories. Although we will soon neglect the center of mass moduli, it will still turn out to be important that we have a $U(2)$ gauge theory, as opposed to an $SU(2)$ one.

We first need to think a bit about the charge lattice, following \cite{kapustin:lines,GMN:framed}. The weight lattice of $U(2)\cong (U(1)\times SU(2))/Z_2$ is spanned by $\gamma_1=(\half,\half)$ and $\gamma_2=(1,0)$, where the first factor labels half of the charge of a $U(1)$ representation and the second labels an $SU(2)$ representation (by its spin). That is, if we parametrize the maximal torus of $U(2)$ via $T=e^{i\alpha/2}\twoMatrix{e^{i\beta/2}}{}{}{e^{-i\beta/2}}$ then a weight $(a,b)$ determines a homomorphism $T\mapsto e^{i(a\alpha+b\beta)}$ from the maximal torus to $U(1)$; this is well-defined (i.e., invariant under simultaneous shifts of $\alpha\mapsto \alpha+2\pi$, $\beta\mapsto \beta+2\pi$, as well as $\beta\mapsto \beta+4\pi$) if $a\equiv b\!\!\mod 1$.\footnote{While this representation of the weight lattice is convenient for us, we note that instead parametrizing the lattice by $(a-b,a+b)$ yields an isomorphism with $\ZZ^2$ and that the homomorphism from the maximal torus to $U(1)$ now takes the form $\twoMatrix{e^{i\phi_1}}{}{}{e^{i\phi_2}}\mapsto e^{i(\phi_1(a+b)+\phi_2(a-b))}$.} Since $U(2)$ maps to itself under S-duality, the magnetic weight lattice is isomorphic to the weight lattice. Now, magnetic weights $(a',b')$ determine homomorphisms from $U(1)$ to the maximal torus defined by $e^{i\phi}\mapsto e^{ia'\phi}\twoMatrix{e^{ib'\phi}}{}{}{e^{-ib'\phi}}$. Composing the homomorphisms associated to a weight $(a,b)$ and a magnetic weight $(a',b')$ gives a homomorphism from $U(1)$ to itself:
\be e^{i\phi}\mapsto e^{ia'\phi}\twoMatrix{e^{ib'\phi}}{}{}{e^{-ib'\phi}} \mapsto e^{2i(aa'+bb')\phi} \ . \ee
This allows us to define an integral pairing $\llangle (a',b'),(a,b)\rrangle = 2(aa'+bb')$. We turn this into a symplectic pairing on the electromagnetic charge lattice by defining
\be \avg{(a,b,a',b'),(c,d,c',d')} = \llangle (a',b'),(c,d) \rrangle - \llangle (c',d'), (a,b) \rrangle \ . \ee
We take $\gamma_1'=(1,0)$ and $\gamma_2'=(\half,-\half)$ as a basis for the magnetic weight lattice; then,
\be \avg{\gamma_1',\gamma_1}=\avg{\gamma_2',\gamma_2}=1\ , \quad \avg{\gamma_1',\gamma_2}=2 \ , \ee
and all other pairings (besides those determined by anti-symmetry of the pairing) vanish.

We now define $\gamma_e = 2\gamma_1 - \gamma_2$ and $\gamma_m = \gamma_1'-2\gamma_2'$, which are associated, respectively, to the adjoint representation of the electric and magnetic $U(2)$ gauge group. These satisfy
\be \avg{\gamma_m,\gamma_e}=2 \ , \quad \avg{\gamma_m,\gamma_1'}= \avg{\gamma_m,\gamma_2} = \avg{\gamma_e,\gamma_1'} = \avg{\gamma_e,\gamma_2} = 0 \ . \label{eq:newInts} \ee
The BPS spectrum is then as follows: for each relatively prime $p,q\in\ZZ$, we have a vector multiplet with gauge charges $p\gamma_m+q\gamma_e$. With these conventions, we have $Z_m=\tau_F Z_e$, where $\tau_F$ transforms in the usual way under S-duality. As usual, we denote $Z_e=a$. The second set of equations in \eqref{eq:newInts} implies that the BPS spectrum does not correct $\Y^{\rm sf}_{\gamma'_1}$ or $\Y^{\rm sf}_{\gamma_2}$. The generalization of \eqref{eq:darboux} is now
\begin{align}
\varpi(\zeta) &= \frac{1}{8\pi^2R} \epsilon^{ij} d\Y_i(\zeta)\wedge d\Y_j(\zeta) \nonumber \\
&= \frac{1}{4\pi^2R} \parens{d\Y^{\rm sf}_{\gamma_1'}\wedge d\Y_{\gamma_1}-2d\Y_{\gamma_2'}\wedge d\Y_{\gamma_1} + d\Y_{\gamma_2'}\wedge d\Y^{\rm sf}_{\gamma_2}} \nonumber \\
&= \frac{1}{8\pi^2R} \parens{d\Y^{\rm sf}_{\gamma_1'}\wedge d\Y^{\rm sf}_{\gamma_2} + d\Y_m\wedge d\Y_e } \ , \label{eq:U2varpi}
\end{align}
where $\epsilon=-\Sigma^{-1}$, $\Sigma_{ij}=\avg{\gamma_i,\gamma_j}$ is the symplectic pairing on the electromagnetic charge lattice, and $i,j$ run over our basis. We take $Z_{\gamma_2}$ and $a$ to be coordinates on the 4d Coulomb branch; the $Z_2$ Weyl group of $U(2)$ (which maps a weight $(a,b)$ to $(a,-b)$, so $\gamma_1\mapsto -\gamma_1+\gamma_2$, $\gamma_2\mapsto \gamma_2$, and $\gamma_e\mapsto -\gamma_e$) implies that these are well-defined up to $a\sim -a$. Since neither $\Y^{\rm sf}_m$ and $\Y^{\rm sf}_e$ nor corrections thereto depend on $Z_{\gamma_2}$, we set the latter to 0.

Lastly, we have the fiber coordinates $\theta_1$, $\theta_2$, $\theta_1'$, and $\theta_2'$. Requiring that the homomorphisms from the maximal torus to $U(1)$ associated, respectively, to $\gamma_1$ and $\gamma_2$ take the forms $e^{i\theta_1}$ and $e^{i\theta_2}$ implies that $\theta_1$ and $\theta_2$ parametrize the maximal torus via $\alpha=\theta_2$, $\beta=2\theta_1-\theta_2=\theta_e$. (As required, these expressions are well-defined up to $\beta\sim \beta+4\pi$, $(\alpha,\beta)\sim (\alpha+2\pi,\beta+2\pi)$. The inverse relations are $\theta_1=\half\alpha+\half\beta$ and $\theta_2=\alpha$, which are well-defined, mod $2\pi$, under these identifications.) That is, $\theta_1$ and $\theta_2$ parametrize flat $U(2)$ connections with holonomy in the fundamental representation of the form $e^{i\oint A}=\twoMatrix{e^{i\theta_1}}{}{}{e^{i(\theta_2-\theta_1)}}$. For the maximal torus of the magnetic $U(2)$ group, we similarly have the coordinates $\theta_m=\theta_1'-2\theta_2'$, $\theta_1'$, which are well-defined up to $\theta_m\sim \theta_m+4\pi$ and $(\theta_1',\theta_m)\sim (\theta_1'+2\pi, \theta_m+2\pi)$. We henceforth adopt the coordinates $(\theta_2,\theta_e,\theta_1',\theta_m)$ and, as with $Z_{\gamma_2}$, focus on the submanifold of moduli space with $\theta_1'=\theta_2=0$ (mod $2\pi$). By using $(\theta_2,\theta_e)\sim (\theta_2+2\pi,\theta_e+2\pi)$ to set $\theta_2=0$, and its magnetic version to set $\theta_1'=0$, we are simply left with the identifications $\theta_e\sim \theta_e+4\pi$ and $\theta_m\sim \theta_m+4\pi$. The Weyl group acts on the moduli space of the 4d theory on a circle via $(a,\theta_m,\theta_e)\sim (-a,-\theta_m,-\theta_e)$.

This $4\pi$-periodicity is at the root of the $Z_2^4$ symmetry of our K3 surfaces, which manifests itself in this field theory example as a $Z_2^2$ symmetry. It is clear at the level of equations, since only the values of $\theta_e$ and $\theta_m$ mod $2\pi$ affect the integral equations determining $\X_e$ and $\X_m$. Physically, the reasons for this symmetry are the spontaneously broken $U(1)$ 1-form global symmetries associated to the centers of the electric and magnetic $U(2)$ gauge groups, which respectively multiply Wilson or 't Hooft lines by a phase. These manifest as symmetries under translating $\theta_1'$ or $\theta_2$ while fixing $\theta_e$ and $\theta_m$. Of particular interest are the $Z_2\subset U(1)$ subgroups generated by translating these angles by $2\pi$, as these act on the submanifold of moduli space with $\theta_1'=\theta_2=0$ mod $2\pi$ by translating $\theta_e$ or $\theta_m$ by $2\pi$. This accounts for the $Z_2^2$ symmetry of this submanifold.

At this point, it is worthwhile to compare with the $SU(2)$ gauge theory. Again, there exist charges $\gamma_e,\gamma_m$ such that $Z_m=\tau_F Z_e$ (where $\tau_F$ transforms as expected under S-duality), $\avg{\gamma_m,\gamma_e}=2$, and the BPS spectrum consists of vector multiplets with charges $p\gamma_m+q\gamma_e$ with $p,q$ coprime. A basis for the lattice of charges consistent with the Dirac quantization condition is given by $\gamma_m,\half\gamma_e$. So, in this theory the equations determining the geometry of the moduli space look identical to those of the $U(2)$ theory (if we ignore the center of mass moduli), but $\theta_m$ is well-defined mod $2\pi$. This corresponds to the fact that the magnetic group is now the centerless $PSU(2)\cong SO(3)$, and so there is no magnetic $Z_2$ 1-form symmetry. So, the (non-center-of-mass part of the) $U(2)$ moduli space is a double cover of the $SU(2)$ moduli space. Similar statements hold for the S-dual $PSU(2)$ gauge theory if we replace electric and magnetic throughout. From the point of view of these theories, the more complicated charge lattice of the $U(2)$ theory has effectively managed the inclusion of both $\half\gamma_e$ and $\half\gamma_m$ while preserving the Dirac quantization condition.

We are finally ready to compare with our $SU(2)$ $N_f=4$ results. We first note that the above discussion motivates defining $\theta_e^{\N=4}=2\theta_e^{N_f=4}$ and $\theta_m^{\N=4}=2\theta_m^{N_f=4}$. We similarly define $R^{\N=4}=2R^{N_f=4}$; then, the semi-flat limit of \eqref{eq:U2varpi} agrees with that of the $N_f=4$ theory. To avoid superscripts, we write $R=2\tilde R$, $\X_\gamma = \tilde \X_\gamma^2$, $\theta_m=2\tilde\theta_m$, and $\theta_e=2\tilde\theta_e$. Note that $\avg{\gamma_m,\gamma_e}=2$ in the $\N=4$ theory, whereas $\avg{\gamma_m,\gamma_e}=1$ in the $N_f=4$ theory. So, we introduce a new symplectic pairing $[\gamma_m,\gamma_e]=1$ on the sublattice generated by $\gamma_m$ and $\gamma_e$. When we write an expression in $N_f=4$ notation, charges will always be valued in this lattice. We observe that $\tilde\X_{\gamma+\gamma'}=(-1)^{[\gamma,\gamma']} \tilde\X_\gamma \tilde\X_{\gamma'}$ and $e^{i\tilde\theta_{\gamma+\gamma'}}=(-1)^{[\gamma,\gamma']}e^{i(\tilde\theta_\gamma+\tilde\theta_{\gamma'})}$ are consistent with $\X_{\gamma+\gamma'}=(-1)^{\avg{\gamma,\gamma'}} \X_\gamma \X_{\gamma'}$ and $e^{i\theta_{\gamma+\gamma'}}=(-1)^{\avg{\gamma,\gamma'}}e^{i(\theta_\gamma+\theta_{\gamma'})}$.

From the $\N=2$ point of view, an $\N=4$ vector multiplet consists of a vector multiplet $(\Omega=-2)$ with R-charge (i.e., charge for the Cartan of $SU(2)_R$) 0 and two half-hypermultiplets $(\Omega=1)$ with respective R-charges $\pm 1$. Introducing the real mass $\theta_R$ associated to $SU(2)_R$, the $\N=4$ integral equation is
\be
\Y_\gamma(\zeta) = \Y_\gamma^{\rm sf}(\zeta) - \frac{1}{4\pi i}\sum_{p,q}\avg{\gamma,(p,q)}\int_{\ell_{p,q}(a)} \frac{d\zeta'}{\zeta'} \frac{\zeta'+\zeta}{\zeta'-\zeta} \log \frac{(1-e^{i\theta_R}\X_{p,q}(\zeta'))(1-e^{-i\theta_R}\X_{p,q}(\zeta'))}{(1-\X_{p,q})^2} \ .
\ee
In $N_f=4$ notation, this is
\be \tilde\Y_\gamma(\zeta) = \tilde\Y_\gamma^{\rm sf}(\zeta) - \frac{1}{4\pi i}\sum_{p,q}[\gamma,(p,q)] \int_{\ell_{p,q}(a)} \frac{d\zeta'}{\zeta'} \frac{\zeta'+\zeta}{\zeta'-\zeta} \log \frac{(1-e^{i\theta_R}\tilde\X^2_{p,q}(\zeta'))(1-e^{-i\theta_R}\tilde\X^2_{p,q}(\zeta'))}{(1-\tilde\X^2_{p,q})^2} \ . \ee
Remarkably, this precisely coincides with the $SU(2)$ $N_f=4$ integral equation \eqref{eq:su2intEqn} when $\theta_1=\half\theta_R$, $\theta_2=0$, $\theta_3=\pi+\half\theta_R$, and $\theta_4=\pi$!

Although this proves that these moduli spaces are the same non-perturbatively, it is interesting to see how this equivalence works in our usual perturbation theory about the orbifold point. The contribution of a vector multiplet with gauge charge $\gamma=(p,q)$ to the approximation \eqref{eq:firstApprox} is
\begin{align}
\varpi^{\rm inst}_\gamma(\zeta) &= -\frac{i}{8\pi^2}d\Y^{\rm sf}_\gamma(\zeta)\wedge \brackets{\sum_{n>0} e^{in\theta_\gamma}2\parens{\cos(n\theta_R)-1} \right. \nonumber \\
&\left.\times\parens{-|Z_\gamma| K_1(2\pi R n |Z_\gamma|) \,d\log(Z_\gamma/\bar Z_\gamma) + K_0(2\pi R n |Z_\gamma|) \parens{\frac{1}{\zeta} dZ_\gamma - \zeta d\bar Z_\gamma} } } \ . \label{eq:vectorApprox}
\end{align}
In terms of the $(N_f=4)$-rescaled variables, this is
\begin{align}
\varpi^{\rm inst}_\gamma(\zeta) &= -\frac{i}{16\pi^2}d\tilde\Y^{\rm sf}_{2\gamma}(\zeta)\wedge \brackets{\sum_{n>0} e^{in\tilde\theta_{2\gamma}}2\parens{\cos(n\theta_R)-1} \right. \nonumber \\
&\left.\times\parens{-|Z_{2\gamma}| K_1(2\pi \tilde R n |Z_{2\gamma}|) \,d\log(Z_{2\gamma}/\bar Z_{2\gamma}) + K_0(2\pi \tilde R n |Z_{2\gamma}|) \parens{\frac{1}{\zeta} dZ_{2\gamma} - \zeta d\bar Z_{2\gamma}} } } \ .
\end{align}
We now expand to order $\theta_R^2$:
\begin{align}
\varpi^{\rm inst}_\gamma &\approx \frac{i\theta_R^2}{16\pi^2} \sum_{n>0} d\tilde\Y_{2n\gamma}^{\rm sf}\wedge e^{i\tilde\theta_{2n\gamma}} \nonumber \\
&\times\parens{-|Z_{2n\gamma}| K_1(2\pi \tilde R |Z_{2n\gamma}|) \,d\log(Z_{2n\gamma}/\bar Z_{2n\gamma}) + K_0(2\pi \tilde R |Z_{2n\gamma}|) \parens{\frac{1}{\zeta} dZ_{2n\gamma} - \zeta d\bar Z_{2n\gamma}} } \ .
\end{align}
As before, we can combine the sums over $(p,q)$ and $n$:
\begin{align}
\varpi^{\rm inst} &\approx \frac{i\theta_R^2}{16\pi^2} \sum_{\gamma\in 2\ZZ^2\backslash(0,0)} d\tilde\Y^{\rm sf}_\gamma\wedge e^{i(p'\tilde\theta_m+q'\tilde\theta_e)} \nonumber \\
&\times\parens{-|Z_\gamma|K_1(2\pi \tilde R|Z_\gamma|) d\log(Z_\gamma/\bar Z_\gamma) + K_0(2\pi \tilde R |Z_\gamma|)\parens{\frac{1}{\zeta} dZ_\gamma - \zeta d\bar Z_\gamma}} \ .
\end{align}
Here, we have written the components of $\gamma$ as $(p',q')=n(p,q)$. We next use
\be \frac{1}{4}\brackets{1+(-1)^{p'}+(-1)^{q'}+(-1)^{p'+q'}} = \piecewise{1}{2|p'\wedge 2|q'}{0}{\rm else} \ee
in order to extend the range of summation:
\begin{align}
\varpi^{\rm inst}&\approx \frac{i\theta_R^2}{64\pi^2} \sum_{\gamma\in \ZZ^2\backslash(0,0)} d\tilde \Y^{\rm sf}_{\gamma} \wedge e^{i(p'\tilde\theta_m+q'\tilde\theta_e)} \brackets{1+(-1)^{p'}+(-1)^{q'}+(-1)^{p'+q'}} \nonumber \\
&\times\parens{-|Z_{\gamma}| K_1(2\pi \tilde R|Z_{\gamma}|) d\log(Z_{\gamma}/\bar Z_{\gamma}) + K_0(2\pi \tilde R |Z_{\gamma}|) \parens{\frac{1}{\zeta} dZ_{\gamma} - \zeta d\bar Z_{\gamma}}} \ .
\end{align}
Comparing with \eqref{eq:combinedSum} and \eqref{eq:thetaFI} shows that this agrees with the $SU(2)$ $N_f=4$ theory if we set $\delta\theta_1=\delta\theta_3=\half\theta_R$ and $\delta\theta_2=\delta\theta_4=0$. Using \eqref{eq:thetaFIfinal}, this translates to $\xi_{00,\RR}^2=\xi_{10,\RR}^2=\xi_{01,\RR}^2=\xi_{11,\RR}^2=\frac{\theta_R^2}{64\pi^4 \tilde R^2}$. That this specialization of FI parameters leads to an effective change in the lattice in which $\gamma$ is valued is sensible from the point of view of Poisson resummation, since now we can resum over all $n\in\Lambda$, rather than over only those in the same equivalence class in $\Lambda/2\Lambda$.

We now return to our original interest, the $\N=(1,1)$ little string theory. We first discuss the global symmetries: there are 4, in addition to $SU(2)_R$. When we think of our little string theory as arising on two type IIB NS5-branes on $T^2$, these are the symmetries associated to conservation of momentum and winding on the 1-cycles of $T^2$. ($\Spin(4)=SU(2)_L\times SU(2)_R$ rotates the transverse $\RR^4$ to the NS5-branes.)

There is also a $U(1)^2$ 1-form global symmetry associated to D-string winding on either 1-cycle, but this is spontaneously broken by Wilson lines of the center of the NS5-brane $U(2)$ gauge group, and so we cannot grade by the charges associated to these symmetries.\footnote{Wound D-strings survive the $g_s\to 0$ limit, as their binding energy with the NS5-branes is nearly identical to their rest mass \cite{w:bound,s:hlst}.} Intuitively, this corresponds to the fact that D-strings can end on NS5-branes, and so there are not separate 1-form global symmetries associated to D-strings and Wilson lines. S-dually, this is clear from the D5-brane DBI action, since the Yang-Mills coupling $\tra(*F\wedge F)$, plus the fact that $F$ only shows up in the combination $F+B$, necessitates a $\tra(*F)\wedge B$ interaction, and $\tra *F$ is the conserved current for the symmetry under which Wilson lines are charged. Another way to see this is by T-dualizing the D5-branes to D3-brane probes of $T^2_B$ in type IIB string theory, or equivalently of $T^2_B\times T^2_F$ in F-theory. In this duality frame, these $U(1)^2$ symmetries simply correspond to translation along the 1-cycles of $T^2_B$, and they are certainly spontaneously broken by the vacuum expectation values of the D3-brane positions.

Having worked out the correspondence between the moduli spaces of the two 4d gauge theories, it is not hard to guess the BPS index of the $\N=(1,1)$ little string theory -- or at least that part which affects the metric at order $\xi^2$. For each relatively prime $p,q\in\ZZ$ and $\tilde n^1,\tilde n^2\in \ZZ$, we have a $1/2$-BPS vector multiplet with gauge charges $(p,q)$ and flavor charges (ignoring R-charge) $(p\tilde n^1,p\tilde n^2,q\tilde n^1,q\tilde n^2)$. In the NS5-brane frame, the first two flavor charges correspond to winding and the last two correspond to momentum.

Next, we account for the $Z_2^4$ symmetries of our K3 surfaces. In the 4d field theory case, these were associated to spontaneously broken $Z_2$ electric and magnetic 1-form global symmetries. Now, we simply have 3 such electric symmetries, since the little string theory is compactified on $T^3$, in addition to the magnetic 1-form global symmetry.

The parameters of the two theories are related as follows. First, note that there is an obvious correspondence between $\ZZ^4$ sublattices of the flavor lattices of the two theories. We denote the corresponding real mass parameters by $\theta_{\gamma^f}=2 \tilde \theta_{\gamma^f}$. The relationship between the other real mass parameters of the two theories is $\theta_{n_B,1}=\half\theta_R$, $\theta_{n_B,2}=0$, $\theta_{n_B,3}=\pi+\half\theta_R$, and $\theta_{n_B,4}=\pi$, for all $\tilde n^1$, $\tilde n^2$. The complex masses are easier to relate: the two theories have the same $\tau_F$ and $\tau_B$. With these identifications, as well as those noted above, such as $R=2\tilde R$, the canonical coordinates of the two theories again agree -- at least at order $\xi^2$.

\subsection{String webs and holomorphic curves with boundary} \label{sec:stringWebs}

\begin{figure}
\begin{center}
\begin{subfigure}[b]{.45\textwidth}
\begin{center}
\includegraphics[scale=0.2]{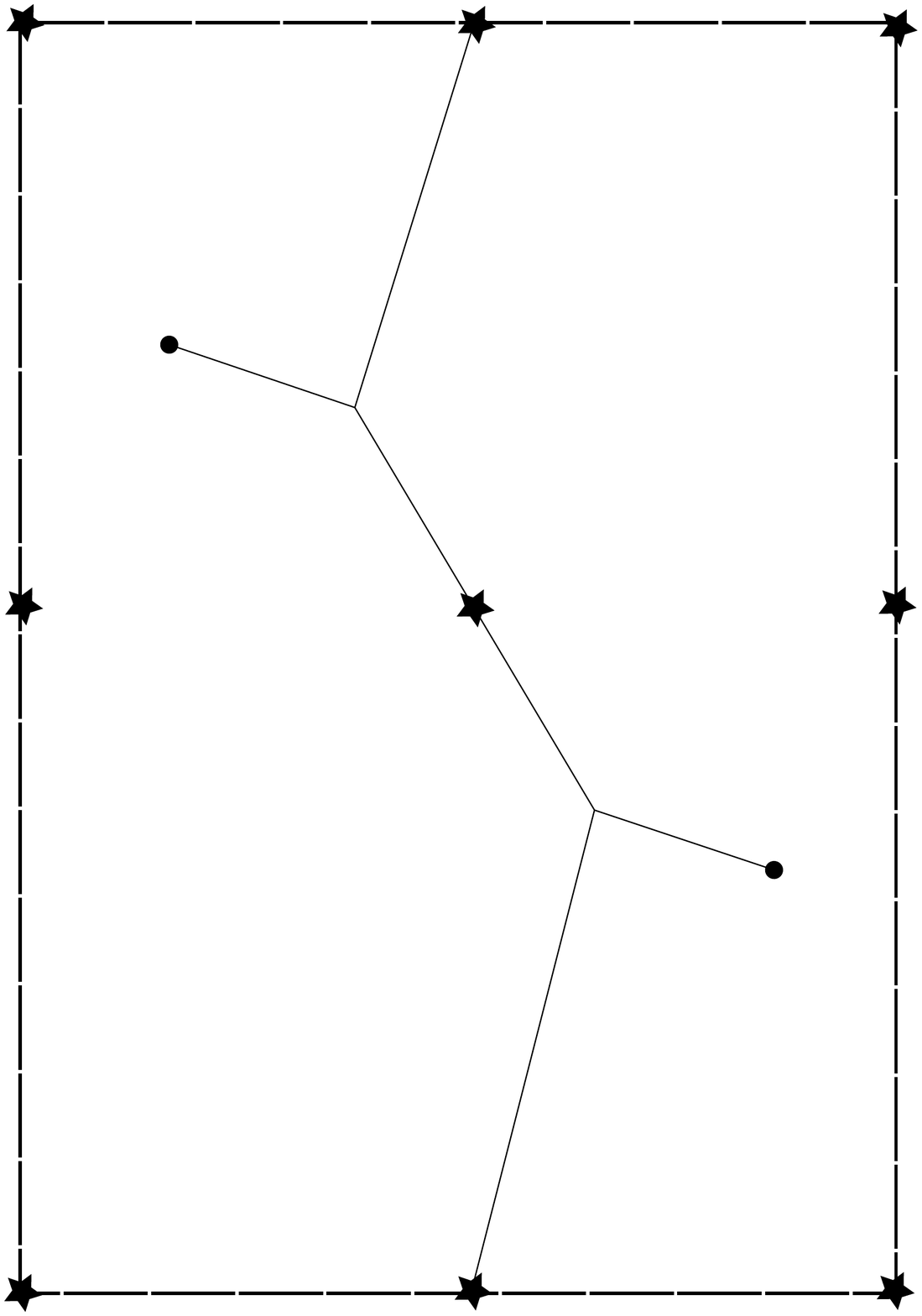}
\caption{}
\label{fig:webOrient}
\end{center}
\end{subfigure}
~
\begin{subfigure}[b]{.45\textwidth}
\begin{center}
\includegraphics[scale=0.2]{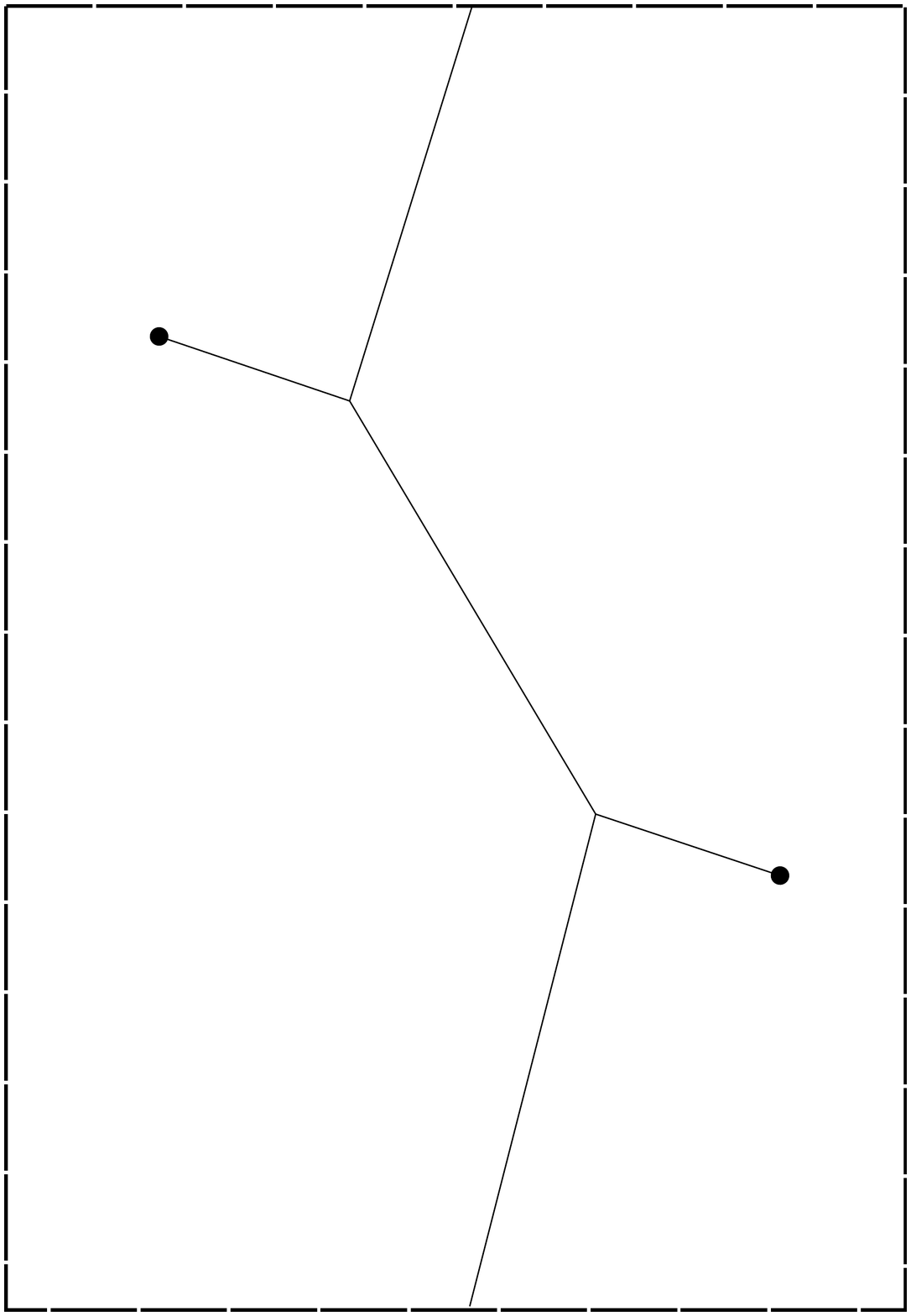}
\caption{}
\label{fig:web11}
\end{center}
\end{subfigure}
\caption{The same string web, regarded, respectively, as a web in the $\N=(1,0)$ or $\N=(1,1)$ litle string theory. In the first diagram, stars correspond to $Z_2$ fixed points of $T^2_B$, at which an O7-plane and 4 D7-branes reside. Note that only half of this diagram is physical -- the other half is the $Z_2$ image. In both diagrams, the circles correspond to D3-branes.}
\label{fig:webs}

\end{center}
\end{figure}

The BPS spectra we have studied in the last few sections have intuitive interpretations in terms of string webs, examples of which are provided in Figure \ref{fig:webs}. Physically, these arise naturally in the duality frame where, for the $\N=(1,0)$ (resp. $\N=(1,1)$) little string theory, one studies a D3-brane probing K3 (two D3-branes probing $T^4$) in F-theory. BPS states in this picture correspond to string webs on the base of the K3 surface ($T^4$) comprised of F- and D-strings which end on D3-brane(s) and 7-branes \cite{sen:network,bergman:FWebs,sethi:FWebs}. By T-dualizing a large transverse circle, one finds an equivalent picture in M-theory on K3 ($T^4$) with vanishingly small elliptic fibers, and with an M5-brane (two M5-branes) wrapping an elliptic fiber (two elliptic fibers). BPS states in this duality frame correspond to M2-branes wrapping special Lagrangian -- or, after hyper-K\"ahler rotation, holomorphic -- curves that terminate on the M5-brane(s) \cite{yi:surfaces,mikhailov:surface}. As explained in \cite{mz:k3}, this frame allows us to relate the Coulomb branch approach to the Strominger-Yau-Zaslow picture of mirror symmetry \cite{strominger:mirrorT}, as upon compactification on $S^1_R$ these M2-branes yield open string worldsheet instantons in a D4-brane worldvolume theory. This open string Gromov-Witten problem for K3, and its string web/tropical limit, has recently been studied in \cite{lin:thesis,lin:walls1,lin:walls2,lin:walls3,lin:walls4,lin:walls5}.

We now interpret our BPS spectra in this picture, starting with that of the $\N=(1,0)$ little string theory. The F-theory compactification of interest is an isotrivial $T^2_F$ fibration over $T^2_B/Z_2$. At each of the four $Z_2$ fixed points of $T^2_B$, there is an O7-plane and 4 D7-branes. The simplest type of string web consists of a string that connects the D3-brane to one of these 7-brane stacks. These are the same configurations that exist in the $SU(2)$ $N_f=4$ limit which obtains by zooming in on one of the fixed points. However, in contrast to the field theory string webs, the little string theory string webs can wind around $T^2_B$ before ending on a 7-brane stack. Since there are such stacks at each of the four $Z_2$ fixed points of $T^2_B$, the string webs can wind any half-integral number of times around each of the two cycles of $T^2_B$ before terminating. It is now clear that the physical interpretation of the $U(1)^4$ symmetries in this frame correspond to F1 and D1 winding around each of the two cycles of $T^2_B$. The remaining $\Spin(8)^4$ symmetry corresponds to the four 7-brane stacks; which $\Spin(8)$ a BPS state is charged under is determined by the stack on which the string ends. This picture explains why we found a tower of copies of the $SU(2)$ $N_f=4$ BPS spectrum. The surprising conclusion of our study of the BPS spectrum of the little string theory is that these simplest strings are the only string webs that affect the metric at order $\xi^2$ in this locus of the little string theory's parameter space where the F-theory description coincides with a $T^2/Z_2$ type IIB orientifold.

The string webs for the $\N=(1,1)$ theory have a nearly identical description. Now, we have two D3-branes probing $T^2_B$ in type IIB string theory. The Coulomb modulus $a$, defined up to $a\sim a+1\sim a+\tau_B\sim -a$, specifies their relative separation. The simplest string webs consist of a string that connects the two D3-branes. The BPS states associated to such a string also exist in the 4d $\N=4$ $U(2)$ field theory that obtains at low energies when $a$ is small. However, now the string can wind. Again, the $U(1)^4$ global symmetry is associated to F1 and D1 winding around each of the two cycles of $T^2_B$. Note that a winding number $n_B=\tilde n_1+\tau_B \tilde n_2$ means that a string starts at $-a$ and ends at $n_B+a$, so on the covering space $\CC$ of $T^2_B$ it passes through $\frac{n_B}{2}$ after undergoing a displacement of $\frac{n_B}{2}+a$. Indeed, this is the only such `half-integral' point that the string passes through on $\CC$, since repeating the displacement that takes one from $-a$ to a point of the form $\frac{n_B}{2}$ takes one to a point of the form $n_B+a$. So, as in the $\N=(1,0)$ theory, winding numbers are naturally specified by half-integral points of $\CC$. This also explains the need for the rescaling $R=2\tilde R$ of \S\ref{sec:11}: the strings in the $\N=(1,1)$ theory are twice as long as the corresponding strings of the $\N=(1,0)$ theory, since when they reach this half-integral point they are only halfway to the other D3-brane.\footnote{Really, we should have rescaled the central charges, e.g. $Z_e=2a$, but we found rescaling $R$ to be less notationally cumbersome.} Lastly, we note that when $a=0$, the D3-branes are not playing much of a role and this open string counting problem can likely be related to the analogous closed string problem studied in \cite{oberdieck:tori}.

Before proceeding, we note that it is also natural to study string webs on the other elliptically fibered manifolds with constant $\tau_F$ studied in \cite{dasgupta:constant}. As stressed in \cite{sethi:FWebs}, the flat metric on the Coulomb branch simplifies the determination of the moduli spaces of string webs, which is a necessary step in determining the multiplets to which the webs correspond. A particularly interesting special case involves manifolds with 12 $II$ fibers, as these yield smooth K3 surfaces even without any real masses turned on. It would be interesting to know if existing results on counts of geodesics on flat surfaces, which as explained in \cite{mz:asymp} characterize the large charge asymptotics of many 4d $\N=2$ BPS spectra, can be generalized to these string webs.

\subsection{Open topological string theory}

Following \cite{vafa:topWalls,vafa:RTwist,lin:thesis,lin:walls4}, we now explain how the picture of the previous section can be related to open string worldsheet instantons in a Calabi-Yau threefold compactification which contribute to a D-brane superpotential. The idea is an open string analogue of \cite{maulik:NL}.

We begin with the picture from the last section of BPS states (in the $\N=(1,0)$ theory, for concreteness) as corresponding to special Lagrangian curves in K3 ending on an elliptic fiber. We note that the phase $\arg Z_\gamma$ of a BPS state corresponds to the phase of the corresponding special Lagrangian curve, i.e. to the relative phase between the restriction of the holomorphic 2-form to the curve and the volume form of the curve. There is an arbitrariness in the choice of the phase of the holomorphic 2-form, but the relative phase of two BPS states is meaningful. We henceforth make a hyper-K\"ahler rotation so that the fibers of the K3 surface, which used to be holomorphic, are now special Lagrangian. There is an $S^1$ freedom in our choice of this hyper-K\"ahler rotation. Depending on this choice, only those BPS states with a particular phase will correspond to curves which are holomorphic in the chosen complex structure (and whose volume form is given by the K\"ahler form, as opposed to its negative). Conversely, every BPS state corresponds to a curve which is holomorphic and calibrated by the K\"ahler form in one of these complex structures.

We now consider a (special Lagrangian fibered K3)-fibered Calabi-Yau threefold with a $\PP^1$ base. We compactify type IIB string theory on this threefold. Suppose that there is a special Lagrangian 3-cycle which is fibered over a loop in the base, and whose fibers consist of the special Lagrangian fibers of the corresponding K3 fibers. We wrap a D3-brane on this cycle. The important observation now is that among the holomorphic curves which correspond to worldsheet instantons in the D3-brane worldvolume will be holomorphic curves which reside entirely in a single fiber of the threefold. However, we only obtain those holomorphic curves of a K3 fiber which are holomorphic in the complex structure induced from the ambient threefold. By choosing the threefold and special Lagrangian 3-cycle wisely, one may be able to extract a solution to the K3 counting problem of interest by studying the corresponding problem for this threefold.

An example was studied in \cite{vafa:topWalls,vafa:RTwist}: the threefold was the twistor space of the manifold (here, K3) of interest and the 3-cycle was the fibration of the elliptic fibers over the equator of the twistor sphere. This 3-cycle is not actually Lagrangian, but it was argued that the A-model with this brane could still be made sense of in a certain limit. This setup has the benefit that each of the holomorphic curves of interest appears as a holomorphic curve in exactly one fiber of this 3-cycle, namely the fiber corresponding to the phase of the BPS state. That is, these holomorphic curves of K3 are splayed out around the equator of the twistor sphere.

It would be interesting to study other examples. In particular, torus orbifolds may yield configurations which are amenable to perturbative methods of the ilk studied in \S\ref{sec:hk}. For, following \cite{douglas:superpot,sk:openSuper,sk:openSuper2}, we can employ mirror symmetry to find a theory whose superpotential is unaffected by worldsheet instanton corrections. Just as a non-renormalization theorem enabled the determination of the metric on moduli space in \S\ref{sec:hk}, an analogous non-renormalization theorem will likely now enable the determination of the superpotential.

Having explained this relationship between metrics on moduli spaces of theories with 8 supercharges and superpotentials of theories with 4 supercharges, let us comment on some reasons that this might yield useful new means of solving the BPS state counting problem of interest. First, the superpotential is strongly constrained by the requirement that it be a (smooth) holomorphic section of a line bundle. (Indeed, the fact that the superpotential is smooth, despite the fact that the instantons which contribute to it undergo wall crossing, was the basis of the argument in \cite{vafa:topWalls} for the wall crossing formula.) By constraining exactly what sort of special function (e.g., automorphic form) one is dealing with, this could lead to powerful constraints on the BPS spectra of 4d $\N=2$ theories. Of course, this same reasoning applies to the holomorphic symplectic form $\varpi(\zeta)$ (which was the basis of the argument in \cite{GMN:walls} for the wall crossing formula), as the latter is a holomorphic section of a line bundle on the twistor space, but the more constrained form of instanton contributions to the superpotential will likely prove convenient. In addition, in many examples D-brane superpotentials have been related to periods of non-compact Calabi-Yau fourfolds (see \cite{aganagic:openClosed} and references therein). This would yield a fascinating means of studying 4d $\N=2$ theories.

\subsection{Geometric engineering}

The last two subsections described problems which were closely related. Now, we turn to something completely different. For a more detailed discussion of geometric engineering of little string theories using F-theory, we refer the reader to \cite{vafa:fClass} and references therein. In particular, the geometry associated to the $\N=(1,0)$ little string theory is discussed in \S9.1 and Appendix F.

We return to the $SO(32)$ heterotic and type IIB NS5-brane descriptions of our little string theories. We compactify two transverse circles, which does not affect the little string theory \cite{s:ns5}. Next, we S-dualize and then T-dualize these circles. The $\N=(1,0)$ $SO(32)$ heterotic little string theory is now related to a D7-brane wrapping $T^2/Z_2$ in a type IIB orientifold. The D9-branes and O9-plane of type I string theory are mapped, by these T-dualities, to 7-branes transverse to $T^2/Z_2$. This configuration can be lifted to F-theory on an elliptic fibration over a blow up of $\PP^1\times \CC$, where the $\PP^1$ is obtained by smoothing out $T^2/Z_2$. The little string is now a D3-brane wrapping this $\PP^1$; the latter has self-intersection 0, so its volume cannot be made to vanish and provides a string scale.

Similarly, starting with two type IIB NS5-branes and repeating these steps yields type IIB with two D7-branes wrapping $T^2$. This lifts to F-theory on the product of the total space of an $I_2$ singular fiber\footnote{That is, an elliptic fibration over a disc with an $I_2$ singular fiber at its center. Of course, this can be embedded into geodesically complete spaces which are smooth away from the $I_2$ singular fiber, such as the Coulomb branch of 4d $SU(2)$ $N_f=4$ on a circle with appropriately tuned complex masses, if one wishes. (The smooth non-compact geometry away from the singularity does not affect the limit in which gravity decouples.) This manifold can be studied by probing two D7-branes with a D3-brane, whose worldvolume is described by a 4d $U(1)$ gauge theory with two hypermultiplets. The natural hyper-K\"ahler metric on this manifold can easily be determined using the approaches of \cite{seiberg:mirrorT,GMN:walls}.} -- also known as a $\hat A_1$ singularity because the 2-cycles of this geometry intersect according to the $\hat A_1$ Cartan matrix, where the cycle with vanishing self-intersection corresponds to the elliptic fiber -- with $T^2$. The F-theory fibers are provided by the fibers of the $\hat A_1$ singularity. The little string is a D3-brane wrapping $T^2$.\footnote{There is another geometric construction of the $\N=(1,1)$ little string theory in terms of type IIA string theory on an $A_1$ singularity. In this construction, the fundamental type IIA string is the little string. However, in order to make contact with Donaldson-Thomas theory, we want the BPS states of the little string theory on $T^2$ to be associated to D-branes, and in this construction the flavor charges correspond to momentum and winding.}

Compactification of either of these little string theories on $T^2$ simply involves replacing F-theory with type IIA string theory (since F-theory on $S^1$ is M-theory, and M-theory on $S^1$ is type IIA string theory). So, we are interested in the BPS spectrum associated to D-branes wrapping this threefold -- i.e., `generalized Donaldson-Thomas theory.'\footnote{As described in \cite{denef:walls}, Donaldson-Thomas theory proper is concerned with bound states of D0- and D2-branes with a single D6-brane in a particular limit of moduli space. We are interested in considering both other charge configurations and other points in moduli space.\label{ft:dt}} Mathematically, these are stable objects in the derived category of coherent sheaves.

Duality chasing allows us to relate the charges in these type IIA pictures to those in the original NS5-brane pictures. For simplicity, we focus on the $\N=(1,1)$ little string theory. Momentum and winding on one circle map, respectively, to D2 charge on the elliptic fiber or $T^2$, while those charges on the other circle map to D0 or D4(fiber$\times T^2$) charge. Electric charge (associated in the NS5-brane picture to D-strings stretched between the NS5-branes) maps to D2 charge on the $\PP^1$ in the base of $\hat A^1$, while magnetic charge (associated to D3-branes wrapping the $T^2$ upon which the NS5-branes are compactified and stretched between the NS5-branes) maps to D4 charge on the product of this $\PP^1$ with $T^2$. Lastly, the little string is now an NS5-brane wrapping the product of an elliptic fiber with $T^2$. (If we go backwards, from type IIA to F-theory, but use $T^2$ as the F-theory fiber, then we find the $\N=(2,0)$ little string theory on $T^2$. This choice of two different elliptic fibrations, once we compactify on a circle to go from F-theory to M-theory, is a manifestation of the T-dual relationship of these little string theories.)

The 4d-5d lift \cite{4d5d} / GW-DT correspondence \cite{vafa:GWDT,maulik:GWDT} relates counts of certain BPS states in these compactifications to BPS state counts of the little string theories on $S^1$ -- i.e., of M-theory on these geometries -- which are computable via a variety of methods. But, as explained in footnote \ref{ft:dt}, this does not suffice to solve our counting problem. Nevertheless, in the case of the $\N=(1,1)$ little string theory one can likely manipulate these results in order to obtain much, and possibly all, of the BPS spectrum. For, one has the benefit of a large U-duality group and the existence of the indices $B_4$ and $B_6$. Concretely, one may consider type IIA on $K3\times T^2$, where the K3 surface is elliptically fibered and has an $I_2$ singular fiber; the BPS spectrum of interest is then a part of the BPS spectrum of this string compactification. And, the indices $B_4$ and $B_6$ are known for all charges, everywhere in moduli space \cite{DVV,Sen:2007vb,sen:2center,Cheng:2007ch,banerjee2008partition,sen:macroN4,Dabholkar:2012zz}. (This last statement is trivial for $B_4$, as it is constant in moduli space.)

One drawback of this approach is that the R-symmetry does not exist in this compactification, and so one cannot flavor by it. It would therefore be quite interesting to generalize the computations of \cite{maulik:GW,maulik:qCoh,maulik:DT} to study the Donaldson-Thomas theory of $\hat A_1\times T^2$ and then to use U-duality and wall crossing formulae to produce the BPS spectrum of the $\N=(1,1)$ little string theory. The results of these references are quite suggestive of an interesting algebraic characterization of the wall crossing formula along the lines of \cite{Cheng:2008fc,nagao:weylWall}.

Lastly, we comment on the possible applicability of a common approach to the study of BPS B-branes on Calabi-Yau threefolds: quiver quantum mechanics \cite{fiol:fractBrane,denef:hall,gukov:refineMotive2,vafa:quiver1,vafa:quivers,Chuang:2013wt}. The appropriate quivers are known for toric threefolds \cite{hanany:toric1,hanany:toric2,vafa:dimer}. Strictly speaking, an $\hat A_1\times T^2$ singularity is not toric. However, by turning on complex mass parameters associated to the R-symmetry we obtain a threefold \cite{vafa:6dgenera} which is toric in a generalized sense discussed in \cite{vafa:vertex}: the web diagram characterizing the manifold is doubly periodic. In \cite{vafa:6dgenera}, it was shown that the mirror of this manifold could be determined using the usual methods for toric threefolds, as long as one accounted for this periodicity by including image branes appropriately. It is therefore natural to hope that one may analogously find a quiver quantum mechanical characterization of the BPS spectrum of this threefold, where the quiver is constructed via the orbifold approach of \cite{wati:dBraneT} from a gauge theory with an infinite-dimensional gauge group.

\subsection{Other problems}

\begin{itemize}

\item \textbf{Elliptic networks}

The mirror threefold mentioned at the end of the last section suggests another approach to determining the BPS spectrum of the compactified $\N=(1,1)$ little string theory. It is characterized in terms of a genus 3 Riemann surface embedded in $T^4$, and type IIB string theory on this mirror threefold is dual to M-theory on $T^4$ with an M5-brane wrapping this Riemann surface \cite{ganor:spectral,vafa:6dgenera,hollowood:6d} via the usual `class S' relationship of \cite{vafa:geodesics}.\footnote{One may also arrive at this curve by adding a D6-brane to the Higgs branch picture appropriate for the $\N=(1,1)$ little string theory -- that is, by considering two D2-branes probing and one D6-brane wrapping $T^4$. As mentioned above, the D6-brane allows for the incorporation, in the Higgs branch picture, of the parameters that resolve the Higgs branch from $\Sym^2(T^4)$ to $\Hilb^2(T^4)$. T-duality relates this configuration to a D4-brane wrapping the genus 3 Riemann surface, which lifts to an M5-brane \cite{ganor:noncomm2}. This picture also suggests that if one is interested in the BPS spectrum of the theory without any complex masses associated to the R-symmetry turned on, then a simpler curve, obtained via the same reasoning without the presence of the D6-brane, should suffice. Indeed, this curve -- comprised of two disconnected tori -- is the boundary of holomorphic cylinders discussed in \S\ref{sec:stringWebs}.} It is therefore natural to suspect that a generalization of the `spectral networks' of \cite{GMN:classS,GMN:networks} exists that will allow for the study of BPS states of compactified little string theories. The first step of this generalization has already been taken: in \cite{walcher:exp,longhi:exp1,longhi:exp2}, the study of `exponential networks' was initiated in order to study the BPS spectra of 5d theories on a circle. In analogy, we suggest that the appropriate generalization to 6d theories (assuming it exists) be termed `elliptic networks.'

\item $L^2$ \textbf{cohomology of soliton moduli spaces}

As we mentioned in \S\ref{sec:11}, the $\N=(1,1)$ little string theory has a weak coupling limit. Since $B_4$ does not wall cross, our results immediately yield predictions for its value at weak coupling. Via the wall crossing formula for $B_6$, one can also relate its value at weak coupling to the BPS spectrum in the locus in moduli space of interest. In this weak coupling limit, many BPS states have a description in terms of $L^2$ harmonic forms on soliton moduli spaces. This description of BPS states is well known from the work of \cite{sen:S} on 4d $\N=4$ gauge theories; see also \cite{moore:semiclassical,moore:semiclassical2} and references therein.

\item \textbf{Holography}

The little string theories discussed in this paper have holographic descriptions \cite{s:ns5holo,kapustin:hlst,narain:hlst}. The BPS spectra of compactified little string theories were recently studied, using these descriptions, in \cite{Harvey:2013mda,Harvey:2014cva,harvey:lstThermo}. Introducing D-branes in the $\N=(1,1)$ theory should be achievable by following \cite{giveon:LSTbrane}.

\item \textbf{Deconstruction}

The $\N=(1,1)$ little string theory on $T^2$ has a number of descriptions using large $N$ limits of 4d $\N=1$ gauge theories \cite{ah:deconstruct20,awa:moose,dorey:deconstruct1,dorey:deconstruct2}. Comments of \cite{ah:deconstruct20} on the appearance of BPS states in deconstruction imply that our results constrain the non-BPS spectra of 4d $\N=1$ gauge theories!

\item \textbf{DLCQ}

Matrix theory descriptions of little string theories were investigated in \cite{sk:decoupling,w:higgs,lowe:dlcq,kachru:dlcq,aharony:LSTmatrix}. 

\end{itemize}

\section{Return of the BPS states} \label{sec:moreStates}

Sections \ref{sec:hk} and \ref{sec:coulomb} wove an enticing story that might lead one to believe that we have found the complete BPS indices of the $\N=(1,0)$ and $\N=(1,1)$ little string theories. However, we will now explain from a few different points of view that this cannot be the case. Conversely, we will also show that the approach of this paper is quite likely to generalize to make feasible the determination of the entire spectra.

The loophole in our determination of the BPS spectra is as follows: perhaps there are BPS states which only make their influence felt in the metric at order $\xi^4$. This might seem like a paranoid concern, since after all why would BPS states not contribute as soon as they are able? However, we will now explain that in the $\N=(1,1)$ theory $1/4$-BPS states are actually not able to contribute at order $\xi^2$. It would be desirable to understand the corresponding physics at work in the $\N=(1,0)$ theory.

As explained in \cite{sen:N4refine}, the contribution of the $1/4$-BPS multiplets of interest to the index $B_2(z)$ takes the form $f(z)\cdot (z-2+z^{-1})^2$. By summing up the contributions to the integral equation \eqref{eq:intEqn} from a set of states associated to the universal factor $(z-2+z^{-1})^2$, one finds that the inside of the logarithm in the integral takes the form
\be \frac{(1-e^{-2i\theta_R}\X_\gamma)(1-\X_\gamma)^6(1-e^{2i\theta_R}\X_\gamma)}{(1-e^{-i\theta_R}\X_\gamma)^4(1-e^{i\theta_R}\X_\gamma)^4} = 1 - \frac{\X_\gamma(1+4\X_\gamma+\X_\gamma^2)\theta_R^4}{(1-\X_\gamma)^4} + \Oo(\theta_R^6) \ . \ee
So, one has no hope of detecting $1/4$-BPS states by working only at order $\theta_R^2$.

Conversely, this provides reason for optimism: there is a good reason that one should need to proceed to order $\theta_R^4$, but no similar justification for needing to continue further. So, while it is logically possible that one requires even higher order results on the Higgs branch side in order to extract the full BPS spectrum, it is sensible to expect that one need only iterate the Higgs branch formalism once more.

One can be confident that one has the full spectrum if it satisfies the wall crossing formula. Indeed, this is a conclusive way to see that our spectrum cannot be the final answer. For instance, consider the $\N=(1,1)$ variant of Figure \ref{fig:webs}. If one brings the two D3-branes toward the center of the diagram, one eventually reaches a wall of marginal stability where a D3-brane sits on top of a 3-string junction. This makes it clear that the $1/4$-BPS state pictured in this figure can be thought of as a bound state of two $1/2$-BPS states. Since this web is not rigid (the string leaving one D3-brane can be shortened as we lengthen the string attaching to the other D3-brane), one might hope that a fermionic superpartner of the bosonic zero mode causes this web to not contribute to the index $B_2(z)$. However, by applying the wall crossing formula \cite{KS:walls} to the two $1/2$-BPS constituents one finds that these states must wall cross to produce a bound state which contributes to the index. It would be interesting to reproduce this result by studying the quantum mechanics on the moduli space of the string web, including the fermi zero modes described in \cite{kol:dyon}.

To see the necessity of $1/4$-BPS states from the point of view of the wall crossing formula, we suppose that the web in Figure \ref{fig:webs} is associated to a state whose charge $\gamma_1$ has $(p,q)=(1,0)$ and winding charges $(\tilde n^1,\tilde n^2)=(0,1)$, and that $\tau_F=i$. The constituent $1/2$-BPS states then respectively have charges $\gamma_2$ and $\gamma_3$ with $(p,q,\tilde n^1,\tilde n^2)=(0,1,0,1)$ and $(1,-1,0,0)$. Using notation explained in \cite{mz:k3}, and denoting the basic unit of R-charge by $\gamma_R$, we define the natural symplectomorphisms
\be \tilde \K_{\gamma_i} = \K_{\gamma_i-\gamma_R} \K_{\gamma_i}^{-2} \K_{\gamma_i+\gamma_R} \ , \quad i=2,3 \ , \ee
which are associated to $1/2$-BPS vector multiplets. We then have not
\be \tilde\K_{\gamma_2}\tilde \K_{\gamma_3} = \tilde\K_{\gamma_3}\tilde \K_{\gamma_2} \ , \nonumber \ee
but rather
\be \tilde\K_{\gamma_2}\tilde K_{\gamma_1}^n \tilde \K_{\gamma_3} ``=" \tilde\K_{\gamma_3}\tilde\K_{\gamma_1}^{n-2} \tilde \K_{\gamma_2} \ , \label{eq:wallCross} \ee
where
\be \tilde \K_{\gamma_1} = \K_{\gamma_1-2\gamma_R} \K_{\gamma_1-\gamma_R}^{-4} \K_{\gamma_1}^6 \K_{\gamma_1+\gamma_R}^{-4} \K_{\gamma_1+2\gamma_R}  \ , \ee
and where $``="$ refers to equality in a truncated algebra, as discussed in \cite{KS:walls,GMN:walls,GMN:classS}. Here, $n$ is an arbitrary integer. Specifically, the actions of the two sides on $\X_{\gamma_2}$ and $\X_{\gamma_3}$ take the form
\be \K \X_{\gamma_i} = S_i \X_{\gamma_i} \ , \ee
where $S_i$ are functions of $\X_{\gamma_i}$, $i=2,3$, and these functions agree at first order in each of the $\X_{\gamma_i}$ (including terms of order $\X_{\gamma_2}\X_{\gamma_3}$). This suffices to prove that these $1/2$-BPS states form a bound state with $B_2(z)=-2(z-2+z^{-1})^2$. Indeed, this is a straightforward modification of the proof in \cite{GMN:walls} of the primitive wall crossing formula \cite{denef:walls}, which for $B_2(z)$ implies \cite{sen:N4refine} that there exists a bound state with\footnote{This is normally written with an extra minus sign and an absolute value sign surrounding the symplectic pairing, corresponding to the fact that $\abs{\avg{\gamma_2,\gamma_3}}$ is the dimension of the $SU(2)$ little group representation associated to angular momentum in the electromagnetic field in the bound state. However, the side of the wall with the bound state is characterized by $\avg{\gamma_2,\gamma_3} \Imag(Z_{\gamma_2}\bar Z_{\gamma_3}) > 0$ \cite{denef:attractor}, and so these different versions of the formula agree, since $\Imag(Z_{\gamma_2} \bar Z_{\gamma_3})<0$ on this side of the wall.}
\be \Delta B_2(z; \gamma_1) = (-1)^{\avg{\gamma_2,\gamma_3}} \avg{\gamma_2,\gamma_3} B_2(z; \gamma_2) B_2(z; \gamma_3) = -2 (z-2+z^{-1})^2 \ . \ee
Dividing both sides by $(z-2+z^{-1})^2$ and taking the limit $z\to 1$ yields
\be \Delta B_6(\gamma_1) = (-1)^{\avg{\gamma_2,\gamma_3}} \avg{\gamma_2,\gamma_3} B_4(\gamma_2) B_4(\gamma_3) = -2 \ , \ee
which is the usual primitive wall crossing formula for $B_6$ \cite{Sen:2007vb,sen:2center,Cheng:2007ch,banerjee2008partition,sen:macroN4}.

The fact that this truncation is required for equality to hold in \eqref{eq:wallCross} indicates that other BPS states with the same phase are present, at least on one side of this wall of marginal stability.

\section{Conclusion} \label{sec:conclusion}

In this paper, we have provided an explicit construction of smooth Ricci-flat metrics on K3 surfaces and related it to the construction of \cite{mz:k3} via Poisson resummation and 3d mirror symmetry. In doing so, we were able to extract part of the BPS spectrum of a compactified little string theory. In \S\ref{sec:counting}, we suggested a number of other approaches for studying this BPS spectrum, while in \S\ref{sec:moreStates} we explained that by repeating our procedure at fourth order in the FI parameters of the Higgs branch formalism it is quite likely that we will be able to extract the full spectrum. We conclude by mentioning a number of other problems that are naturally suggested by this work, some of which we hope to address in \cite{mz:K3HK2,mz:K3HK3} and other future work:
\begin{itemize}
\item The most immediate generalization of this work is to study the analogous hyper-K\"ahler quotient constructions of K3 surfaces near the other torus orbifold loci. In the cases where these loci admit an elliptically fibered sublocus \cite{dasgupta:constant}, we expect to be able to Poisson resum the K\"ahler forms and extract BPS spectra of the heterotic little string theory studied in this paper at other points in its parameter space. This should yield the BPS spectra of the Minahan-Nemeschansky SCFTs \cite{minahan:E6,minahan:En}.
\item More generally, (if possible) we would like to give hyper-K\"ahler quotient constructions of all known compact hyper-K\"ahler manifolds, and potentially some new ones. We can obtain $\Hilb^N(T^4)$ and $\Hilb^N(K3)$ by adding a single D6-brane, using ideas of \cite{raamsdonk:d2d6} (see also \cite{kapustin:impurity}). This essentially (up to the difference between $\Hilb^N(T^4)$ and generalized Kummer varieties discussed in \S\ref{sec:11}) accounts for almost all known compact hyper-K\"ahler manifolds. Other generalizations of our hyper-K\"ahler quotient may yield other smooth hyper-K\"ahler manifolds or BPS spectra of interesting theories. For instance, in addition to arbitrary numbers of D2- and D6-branes one may add O2- and O6-planes, B-field \cite{douglas:noncomm,douglas:nonComm2,cheung:d0nc}, and fractional branes \cite{polchinski:k3Orient,douglas:fract}.
\item We would like to study an analogue of our perturbation theory in \S\ref{sec:hk} that is valid near the fixed points of the $Z_2$ action on $T^4$. Rather than expanding about a flat metric, this expansion should be about the Eguchi-Hanson metric. This should yield a linear algebraic solution to a gluing problem -- i.e., it should specify how to correct the Eguchi-Hanson metric in order to glue it into a K3 surface.
\item We studied two constructions of K3 metrics in this paper, each of which involves an iterative procedure. However, we did not directly relate these two procedures -- rather, they are related via the holomorphic symplectic form. Thanks to the non-uniqueness of canonical coordinates, this means that our Higgs branch procedure does not immediately yield a linear algebraic solution to the integral equation. It would therefore be of interest to identify the canonical coordinates of \cite{GMN:walls} in the language of \S\ref{sec:hk} and directly relate the two iterative procedures.
\item Now that we have smooth K3 metrics, it might be enjoyable to explicitly see theorems concerning them in action, such as those of \cite{yau:geodesic,bourguignon:geo2} concerning geodesics or those of \cite{wolfson:areaK3} concerning minimal surfaces.
\item Another application is to the study of higher-dimensional Calabi-Yau metrics:\footnote{We note that a direct generalization of our Higgs branch approach does not work \cite{douglas:orbMetric}, since the non-renormalization theorem that we relied on to protect the metric does not hold with less supersymmetry. The Coulomb branch approach also crucially relies on hyper-K\"ahler geometry.} by analogy to the SYZ semi-flat limit, where one has a torus fibration with small fibers, and the induced metric on the fibers is the usual flat metric, it would be interesting to study a `semi-K3' limit of K3-fibered manifolds with small Ricci-flat fibers. This should allow one to study analytic approximations to K3-fibered Calabi-Yau metrics. Interesting work along these lines was recently pursued in \cite{Li:semiK3}. This semi-K3 limit is amenable to the adiabatic argument, so one might try to exploit fiberwise dualities (such as mirror symmetry and heterotic-type II duality) to extract interesting implications of our metrics.
\item Around \eqref{eq:N1fourier}-\eqref{eq:N4fourier}, we noted the existence of additional equivalent expressions for the K3 metrics (at least at order $\xi^2$) and suggested that they might yield interesting expansions about non-semi-flat limits in K3 moduli space. It is clearly of interest to understand these expansions -- i.e., what the formulae for the metric look like and what instanton corrections they encode. A natural guess is that they are useful when a K3 surface is roughly of the form $(T^d\times T^{4-d})/Z_2$ with $T^d$ much smaller than $T^{4-d}$, where $d$ is the number of components of $n$ over which we Poisson resum. (When $d=4$, perhaps this means that the blown-up 2-cycles are comparable in size to the rest of the manifold?) This is motivated by the generalization of the SYZ picture to $d$ T-dualities. This may explain the similarities between \eqref{eq:N2fourier} and \eqref{eq:N4fourier} and between \eqref{eq:N1fourier} and \eqref{eq:N3fourier}. For, when $d$ is odd the $Z_2$ orbifold in the T-dual frame is not merely by inversion, but instead by its combination with $(-1)^{F_L}$ (see, e.g., \cite{dabholkar:orientifolds}).
\item There may be another natural approach to obtaining K3 surfaces from infinite-dimensional hyper-K\"ahler quotients. Namely, it is well-known that moduli spaces of Einstein-Hermitian connections on K3 (and hyper-K\"ahler manifolds more generally) are themselves hyper-K\"ahler, since the defining conditions of such a connection yield natural moment maps for a hyper-K\"ahler quotient. Of course, these moment maps require a K3 metric as input, so this procedure may not be so useful, but it is nevertheless of interest to know if such a moduli space is ever a K3 surface. A natural approach to demonstrating this is to study a sheaf moduli space which is known to be a K3 surface and show that every point in the moduli space corresponds to a stable holomorphic vector bundle. (Physically, this roughly means that there are no small instanton singularities.)
\item As we noted at the end of \S\ref{sec:hk}, our results also yield expressions for metrics on non-compact gravitational instantons. It may be fruitful to compare them with the ALF results in \cite{cherkis:dkGravInst,cherkis:singMonopolesDk,cherkis:dkInst} and the ALG results in \cite{rafeLaura:Nf4,kapustin:singularPeriodicMono}.
\item At multiple points in this paper, we have noted severe cancellations of instanton effects. It may be interesting to study their origins (e.g., fermi zero modes) from the geometric points of view described in \S\ref{sec:counting}, as well as from gauge theory limits.
\item Once one has the full BPS spectra of the little string theory on $T^2$ at the points in parameter space with $SO(8)^4$ symmetry that we have focused on, it may be enjoyable to utilize the wall crossing formula in order to determine the BPS spectra at other points in parameter space and compare the resulting metrics of the Coulomb branch formalism with the predictions of our Higgs branch formulae.
\end{itemize}

\newpage
\section*{Acknowledgements}
We thank M. Aganagic, J. Athreya, N. Benjamin, C. Doran, L. Fredrickson, P. Jefferson, Y.-S. Lin, R. Mazzeo, G. Moore, D. Morrison, A. Neitzke, G. Oberdieck, M. Ro\v{c}ek, J. Sawon, W. Taylor IV, and S.-T. Yau for enjoyable conversations on related subjects. The research of S.K. was supported in part by a Simons Investigator Award and the National Science Foundation under grant number PHY-1720397.  The research of A.T. was supported by the National Science Foundation under NSF MSPRF grant number 
1705008. This publication is funded in part by the Gordon and Betty Moore Foundation through grant GBMF8273 to Harvard University to support the work of the Black Hole Initiative. This publication was also made possible through the support of a grant from the John Templeton Foundation.

\newpage
\appendix

\bibliography{Refs}

\end{document}